\documentclass[longauth]{aa}
\usepackage{graphicx}
\usepackage[]{natbib}
\usepackage{amsmath}
\usepackage{xcolor}
\usepackage{amssymb}
\usepackage{soul}

\bibpunct{(}{)}{;}{a}{}{,} 

\def\kms{km~s$^{-1}$}
\def\teff{\textit{T}_{\text{eff}}}
\def\logg{\text{log}(\textit{g})}
\def\mh{[\text{M}/\text{H}]}
\def\feh{[\text{Fe}/\text{H}]}
\def\mgfe{[\text{Mg}/\text{Fe}]}
\def\sife{[\text{Si}/\text{Fe}]}
\def\alpham{[\alpha/\text{M}]}
\def\alphafe{[\alpha/\text{Fe}]}
\def\nife{[\text{Ni}/\text{Fe}]}
\def\alfe{[\text{Al}/\text{Fe}]}
\def\snr{S/N}

\def\cms{cm~s$^{-2}$}

\begin{document}

\title{The RAdial Velocity Experiment (RAVE):\\
Parameterisation of RAVE spectra based on \\
convolutional neural networks
\thanks{The catalogue of atmospheric parameters and chemical abundances 
presented in Section 10 is publicly available on the RAVE website: 
https://doi.org/10.17876/rave/dr.6/020}}

\author{G. Guiglion \inst{1}, 
G. Matijevi\v c \inst{1}, 
A. B. A. Queiroz \inst{1}, 
M. Valentini \inst{1}, 
M. Steinmetz \inst{1},
C. Chiappini \inst{1}, 
E. K.\ Grebel \inst{2}, 
P. J. McMillan \inst{3}, 
G. Kordopatis \inst{4}, 
A. Kunder \inst{5}, 
T. Zwitter \inst{6}, 
A. Khalatyan \inst{1}, 
F. Anders \inst{7,1}, 
H. Enke \inst{1}, 
I. Minchev \inst{1}, 
G. Monari \inst{8,1}, 
R. F.G.\ Wyse \inst{9,10}, 
O. Bienaym\'e \inst{8}, 
J. Bland-Hawthorn \inst{11}, 
B. K. Gibson \inst{12}, 
J. F. Navarro \inst{13}, 
Q. Parker \inst{14,15}, 
W. Reid \inst{16,17}, 
G. M. Seabroke \inst{18}, 
A. Siebert \inst{8}}

\authorrunning{G. Guiglion et al.}
\titlerunning{Parameterisation of RAVE spectra based on Convolutional Neural-Network}

\institute{Leibniz-Institut f{\"u}r Astrophysik Potsdam (AIP), An der Sternwarte 16, 14482 Potsdam, Germany \\ \email{gguiglion@aip.de}
\and{Astronomisches Rechen-Institut, Zentrum f\"ur Astronomie der Universit\"at Heidelberg, M\"onchhofstr.\ 12--14, 69120 Heidelberg, Germany}
\and{Lund Observatory, Department of Astronomy and Theoretical Physics, Lund University, Box 43, 22100 Lund, Sweden} 
\and{Universit{\'e} C{\^o}te d'Azur, Observatoire de la C{\^o}te d'Azur, CNRS, Laboratoire Lagrange, France} 
\and{Saint Martin's University, 5000 Abbey Way SE, Lacey, WA, 98503, USA} 
\and{University of Ljubljana, Faculty of Mathematics and Physics, Jadranska 19, SI-1000 Ljubljana, Slovenia} 
\and{Institut de Ci\`encies del Cosmos, Universitat de Barcelona (IEEC-UB), Mart\'i i Franqu\`es 1, 08028 Barcelona, Spain} 
\and{Observatoire astronomique de Strasbourg, Universit\'e de Strasbourg, CNRS, 11 rue de l'Universit\'e, F-67000 Strasbourg, France} 
\and{The Johns Hopkins University, Department of Physics and Astronomy, 3400 N. Charles Street, Baltimore, MD 21218, USA} 
\and{Kavli Institute for Theoretical Physics, University of California, Santa Barbara, CA 93106, USA} 
\and{Sydney Institute for Astronomy, School of Physics, The University of Sydney, NSW 2006, Australia} 
\and{E.A. Milne Centre for Astrophysics, University of Hull, Hull, HU6 7RX, United Kingdom} 
\and{Department of Physics and Astronomy, University of Victoria, Victoria, BC, Canada V8P5C2} 
\and{CYM Physics Building, The University of Hong Kong, Pokfulam, Hong Kong SAR, PRC} 
\and{The Laboratory for Space Research, Hong Kong University, Cyberport 4, Hong Kong SAR, PRC}  
\and{Department of Physics and Astronomy, Macquarie University, Sydney, NSW 2109, Australia} 
\and{Western Sydney University, Locked bag 1797, Penrith South, NSW 2751, Australia} 
\and{Mullard Space Science Laboratory, University College London, Holmbury St Mary, Dorking, RH5 6NT, UK}}

\date{Received 27/04/2020 ; accepted 23/09/2020}

\abstract{Data-driven methods play an increasingly important 
role in the field of astrophysics. In the context of large spectroscopic 
surveys of stars, data-driven methods are key in deducing physical parameters for 
millions of spectra in a short time. Convolutional neural networks (CNNs) 
enable us to connect observables (e.g. spectra, stellar 
magnitudes) to physical properties (atmospheric parameters, 
chemical abundances, or labels in general).}
{We test whether it is possible to transfer the labels 
derived from a high-resolution stellar survey 
to intermediate-resolution spectra of another survey by using a CNN.}
{We trained a CNN, adopting stellar atmospheric 
parameters and chemical abundances from APOGEE DR16 (resolution $R=22500$) 
data as training set labels. As input,  we used parts of the 
intermediate-resolution RAVE DR6 spectra ($R\sim7\,500$) overlapping 
with the APOGEE DR16 data as well as broad-band ALL\_WISE and 2MASS 
photometry, together with Gaia DR2 photometry and parallaxes.}
{We derived precise atmospheric parameters $\teff$, $\logg$, and $\mh,$ along with the 
chemical abundances of $\feh$, $\alpham$, $\mgfe$, $\sife$, 
$\alfe$, and $\nife$ for $420\,165$ RAVE spectra. The precision 
typically amounts to 60\,K in $\teff$, 0.06 in $\logg$ and 
0.02-0.04 dex for individual chemical abundances. Incorporating photometry 
and astrometry as additional constraints substantially improves the results 
in terms of the accuracy and precision of the derived labels, as long as we 
operate in those parts of the parameter space that are well-covered by the training 
sample. Scientific validation confirms the robustness of the CNN results. 
We provide a catalogue of CNN-trained atmospheric parameters and abundances along 
with their uncertainties for $420\,165$ stars in the RAVE survey.}
{CNN-based methods provide a powerful way to combine
 spectroscopic, photometric, and astrometric data 
without the need to apply any priors in the form of stellar evolutionary models. 
The developed procedure can extend the scientific 
output of RAVE spectra beyond DR6 to ongoing and planned surveys such as 
Gaia RVS, 4MOST, and WEAVE. We call on the community to place a particular collective emphasis and on efforts to create unbiased training samples 
for such future spectroscopic surveys.}


\maketitle


\section{Introduction}

Stellar chemical abundances are key tracers of the star formation history 
of the Milky Way and they are indicators of the timing of successive star 
formation events. The relative chemical abundances of stars thus 
allow us to disentangle stellar populations and to put constraints 
on the nucleosynthetic origin of the respective elements 
\cite{yoshii1981, freeman2002}. It allows us to constrain the composition 
of the gas cloud from which a star was 
formed and the variations of the initial mass function, 
particularly at the high-mass end \citep{wyse1988, matteucci1989}. 
However, in order to perform this 
exercise on the scale of the Galaxy, it is necessary to observe and reduce 
spectra for some hundreds of thousands of long-lived stars 
that are representative of the broad kinematic, chemical, and age distributions 
of Galactic populations \citep{hayden2015, buder2019}.

Over the last two decades, multiple efforts have been undertaken to 
provide the community with high-quality stellar spectra, largely drawn from 
dedicated spectroscopic surveys. The RAdial Velocity Experiment (RAVE) was 
the first systematic spectroscopic Galactic archaeology survey 
\citep{RAVE,steinmetz2020a}, targeting half a million stars. 
While the initial aim was to measure radial 
velocities of stars \citep{steinmetz2006}, RAVE data processing was later extended to 
include stellar atmospheric parameters \citep{zwitter2008, kordopatis2013}, 
chemical abundances \citep{boeche2011, steinmetz2020b}, and Gaia proper motions 
\citep{kunder2017}, thus enabling chemo-dynamical applications 
\citep{ruchti2010, ruchti2011, boeche2013a, boeche2013b, 
boeche2014, minchev_2014, kordopatis2015, antoja2017, minchev2019}. 
Together with RAVE, the Geneva-Copenhagen survey (GCS, \citealt{nordstrom2004}) 
yielded pioneering work in the comprehension of our Galaxy, solely based on $\sim17\,000$ nearby stars. The RAVE and GCS surveys were followed 
by numerous spectroscopic surveys with a broad variety of spectral resolving power. 
The Sloan Extension for Galactic Understanding and Exploration survey
(SEGUE, \citealt{yanny2009}) obtained roughly $240\,000$ 
low-resolution spectra (R=1\,800). The Gaia-ESO survey carried out a high-resolution 
investigation of $10^5$ stars, based on the UVES (Ultraviolet and Visual 
Echelle Spectrograph, R=$48\,000$) and GIRAFFE (R=$16\,000$) spectrographs 
of the Very Large Telescope (VLT, \citealt{GES}). At a lower resolution (R=$1\,800$), the 
ongoing Large sky Area Multi-Object Fibre Spectroscopic Telescope (LAMOST) observed 
about one million stars in the northern hemisphere \citep{zhang2019}. The 
ongoing Apache Point Observatory Galactic Evolution Experiment (APOGEE) just 
released their Data Release 16 \citep{apogeedr16, jonsson2020}. This survey observed 
$\sim400\,000$ stars in both hemispheres using a high-resolution 
near-infrared spectrograph ($R\sim22\,500$).  The Galactic ArchaeoLogy with HERMES project (GALAH), an ongoing survey dedicated to chemical tagging, 
has targeted nearly $350\,000$ stars at high 
resolution ($R\sim28\,000$, \citealt{buder2018}) to provide detailed 
chemical abundances. A common feature of all these endeavors is that automated and eventually unsupervised 
data reductions and parameter determination algorithms have to be employed, owing to 
the sheer number of spectra.

In the near future, the WHT Enhanced Area Velocity Explorer (WEAVE, \citealt{WEAVE}) 
and the 4-metre Multi-Object Spectroscopic Telescope (4MOST, \citealt{4MOST}) 
will deliver intermediate and high-resolution observations of several 
millions of stars (see \citealt{chiappini2019} and 
\citealt{bensby2019} for details on the 4MOST low- and 
high-resolution surveys of the bulge and discs, respectively). The need for automatic 
and fast software for the parameterisation of stellar spectra 
will become even greater.

To derive atmospheric parameters and chemical abundances, 
standard pipelines usually compare spectral models to observations, 
either localised around selected spectral lines or, alternatively, 
over a broader wavelength range. Methods range from the curve-of-growth 
fitting of spectral lines (e.g. \citealt{boeche2011}, 
SP\_Ace \citealt{boeche2018}), on-the-fly spectrum syntheses 
such as Spectroscopy Made Easy  (SME, \citealt{valenti1996}), 
on-the-fly flux ratios 
such as A Tool for HOmogenizing Stellar parameters (ATHOS, \citealt{hanke2018}), 
or a comparison based on a synthetic spectra grid 
(FERRE \citealt{allendeprieto206}; MATISSE \citealt{matisse}; 
GAUGUIN \citealt{GAUGUIN12, guiglion2016}). These 
methods have shown their efficiency in deriving precise and 
accurate abundances \citep{jofre2019} for various spectral ranges and spectral 
resolutions in the context of the major current spectoscopic 
surveys, such as the Gaia-ESO Survey,  
APOGEE, GALAH, and RAVE. 
These families of standard pipelines are essential because they are 
based on the physics of stellar interiors, deriving atmospheric parameters and chemical 
abundances that can be used as stellar labels in the context of 
data-driven methods.

Indeed, data-driven approaches have started to play an important role 
in estimating these stellar labels. 
Such methods transfer the knowledge from a reference set of data, 
so-called training samples, to infer stellar labels. 

The Cannon \citep{ness2015} is one of the pioneering data-driven analysis 
packages and its reliability was demonstrated through applications to 
spectroscopic surveys such as APOGEE and RAVE \citep{casey2016,casey2017}. 
The Payne \citep{ting2019} recently demonstrated that it is possible to 
couple stellar spectra modeling and a model-driven approach 
to reflect stellar labels. We note that the Cannon uses observed spectra 
(with the same set-up, but higher signal-to-noise than the survey) as the training 
data, whereas the Payne uses synthetic spectra as its training set. 

A few recent studies have used convolutional neural networks (CNNs)
to infer atmospheric parameters and chemical abundances from 
high-resolution spectra. \citet{leung2019a} 
derived 22 stellar parameters and chemical abundances 
based on APOGEE DR14 spectra and labels, utilising 
their astroNN tool and purely observational data. 
On the other hand, \citet{fabbro2018} developed 
the StarNet pipeline, which is based on a CNN 
and an input synthetic spectra grid. They applied their StarNet 
to high-resolution data of APOGEE and, more recently, to
Gaia-ESO Survey UVES spectra \citep{bialek2019}. 
\citet{zhang2019} used StarNet to estimate 
atmospheric parameters and chemical abundances 
of LAMOST spectra, based on APOGEE results.

Combining spectoscopy and photometry has been 
explored by \citet{schonrich2014} with physical modelling and a 
Bayesian approach on SEGUE data. The goal of the present paper is to show that 
a CNN-based approach can 
be employed for an efficient transfer of stellar labels from high 
resolution spectra to intermediate-resolution spectra. This is 
done in conjunction with additional observables in the form of 
stellar magnitudes and parallaxes. 
We aim to derive atmospheric parameters and chemical abundances from 
intermediate-resolution RAVE DR6 spectra, based on a training sample 
of common stars with higher resolution APOGEE DR16 \citep{apogeedr16} spectra. 
We also show that using broad-band 
infrared photometry and parallax measurements as an extra set of 
constraints during the training phase improves the atmospheric parameters 
considerably. This study represents a complementary 
approach to the RAVE project's main parameter pipeline, 
and enhances the scientific output of the RAVE spectra. 
This work also has a good synergy with the next full
Gaia release (Gaia DR3), which will provide spectra from 
the Radial Velocity Spectrometer (RVS), which are 
very similar to RAVE spectra in terms of wavelength coverage and resolution.

The paper is laid out as follows. In Sect.~\ref{observations}, we present the data 
we used to build the training sample. In Sect.~\ref{training_phase}, 
we present the main features of the CNN and 
provide details of the training phase. In 
Sect.~\ref{prediction_observed_sample}, we deduce the
atmospheric parameters and chemical abundances 
for more than $420\,000$ RAVE spectra, with 
the error budget treated in Sect.~\ref{errors_section}. 
In Sect.~\ref{validation}, we compare and 
validate the tests with respect to external data sets. 
The scientific verification for some typical Galactic 
archaeology applications is presented in 
Sect.~\ref{science_verification}.


\begin{figure}[ht]
\centering
\includegraphics[width=0.8\linewidth]{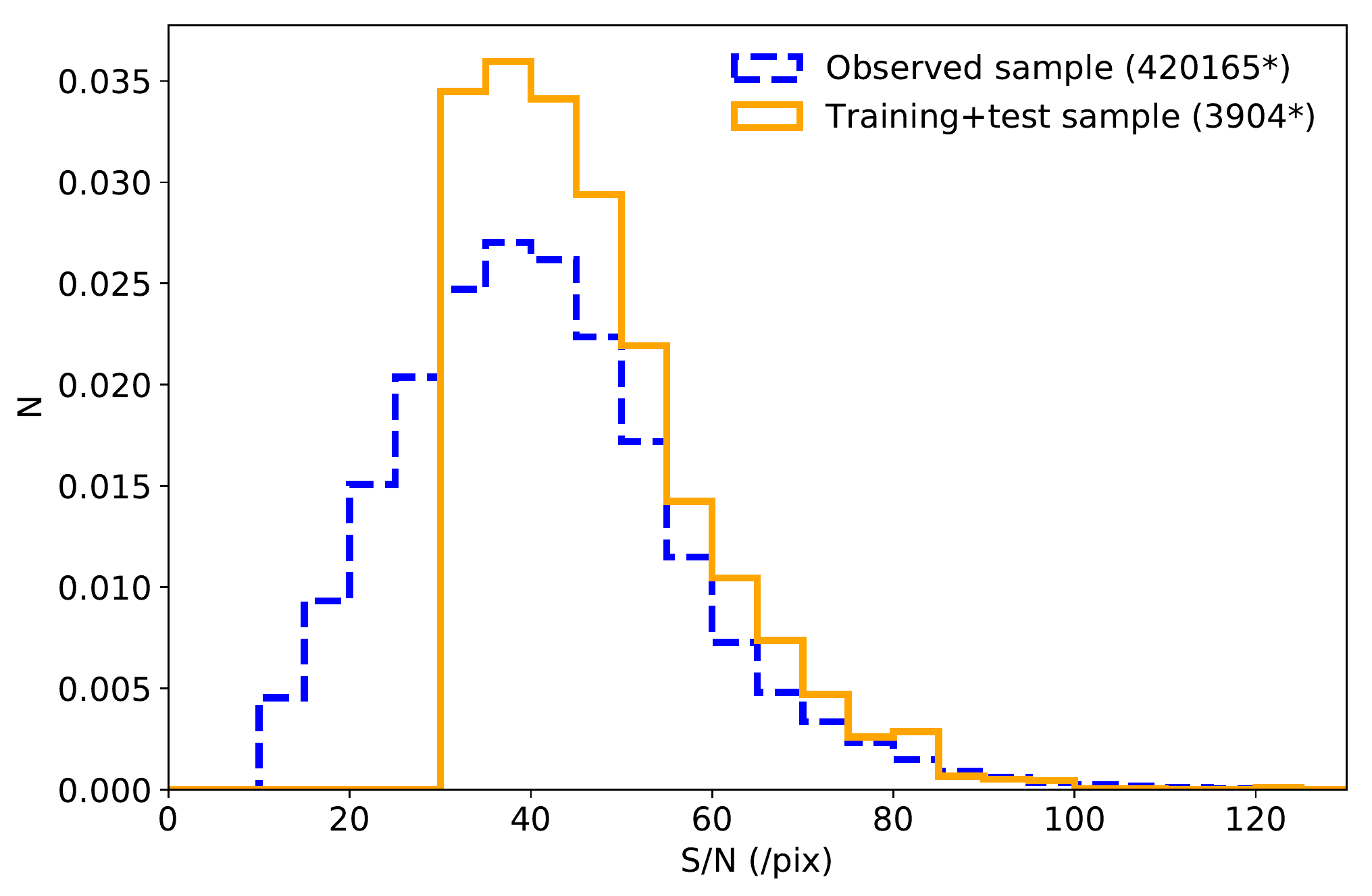}
\caption{\label{snr_distribution}Normalised distribution 
of \snr\  of RAVE DR6 spectra in the observed sample
(blue dashed line, $420\,165$ stars) and in the training and test samples 
(solid orange line $3\,904$ stars in common between RAVE DR6 and APOGEE DR16).}
\end{figure}

\begin{figure*}[ht]
\centering
\includegraphics[width=1.0\linewidth]{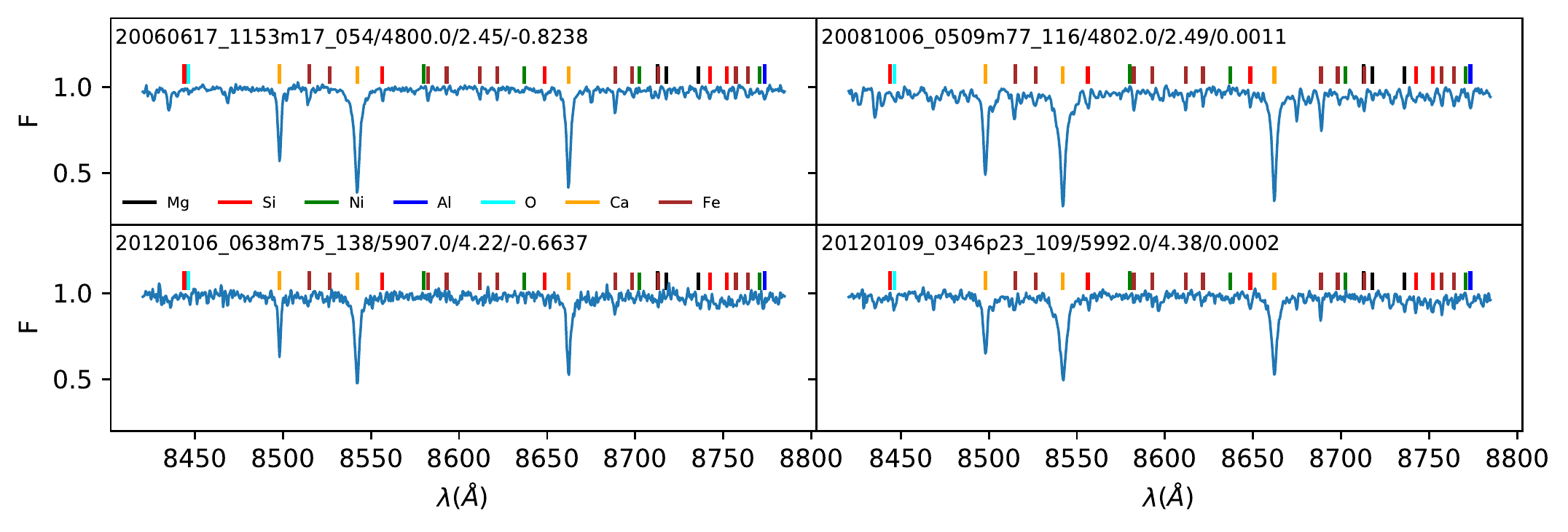}
\caption{\label{rave_spectra}Example of four typical metal-poor 
and metal-rich dwarfs and giants RAVE spectra from the training sample. 
The RAVE\_OBS\_IDs and the atmospheric parameters are indicated in 
the top left corner of each panel. Apart from the prominent 
CaII triplet lines, the RAVE spectra also 
show a variety of more subtle spectral features (main chemical 
abundance diagnostic lines are over-plotted).}
\end{figure*}

\section{Training sample}\label{observations}

One of the main goals of this study is to show that
high-resolution stellar labels can be used to deduce atmospheric parameters and 
chemical abundances from lower resolution spectra. For this purpose, we 
need to build a training set that contains the labels - namely, the parameters we 
wish to derive (in our case, the atmospheric parameters and chemical abundances) 
and the observables (the spectra and photometric measurements). 
Here, we chose to work with labels provided by the APOGEE survey and 
observables from the RAVE spectroscopic survey, complemented by 2MASS \citep{2MASS}, 
Gaia DR2 \citep{babusiaux2018}, and ALL\_WISE photometry \citep{ALLWISE} 
as well as Gaia DR2 astrometry \citep{lindegren2018}. 
Since the APOGEE survey, on average, offers higher resolution and higher 
signal-to-noise ratios (S/N) than the RAVE survey, 
we can translate the higher quality of the derived APOGEE labels to RAVE.

We take advantage of the latest release of APOGEE, namely, DR16 
\citep{apogeedr16, jonsson2020}, which provides high-quality atmospheric parameters and chemical abundances. The
APOGEE spectra are taken at near-infrared wavelengths with high resolution 
($R=22\,500$ and $\lambda \in [1.5-1.7]\mu$m).
The RAVE DR6 spectra have a spectral resolving power of $R\sim7\,500$. We 
re-sampled the spectra to a common wavelength coverage of 
$\lambda \in [8\,420-8\,780]$\AA, with equally spaced 0.4\AA pixels.

We performed a cross-match based on the Gaia DR2 Source IDs 
between the $518\,387$ RAVE DR6 observations and the $473\,307$ 
observations of APOGEE DR16, resulting in a sample of $\sim7\,000$ 
sources. In order to build a clean and coherent training sample 
based on APOGEE stellar labels and RAVE spectra, we 
cleaned this cross-matched sample in the following way.

Firstly, we required that a given star has available measurements of 
$\teff$, $\logg$, $\mh$, 
$\feh$, $\alpham$, $\mgfe$, $\sife$, $\alfe$, $\nife$ and their 
associated errors in the APOGEE set. We excluded parameters for stars 
with \snr\_APOGEE<60 (per pixel) and required the 
ASPCAP\footnote{APOGEE Stellar Parameter and Chemical 
Abundance Pipeline, \citealt{garciaperez2016}} 
parameterisation flag to be aspcap\_flag=0. The mean APOGEE $\snr$ 
of the sample is 420 per pixel. We filtered stars with a bad flag on 
chemical abundances, that is, selecting only X\_Fe\_FLAG=0.
The ASPCAP pipeline uses spectral templates 
for matching any observations. Such procedures can lead to 
systematics (due for example to incomplete line list) that 
will be transferred by the CNN.

Secondly, we adopted the normalised, radial-velocity-corrected spectra 
from the DR6 of RAVE. The normalisation has been performed by the 
RAVE survey, with an iterative second-order polynomial fitting procedure 
(see \citealt{steinmetz2020a} for more details). We required that the 
spectra have at least 
\snr>30 per pixel. We excluded spectra showing signs of binarity or continuum issues 
('c', 'b', and 'w' according to the RAVE DR6 classification scheme, 
see \citealt{steinmetz2020a}).
    
Finally, as detailed in Sect~\ref{section_photometry}, we used absolute magnitudes 
during the training process. We required that a star have an apparent 
magnitude available in the 2MASS $JHK_s$, ALL\_WISE W1\&2 pass-bands, 
and Gaia DR2 $G$, $G_{BP}$  $G_{RP}$, and Gaia parallaxes (with parallax 
errors $e_p<15\%$). As such apparent magnitudes can suffer from dust 
extinction, we took advantage of the StarHorse catalogue, which provides 
improved extinction measurements based on RAVE and Gaia DR2 data 
(\citealt{queiroz2019}, see also \citealt{santiago2016,queiroz2018} 
for details on the method). We required that all spectra have 
an available StarHorse extinction  ($A_V$).

The resulting common sample between APOGEE DR16 and RAVE DR6 
consists of $3\,904$ high-quality RAVE spectra and high-quality 
atmospheric parameters and chemical abundances. 
The RAVE \snr\ distribution of this sample 
is presented in \figurename~\ref{snr_distribution}. 
We carefully checked the spectra of the 
$3\,904$ stars of the sample in order to reject any misclassified stars, 
possibly having a very low \snr. Some examples of RAVE spectra are presented in 
\figurename~\ref{rave_spectra}, for typical metal-poor and metal-rich 
dwarfs and giants. Kiel diagrams of the $3\,904$ targets 
based on APOGEE DR16 parameters are presented in the left panels of 
\figurename~\ref{training_test_sample_kiel_diagrams}.


\section{Training the network }\label{training_phase}

An artificial neural network consists of several layers of neurons 
that are interconnected. The strength of connections between the 
neurons is governed by the weight of each connection. This feature 
enables the network to translate the input data vector to the desired 
output labels. The weights need to be set to values with which the 
translation becomes meaningful. For example, a stellar spectrum sampled 
at $N$ wavelength points is fed into a neural network with $N$ input 
neurons and the network produces an output in the form of, for instance, effective 
temperature. The setting of weights is done through training. This is a 
process of passing a limited set of data vectors through the network and 
gradually adjusting the weights so that the output matches the 
pre-determined labels of the data vectors. Each passing of the input 
data and adjustment of the weights is known as an epoch and many epochs 
are needed 
to successfully train the network. Once this is done, a new data vector 
can be passed through the network and we obtain its label as a result.
We note that convergence is reached when the error from the model 
has been sufficiently minimised. In theory, it could be the case that 
the desired level of error minimisation is never reached and 
the network would run indefinitely. We detail in 
Sect.~\ref{sub_sect_training_phase} how we stop the training in such cases.

\subsection{Architecture of the CNN}

\begin{figure}[h]
\centering
\includegraphics[width=0.9\linewidth]{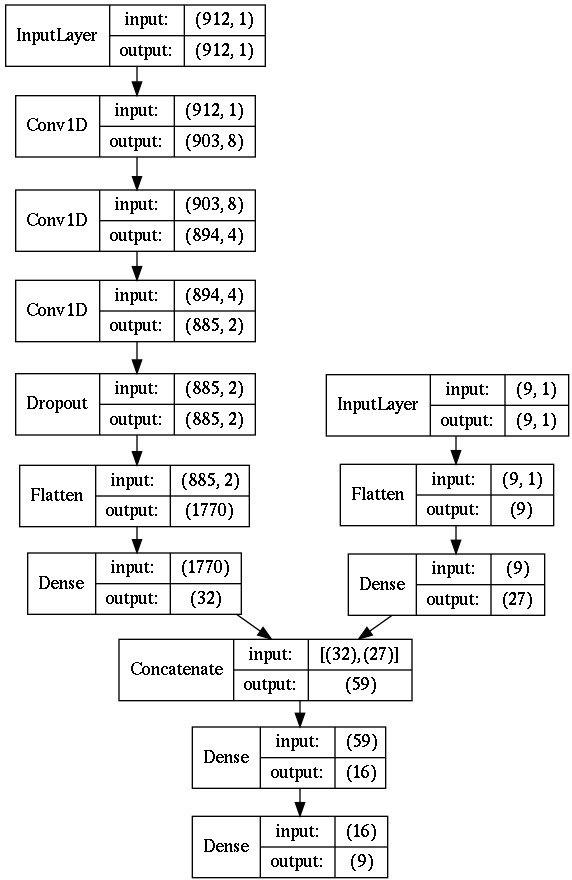}
\caption{\label{model_plot}Representation of the architecture 
of the Keras model used in this study. The input layer (the spectra) 
is passed through three steps of convolution (Conv1D). 
Then, we randomly drop 20\% of the neurons at each epoch of training 
with the dropout layer in order to prevent overfitting. We then 
flatten the output for the next dense layer (also called 
a fully connected layer). As an additional input, we include eight 
absolute magnitudes (2MASS $JHK_s$, ALL\_WISE W1\&2, 
and Gaia DR2 $G$, $G_{BP}$  $G_{RP}$ passbands) and one $A_V$ correction 
(input layer with shape of 9). We concatenate it to the main part of 
the network in the form of 27 neurons. The fully connected part of the 
network is then composed of two dense layers. The output is an array of 
nine parameters (atmospheric parameters and six chemical abundances).}
\end{figure}

In \figurename~\ref{model_plot}, we present the architecture of the 
neural network used in this study. It is composed of three 
convolutional layers and two fully connected or dense layers. In the subsection below, 
we justify the reason for utilizing these features. We used the 
Keras python libraries for coding the network 
\citep{chollet2015keras}. The stellar labels are 
normalised, ranging from 0 to 1 by using a Min-max normalisation.

\subsubsection{Convolution and dense layers}

Convolution layers are the key for detecting patterns and 
features in images (see e.g. \citealt{ciresan2011} for 
more details on this topic).
In the present study, we work with one-dimensional normalised stellar spectra
characterised by spectral line features. Such spectral features are 
indicators of the physical properties of the stars (temperature, gravity, 
chemical composition, etc). The ability to capture the relations between 
the different wavelength pixels in a spectrum, as opposed to treating them 
as independent entities, 
is the key to improved performance and this is provided by the convolutional layers.

To understand the impact 
of these types of layers we experimented with training our network 
with and without the convolution stage. In comparison to the network 
with the convolution stage, the training phase to find a stable solution 
is three to four times longer for the non-convolutional network. In addition 
to a lengthier training period, the output parameters 
are not recovered as precisely. This applies in particular to  chemical abundances. 
After trying many different layouts, we adopted a network with three 
convolution layers that contain eight, four, and two filters, respectively (as 
shown in \figurename~\ref{model_plot}).
We adopted a kernel size of ten pixels for 
all three layers. Tests revealed that kernel sizes between 5 to 20 pixels 
tend to extract features efficiently. Much larger kernels (>40 pixels) 
degrade the performance. \footnote{We note that the performance of the 
network is not impacted by a random uniform shift of a spectrum's continuum 
of up to 20\% in flux. This implies that 
that the network does not extract information from the overall level of the continuum.}

Between the convolution layers and the fully connected part of the network, 
we used a dropout layer that ensures  that a certain randomly chosen fraction 
of the neurons are not used at each of the epochs during the training phase. This 
type of regularisation prevents the over-fitting the network and also
prevents the algorithm from relying on a smaller part of the network alone. 
We tested a range of fractions from 10 to 30\%, with no major change in the 
training phase. We adopted 20\% for the final analysis.

The fully connected layers (also called 'dense' layers) following the 
convolutional stage are a more common type of neural network layers. 
They receive the output of the 
convolutional stage in the form of learned spectral features and convert 
them to the output labels (atmospheric parameters, abundances) that are sought. We must allow enough complexity in the network at this stage for 
it to be able to model the non-linear relations between features and labels.
We adopted the Leaky Rectified Linear Units (Leaky ReLU) 
activation function instead of Rectified Linear Units (ReLU), allowing us to 
face the dead ReLU problem (i.e. null or negative ReLU leading to no learning 
in the layers below the dead ReLU). We are, thus, less sensitive to the initialisation 
of the network.

\subsubsection{Initialisers and cost function}

The weights of the CNN must be 
initialised prior to the training. The choice of how we initialise them 
can influence the performance of the network. 
We adopted the default initialiser for our convolution 
and dense layers, namely, the 'glorot\_uniform' and the default bias initialiser, 'zeros', 
meaning that the weights prior to training are drawn from a uniform distribution 
within a certain range.

To train the network, we need a cost function that evaluates how good 
the network's performance is at each iteration and which would also allow us to compute 
the gradient in the weight space so the difference between the output and 
pre-determined labels can be minimised. The choice of this function is important. 
We experimented with a simple mean-squared error loss-function and a 
negative log-likelihood criterion. Tests performed on the 
negative log-likelihood criterion revealed that such a criterion 
appears  to be inferior for our science case, and it adds too much complexity 
to the framework.

\subsubsection{Effect of noise in the training phase}

The training and test samples include in total $3\,904$ stars 
with $\snr>30$ per pixel. As a test, we constrained this range to $\snr>40$ 
($2\,529$ stars) and $\snr>50$ ($1\,289$ stars). With a lower number of stars, 
the performance naturally tends to degrade. We believe, however, that 
this lack in performance is only due to the fact that we have a limited 
common sample with APOGEE. In general, high $\snr$ data and 
sufficient statistics lead to a better training phase, but 
lower $\snr$ spectra also come with a higher degree of correlated 
noise, which the network is likely to learn.

As another check, we extended the $\snr$ range to 
$\snr>20$, $\snr>15$ and $\snr>10$ per pixel, leading to $4\,802$, $5\,023$, 
and $5\,136$ stars in the training sample. 
We concluded that including such low-$\snr$ data in the training 
phase tends to reduce the quality of the training and degrades 
the overall performance.

We tried to train a network with a sample composed only of stars 
with $\snr<30$, finding that no robust solution could be reached, 
probably owing to 
to the spectral information being too hidden by noise. Especially 
for the chemical abundances, the network is unable to reproduce 
the main Galactic trends and basically fits a straight line in the 
$\alpham$ versus $\mh$ plane instead of reproducing the $\alpha-$rich 
and $\alpha-$poor sequences. A similar finding also holds for other elements. 
Our conclusion is that an efficient training cannot be performed if 
only low $\snr$ stars are present in the training set.

We recommend that for future spectroscopic surveys particular attention 
should be given when defining the training sample $\snr$ range, 
because too low $\snr$ spectra lead to worse training and performance 
for the CNN.

\subsection{Feeding absolute magnitudes to the neural network}\label{section_photometry}

\begin{figure}[ht]
\centering
\includegraphics[width=1.0\linewidth]{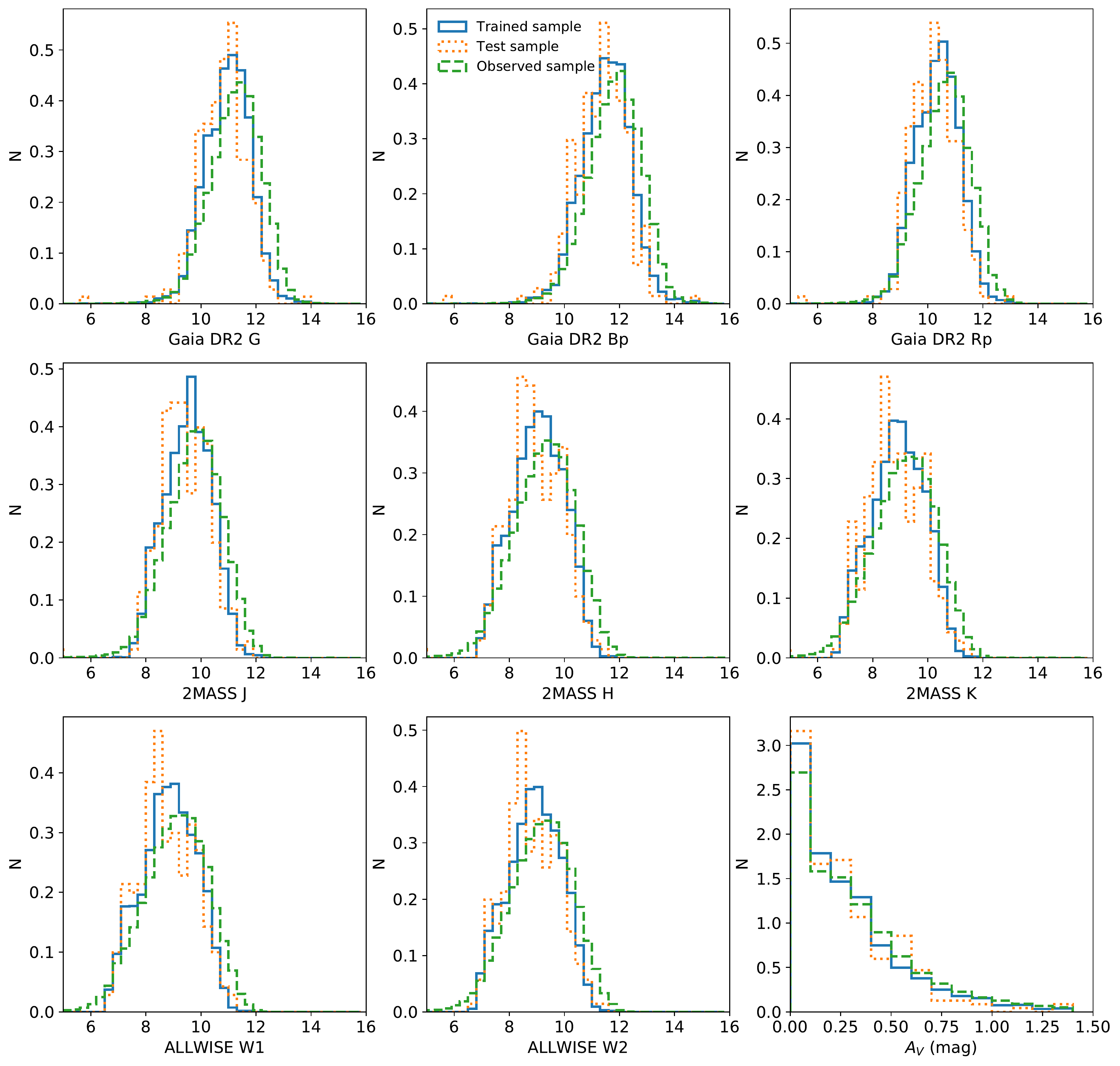}
\caption{\label{apparent_magnitudes_And_A_V_distributions}Normalised distribution of 
Gaia, 2MASS, ALL\_WISE apparent magnitudes and extinction ($A_V$) for the training 
sample (blue, solid), the test sample (orange, dotted), and the observed sample 
(green, dashed). Those magnitudes are converted to absolute magnitudes and are used during 
the training phase.}
\end{figure}

In addition to spectra, our input includes 
broad-band photometry. Absolute magnitudes provide strong 
constraints on the effective temperature and the surface 
gravity of a star. We adopted the 2MASS apparent magnitudes 
$m$ in the passbands $JHK_s$ (1.235, 1.662, and $2.159\,\mu$m, respectively), 
ALL\_WISE W1 and W2 pass-bands (3.4, and $4.6\,\mu$m) 
and Gaia DR2 $G_{BP}$ (328.3-671.4 nm), and $G_{RP}$ 
(629.6-1\,063.7 nm) and $G$ (332.1-1\,051.5 nm) bands, 
using the cross-matches provided in RAVE DR6 \citep{steinmetz2020a}. 
The distributions of these apparent magnitudes are shown in 
\figurename~\ref{apparent_magnitudes_And_A_V_distributions}. 

We computed absolute magnitudes, $M,$ using the
parallaxes ($p$) from the second data release of the Gaia satellite 
\citep{GaiaDR2}, using $M=m+5\times[\log_{10}(p)+1]$. We selected 
the best measurements for which we required 
the errors on the parallax, $e_p$, to be better that $20\%$ (96.5\% 
of the spectra of the initial cross-match with APOGEE DR16 fulfil 
this criterion). We discuss the performances of 
the CNN parameterisation for stars with parallax errors larger than 
$20\%$ in Appendix~\ref{section_parallax_errors}.

As stellar magnitudes can suffer from dust extinction 
even in the infrared passbands, we adopted the extinction correction 
$A_V$ from StarHorse (see \cite{queiroz2018, anders2019} for more details). 
The distributions of $A_V$ for the training, test, and observed sample 
are presented in \figurename~\ref{apparent_magnitudes_And_A_V_distributions}. 
We find that 78\% of our stars have an extinction lower than $A_V=0.5$~mag. 
Our tests found that stars with $A_V>0.8$ show a smaller error in 
$\teff$ by 20\,K if we include this correction.

Our choice to compute absolute magnitudes from parallaxes instead of, 
for example, StarHorse distances was motivated by the fact that we 
want to restrict our model dependency as much as possible. As a test, we computed 
absolute magnitudes using StarHorse distances, but no notable difference  
in the training was measured.

The eight absolute magnitudes and the extinction corrections were then added 
smoothly to the CNN architecture, 
directly in the fully connected part, as 27 neurons 
(see scheme in \figurename~\ref{model_plot}). We tested several 
layer sizes for this part: below 27, the performances tended 
to degrade and above 27, no further improvement was notable. 
We note that we did not directly apply the $A_V$ correction 
to the absolute magnitudes, thus leaving the network with more 
flexibility to learn from it.

It has been shown that Gaia DR2 astrometric measurements have 
small systematic errors, in particular, an 
offset of the parallax zero-point that varies across the sky.
This parallax zero-point offset is dependent on magnitude and colour 
\citep{lindegren2018, arenou2018}. This offset is roughly of 
the order of 50 $\mu\text{as}$. Following the way we compute 
our absolute magnitudes, this parallax offset translates into a shift 
of the order of 0.01 magnitudes. In the context of this study, 
this offset is negligible.
We refer the reader to Sect.~\ref{section_phot_vs_no_phot} for a discussion 
on the advantage of adding photometry during the training process.

\begin{figure}[ht]
\centering
\includegraphics[width=1.0\linewidth]{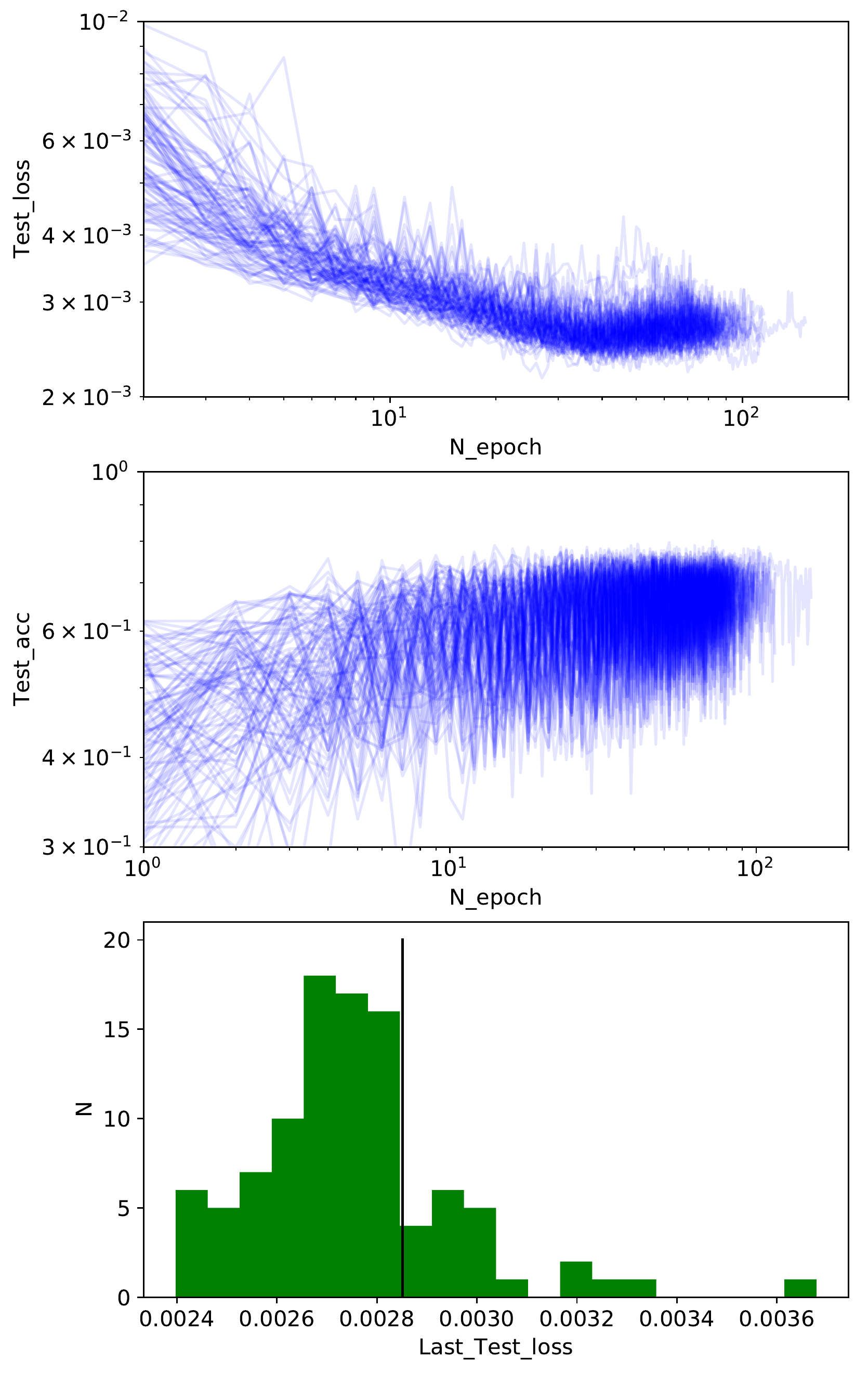}
\caption{\label{loss_figure}
Top: Value of the cost function for the test 
sample (Test\_loss) for the 100 CNN runs as a function of the epoch. Middle: 
Accuracy computed on the test sample (Test\_acc) as a function of the epoch. 
Bottom: Distribution of 100 values of Test\_loss after the training was completed. 
Vertical black line indicates the 80th percentile of the distribution.}
\end{figure}

\subsection{Training an ensemble of 100 CNNs}\label{sub_sect_training_phase}

From the quality cuts and selection process detailed above, 
our starting sample is thus composed of $3\,905$ stars, with stellar 
labels corresponding to atmospheric parameters and chemical abundances.
Before training the CNN, we split the data into 
a training sample and a test sample, as is a common practice in the 
machine-learning community. We adopted a fraction of 6\% for the test sample to retain a large the training sample. This led to $3\,669$ 
stars in the training sample and 235 stars in the test sample. We tested several 
test and training fractions, from 3 to 40\%, with no major 
difference in terms of training.
In order to provide stable results and errors, we built an ensemble 
of 100 trained CNNs, all of them initialised differently.
A similar method was recently used by \citet{bialek2019}.

One challenge while using a CNN is to stop the learning phase at the right time. 
The model can under-fit the training and test samples in case of insufficient training. 
On the other hand, in cases of over-fitting, the training sample will  perfectly fit
the model, but the performances on the test sample will degrade drastically 
(which is the main reason behind the training-test split). One solution is to stop the training phase 
when the performance on a validation dataset starts to degrade. In this context, 
we adopted the commonly used early-stop procedure. If after 40 epochs (the so 
called patience period), the solution does not improve, we stop the training. 
We tried different levels of patience, 
finding that 40 epochs provide the best compromise between final accuracy 
and computation time.

Typical curves of the cost functions 'Test\_loss' for the test sample are presented in 
\figurename~\ref{loss_figure} for the 100 runs, as well as the accuracy Test\_acc. 
It is clear that the training phase takes no more than 120 epochs. 
Training the CNN takes between 70 to 90 seconds per run. 
We can also see that the last value of the cost function of the test sample 
(Last\_Test\_loss) can vary from one run to another. We plot such values in the bottom 
panel of \figurename~\ref{loss_figure}. We excluded networks with 
too large a value of Last\_Test\_loss (everything inside of the lower 20th percentile 
of the distribution).\\


\begin{figure*}[h]
\centering
\includegraphics[width=0.3\linewidth]{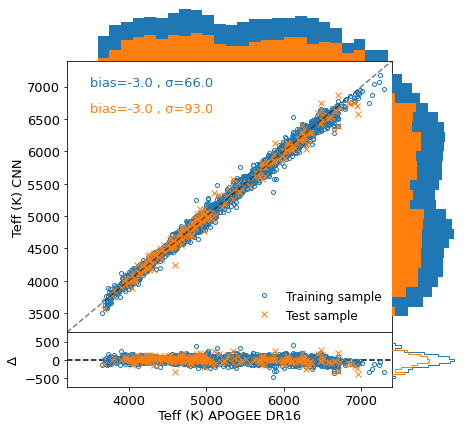}
\includegraphics[width=0.3\linewidth]{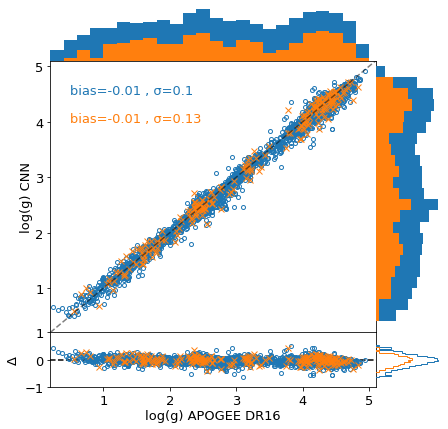}
\includegraphics[width=0.3\linewidth]{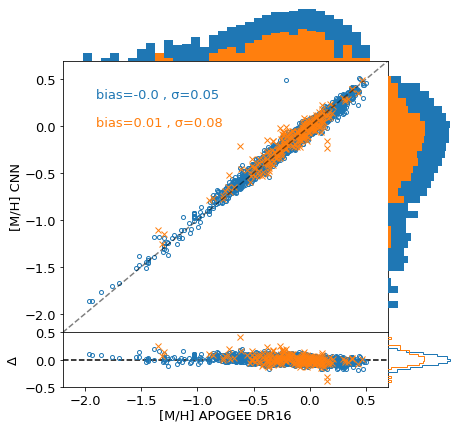}
\includegraphics[width=0.3\linewidth]{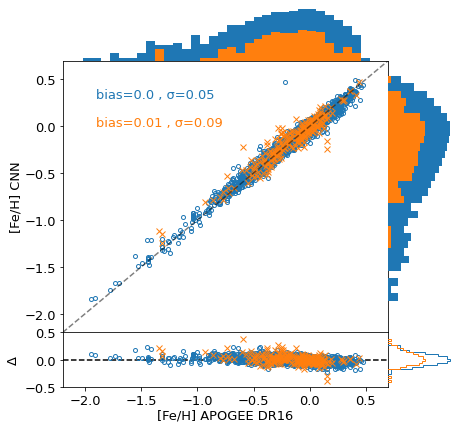}
\includegraphics[width=0.3\linewidth]{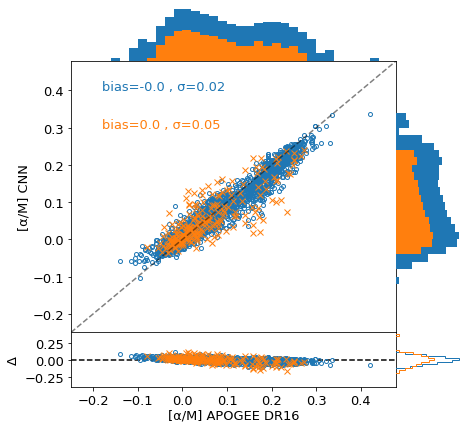}
\includegraphics[width=0.3\linewidth]{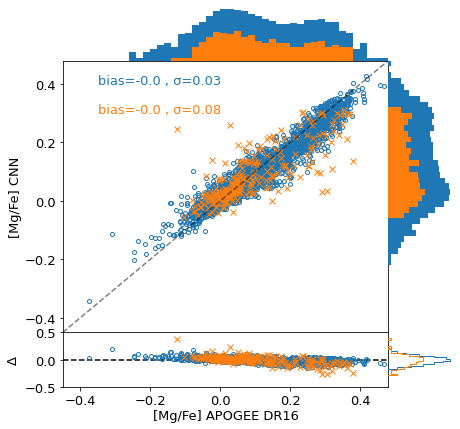}
\includegraphics[width=0.3\linewidth]{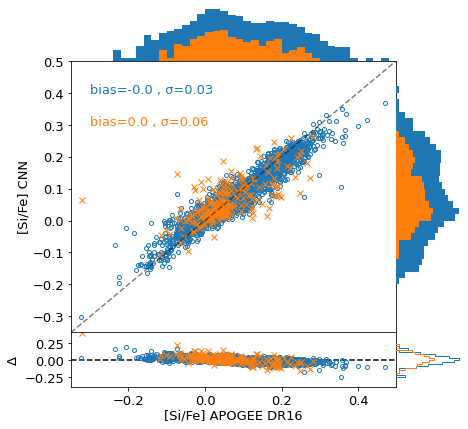}
\includegraphics[width=0.3\linewidth]{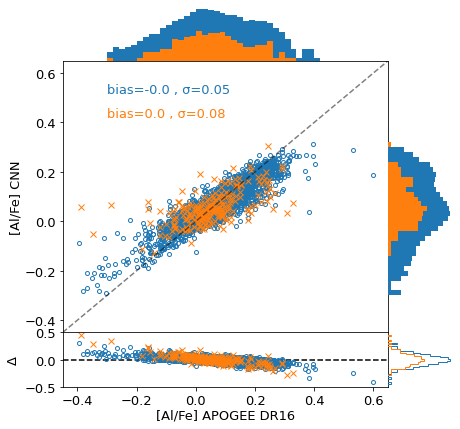}
\includegraphics[width=0.3\linewidth]{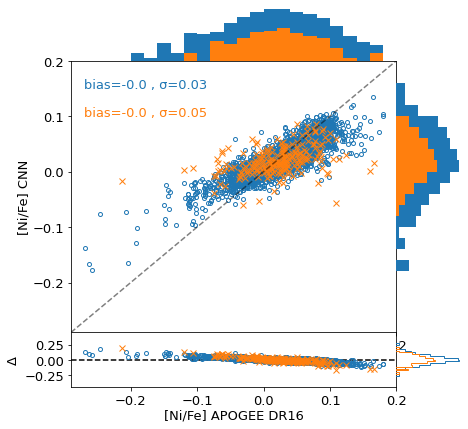}
\caption{\label{training_test_sample_one_to_one}One-to-one relation between 
the CNN trained labels (y-axis) and the input labels (x-axis, APOGEE DR16 data). 
The training sample is plotted with blue circles, while the test sample 
is shown with orange crosses. The x- and y-axis parameters 
are presented as histograms with a logarithmic scale. For each parameter, 
a typical mean difference and scatter are computed in both sets. 
We plotted the difference $\Delta$ between the CNN trained labels and 
the APOGEE DR16 input labels with the same symbols and colours, 
and its histogram with a logarithmic scale.}
\label{fig:fig}
\end{figure*}

\begin{figure*}[h]
\centering
\includegraphics[width=1.0\linewidth]{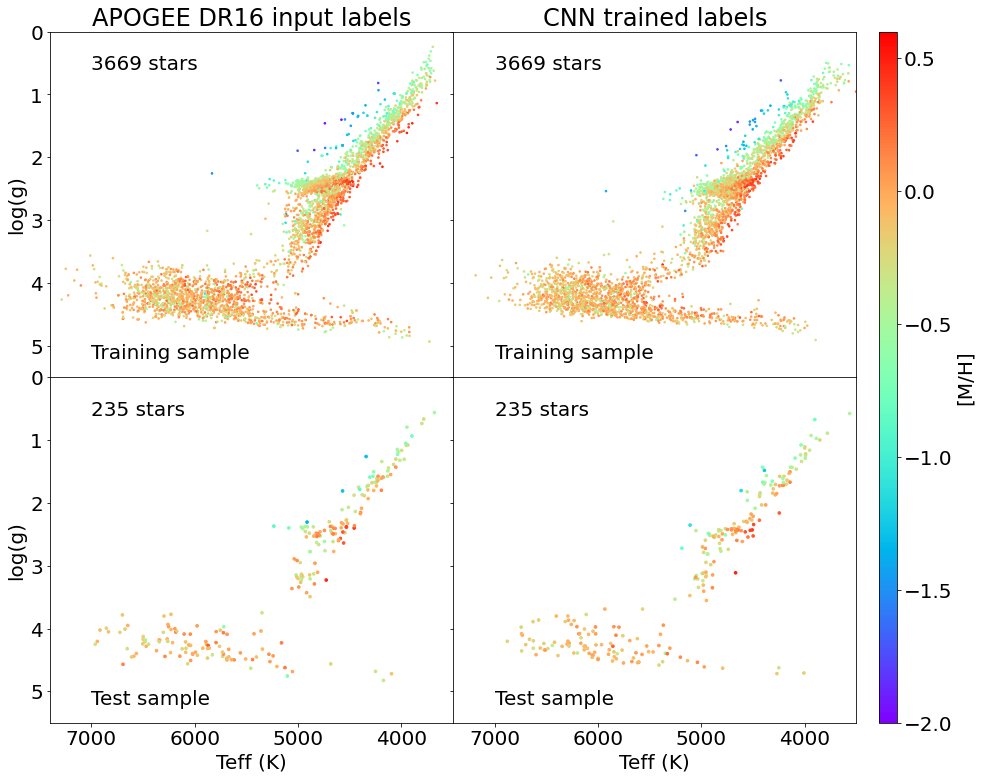}
\caption{\label{training_test_sample_kiel_diagrams}Top left: Kiel diagram 
of the APOGEE DR16 stars (used in the training sample), colour-coded with overall $\mh$. 
Top right: For the same stars, trained labels, averaged over 80 trained CNN. 
Bottom left: APOGEE DR16 parameters of the test sample. Bottom right: 
Trained labels, averaged over 80 trained CNN, for the same test sample. 
The right panels correspond to what the network learns from 
the APOGEE parameters (left panels).}
\end{figure*}

\begin{figure}[h]
\centering
\includegraphics[width=1.0\linewidth]{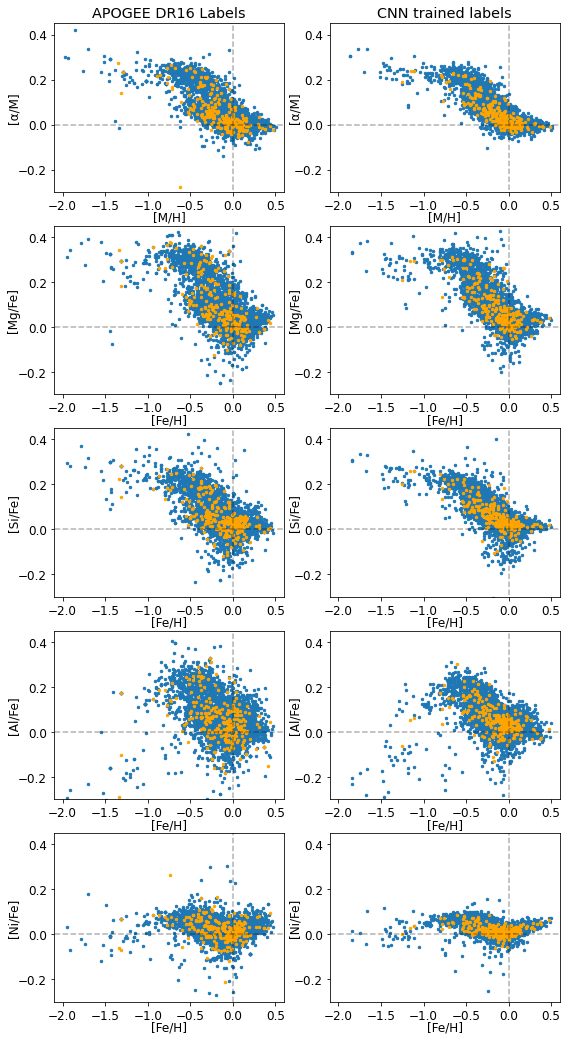}
\caption{\label{training_sample_abundances}Left panels: Abundance patterns 
of the APOGEE DR16 labels used as input for our CNN, for the training sample (blue) 
and for the test sample (orange). Right panels: Abundance patterns 
of the averaged labels trained over 80 CNNs.}
\end{figure}

\begin{figure*}[h]
\centering
\includegraphics[width=1.0\linewidth]{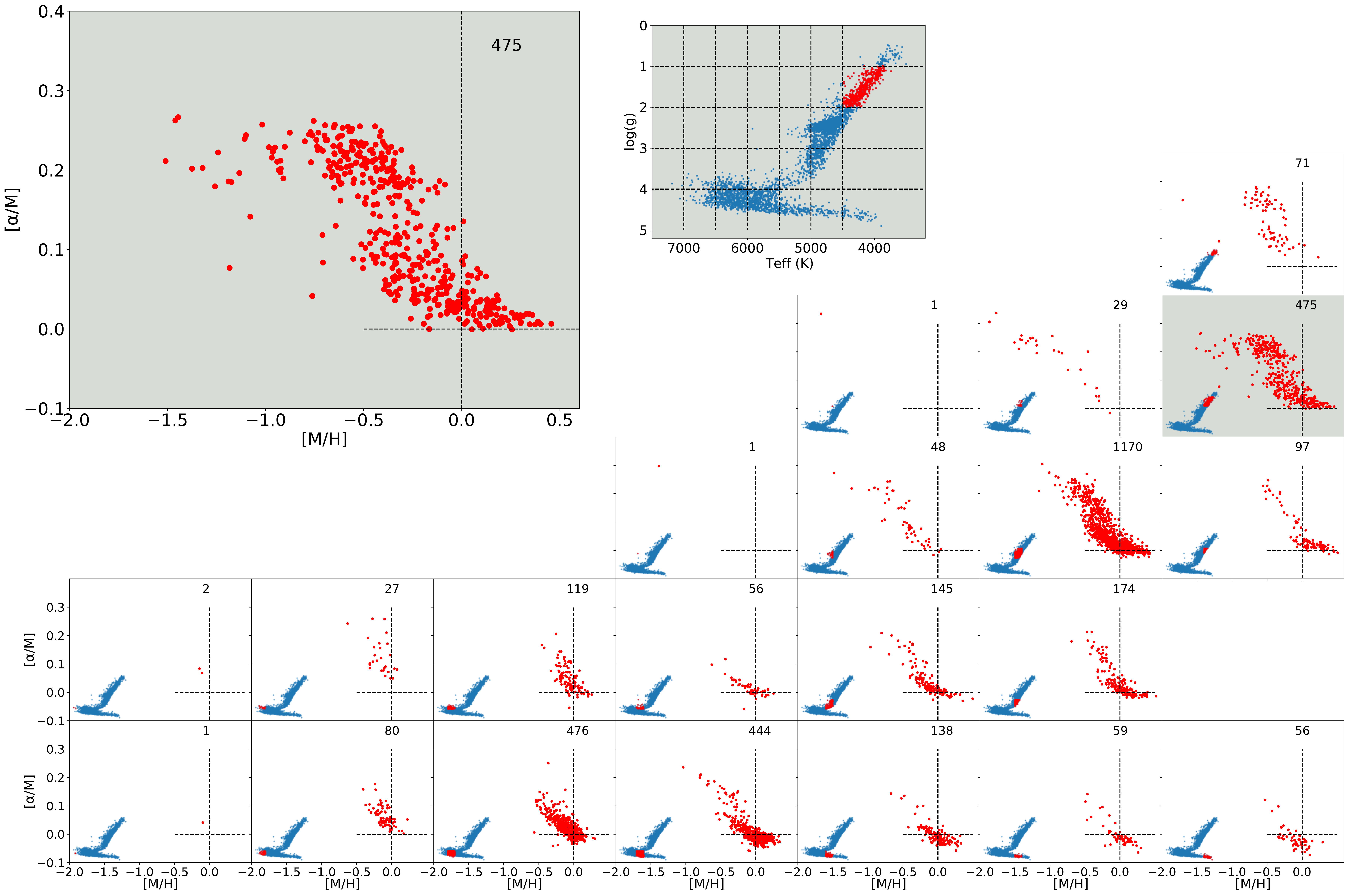}
\caption{\label{training_sample_alphaM_Kiel}Averaged trained abundance 
patterns $\alpham$ vs. $\mh$ for the training sample (red dots). 
Trends are shown for sub-samples in $\teff$ (500\,K bins) and $\logg$ (1 dex bins) 
shown as in-set Kiel diagrams, where the overall stellar distribution is plotted 
in blue with the selected subsample highlighted in red. 
The number of stars is indicated in the top right corner of each panel.}
\end{figure*}

\subsection{Result of the training}\label{results_training}

In \figurename~\ref{training_test_sample_one_to_one}, we compare the 
labels used as input of our CNN (from APOGEE DR16)
to those trained by the network (averaged over the 80 runs). The network is able to 
learn a significant amount of information about the main atmospheric 
parameters $\teff$, $\logg$, $\mh$ as well as $\feh$. No obvious 
systematic trends are visible 
while the dispersion is low, for both training and test samples. 
The mappings of $\teff$ and $\logg$ are very similar between 
the training and test samples, as seen in the distributions. Abundances 
$\alpham$, $\sife$, $\mgfe$, $\alfe$ and $\nife$ compare well with 
the input labels. Because of the poor mapping of the parameter space, the stars 
with very low or very high abundance ratios can suffer from systematic trends, 
especially in the metal-poor regime. It is, for example, visible for the 
[Al/Fe]-poor tail. In general, the dispersion in the test 
sample is similar to the one in the training sample, indicating that 
we do not over-fit our data. Finally, we note that for $\alfe$ and $\nife$, 
the comparison with the input APOGEE DR16 labels does not track the 1-to-1 
relation, even for the bulk of the data, meaning that the model predicted 
during the training could suffer from systematic trends for those two elements. 
In general, we warn the reader that systematics a low S/N, 
typically $\snr<30$, can be present in the data. The abundances for those stars 
should be thus used with caution.

In \figurename~\ref{training_test_sample_kiel_diagrams}, we present 
a Kiel diagram of $\teff$ and $\logg$ from the training sample (left columns), 
for the training (top) and test (bottom) samples. In the right columns, 
we present the labels as trained by the CNN. The main features 
in the Kiel diagram are well recovered in both training and 
test samples: the position and inclination 
of the red clump, the giant branch with a smooth metallicity sequence, 
the turn-off sequence. The sequence of the very cool dwarfs spans a large 
$\teff$ range, and shows low scatter even in the very cool regime.

In the left panels of \figurename~\ref{training_sample_abundances}, we present 
the abundance patterns used as input for our CNN, for both training 
and test samples. We recall that those labels ($\feh$, $\alpham$, $\sife$, $\mgfe$, 
$\alfe$, $\nife$) are derived by APOGEE DR16. In the right panels, we present 
the labels as trained by our CNN, averaged over 80 runs. 
The chemical patterns of the trained labels, in particular $\alfe$, show 
slightly less scatter than the original labels (around 0.05 dex). This effect 
comes mainly from the fact that during the training, the neural network 
values tend to stay within the boundaries of the data. In spite of the poor mapping of 
the parameter space in the metal-poor regime, the network is still able to 
provide robust output in that metallicity regime.

In \figurename~\ref{training_sample_alphaM_Kiel}, we present the averaged 
$\alpham$ ratios of the training sample, as a function of $\mh$, for 
different bins of $\teff$ and $\logg$. One can see that the 
$\alpham-$rich sequence is mainly composed of red giant branch stars, 
while only a few stars are dwarfs. Similar plots are presented 
in Appendix~\ref{annex_patterns} 
for $\mgfe$, $\sife$, $\alfe$, and $\nife$.


\section{Estimation of atmospheric parameters 
and abundances of RAVE DR6 spectra}\label{prediction_observed_sample}

In this section, we provide details of the way we built an observed sample 
of stars based on RAVE DR6 spectra, then we present the predicted 
atmospheric parameters and chemical abundances of this observed sample.

\subsection{Creation of the observed sample}\label{observed_sample_section}

Our observed sample is based on RAVE DR6 normalised 
radial-velocity-corrected spectra \citep{steinmetz2020a}. We required 
that a spectrum have ALL\_WISE W1\&2, 2MASS $JHK_s$ photometry and Gaia DR2 
$G$, $G_{BP}$, $G_{RP}$ bands available as well as its Gaia DR2 
parallax (no cut on parallax errors). 
We checked that all spectra have StarHorse extinction measurements 
($A_V$, \citealt{queiroz2019}). 
Finally, we restricted our observed sample to a range of \snr>10 per pixel 
(as determined by RAVE DR6), removing stars 
with problematic spectra ("c" and "w" according to the RAVE DR6 classification). 
This leads to an observed sample composed of $420\,165$ stars with $\snr>10$ per pixel. 
The \snr\ distribution of the observed sample is presented in 
\figurename~\ref{snr_distribution}.

Adopting the orbital data from \citet{steinmetz2020b}, 
we carefully checked that both the training and observed samples 
probe the same Galactic volume, in terms of mean Galactocentric radii 
and height above the Galactic plane. Also, as the 
stellar age distribution can vary from one sample to another 
we took advantage of the StarHorse ages of \citet{queiroz2019} 
to check the age distributions of both the training and observed 
samples. The age distributions cover the same range and their shapes 
are consistent. Tests performed with BDASP ages from 
\citet{steinmetz2020b} have led to the same conclusion.

\subsection{Prediction of atmospheric parameters and abundances}

Once a given CNN is trained, we can predict atmospheric parameters 
and chemical abundances for the entire observed sample. 
Predicting nine parameters for $420\,165$ stars is quick, 
lasting ten seconds on a simple GPU unit. Thus, estimating parameters 
for 80 CNN runs does not take more than 15 minutes. 
We then computed a set of parameters 
averaged over the 80 runs, as well a typical dispersion 
used as error (see Sect.~\ref{errors_section}).

\subsubsection{Atmospheric parameters}

\begin{figure}[h]
\centering
\includegraphics[width=1.0\linewidth]{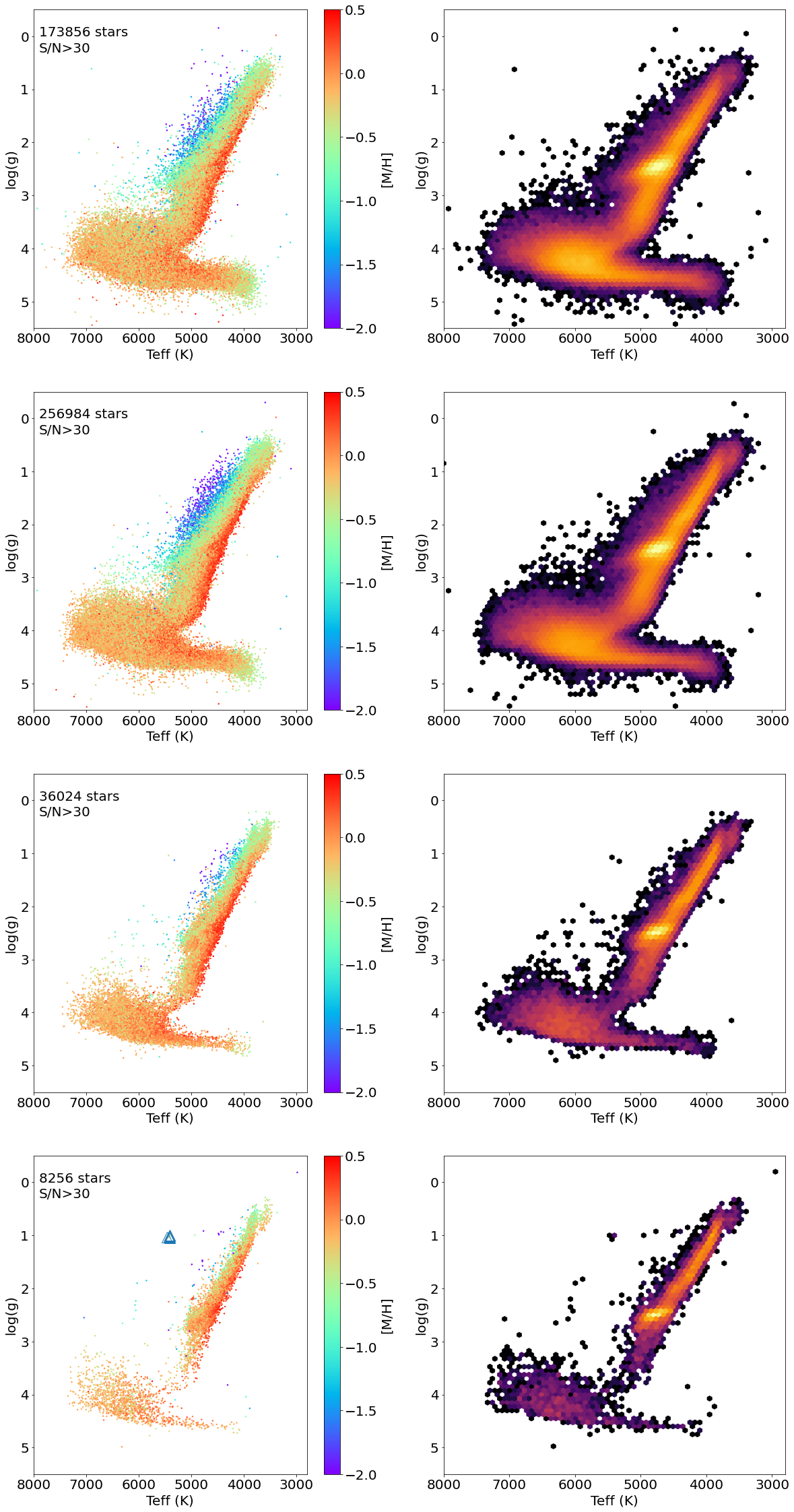}
\caption{\label{observed_kiel_diagram_snr_cuts}Kiel diagram of $371\,967$ stars 
of the observed sample, sliced in \snr, colour-coded by $\mh$ (left column) and 
plotted as a normalised density map (right column). 
Only stars with 'n' and 'o' classification (normal and hot stars), 
and parallax errors better than $20\%$ are plotted. The main features of the Kiel 
diagram are well recovered in the observed sample. The 6 blue triangles 
in the bottom panel correspond to the 
yellow supergiant Gaia "5983723702088571392", discussed 
in Section~\ref{section_outliers}.}
\end{figure}

In \figurename~\ref{observed_kiel_diagram_snr_cuts}, we present 
a Kiel diagram of the observed sample, sliced in \snr, 
for $371\,967$ stars with $\snr>20$ per pixel, and parallax errors better 
than $20\%$. We 
plotted such a diagram in two different fashions: 
colour-coded with overall metallicity, and normalised-density map. 
For such a plot, we selected normal and hot stars ('n' and 'o') according 
to the RAVE DR6 classification scheme \citep{steinmetz2020a}.

At low \snr, we recover the main features of a typical Kiel diagram, 
especially the cool main sequence and the location of the red clump. 
The bottom of the cool main sequence shows a gradient in metallicity, 
while the turn-off shows no clear gradient. For very high \snr, 
the cool dwarf sequence is very narrow, while the red giant branch 
shows a slight warp as in the training sample. At low temperatures 
($\teff<4300\,K$), we are able to properly characterise giants and 
dwarfs, putting them on the right sequence, with no degeneracy observed.

\begin{figure}[h]
\centering
\includegraphics[width=1.0\linewidth]{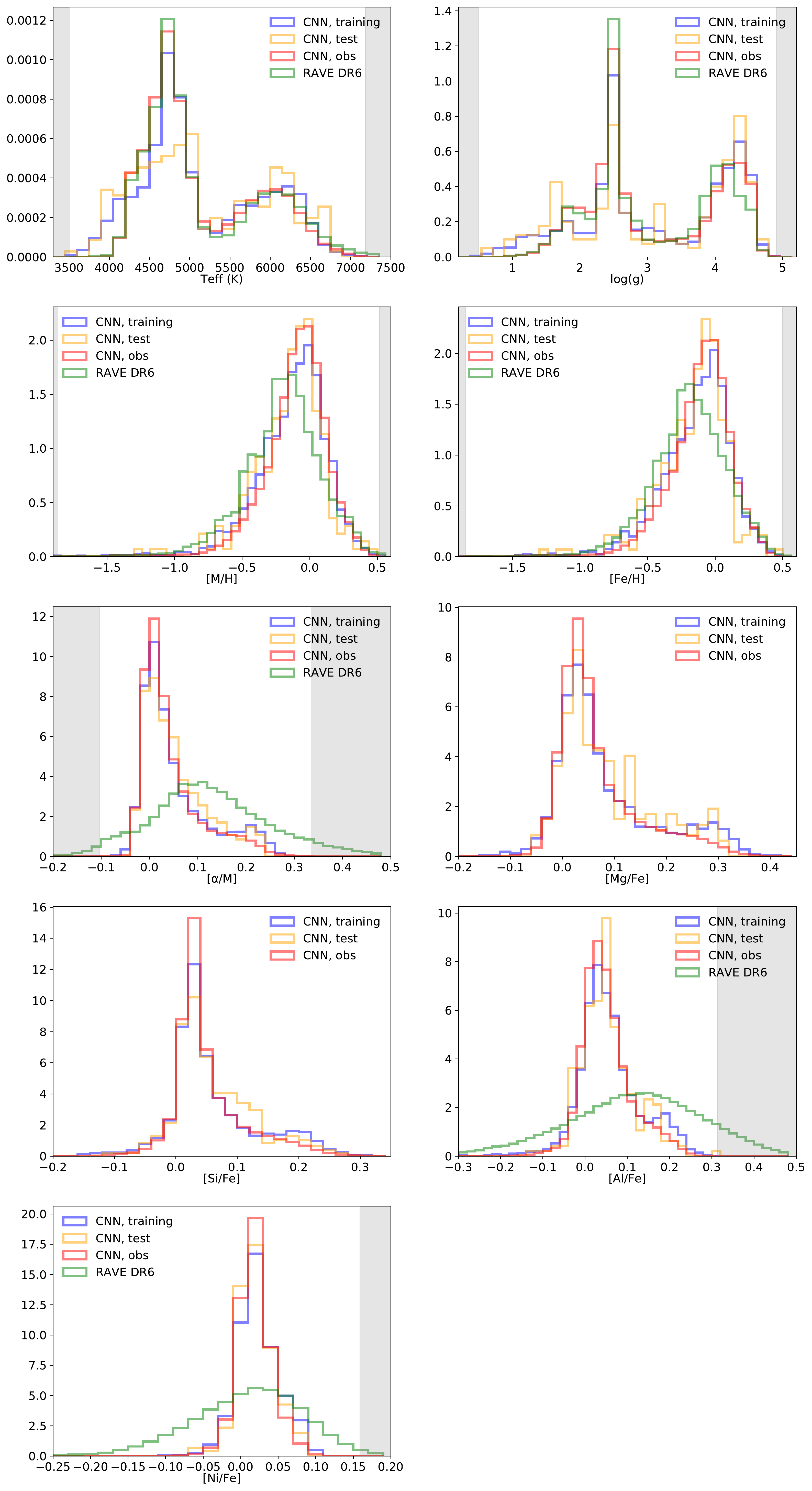}
\caption{\label{parameters_and_abudances_histo_plus_rave_dr6}Normalised 
distribution of atmospheric parameters and abundances in the training sample (blue), 
in the test sample (yellow), and the observed sample (red). For the same stars of the 
observed sample, we show a normalised distribution of the corresponding RAVE DR6 
parameters (taken from \citet{steinmetz2020b}). The grey areas define the zones 
outside the limits of the training sample parameters space.}
\end{figure}

In \figurename~\ref{parameters_and_abudances_histo_plus_rave_dr6}, 
we present normalised distributions of $\teff$, $\logg$, $\mh$, and $\feh$ of the training, 
test, and observed sample, for $\snr>40$. We also added distributions of RAVE DR6 
parameters for the same stars (with algo\_conv\_madera=0, corresponding to the best 
solutions, see \citealt{steinmetz2020b} for more details). We first see that the 
training and test sample distributions tend to track each other very well and that 
the observed sample is well defined in the training and test sample limits (defined 
by the grey areas). The same behaviour is observed for $\feh$, because APOGEE DR16 
$\feh$ and $\mh$ tend to track very well each other \citep{jonsson2020}. 
Both $\teff$ and $\logg$ from RAVE DR6 track pretty well the CNN distributions. 
In addition, both RAVE DR6 $\mh$ and $\feh$ present a metallicity-dependent shift with respect 
to our study, varying basically for zero in the metal-rich regime to roughly 0.1 dex 
in the metal-poor regime. It is a known systematic shift between RAVE DR6 and APOGEE DR16; see, for 
example, Figure 22 in \citet{steinmetz2020b}.

\subsubsection{Individual chemical abundances}

\begin{figure*}[h]
\centering
\includegraphics[width=1.0\linewidth]{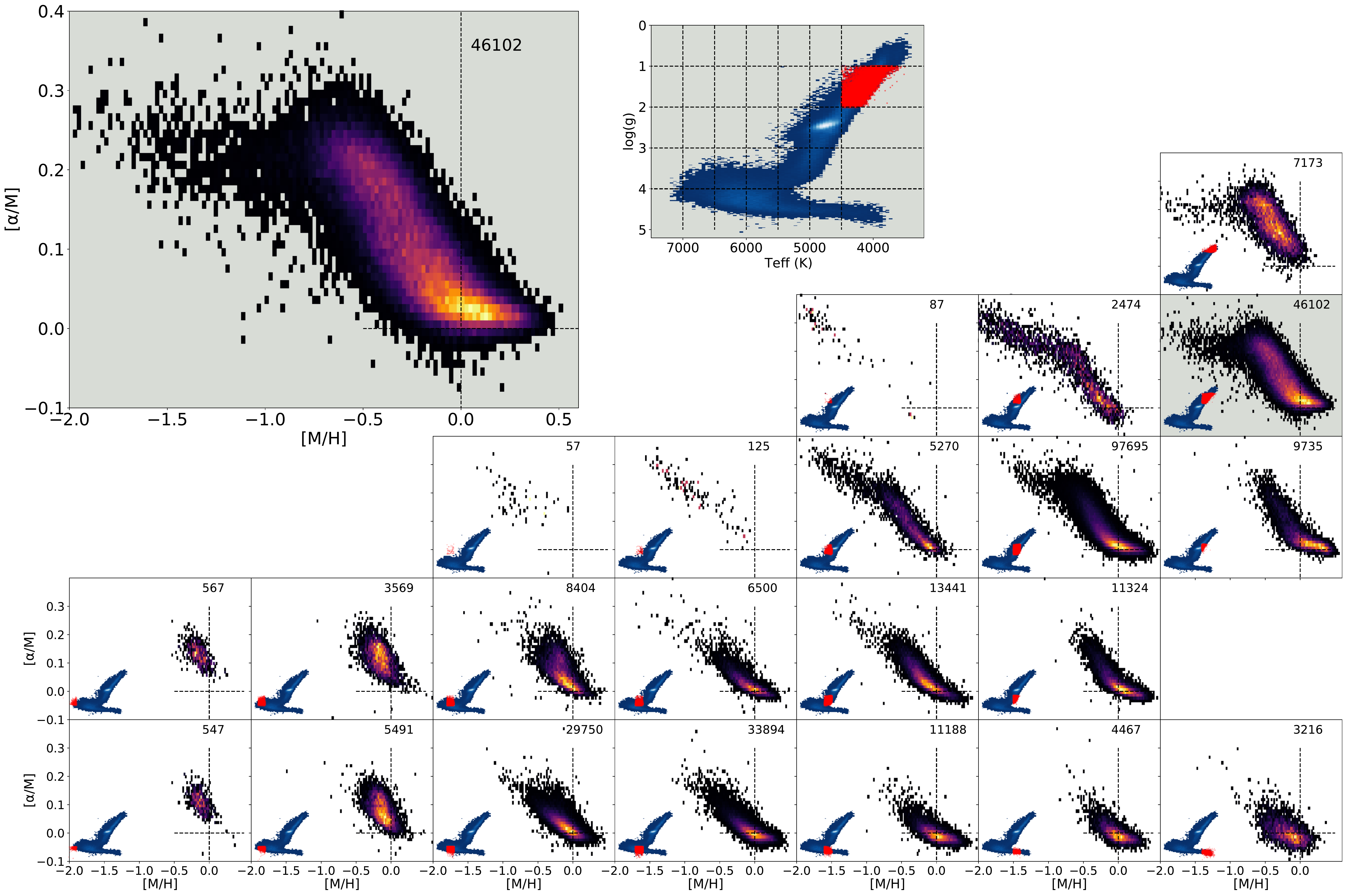}
\caption{\label{kiel_observed_AlphaM}The $\alpham$ vs. $\mh$ for 
$301\,076$ stars of the observed sample with \snr>30 per pixel, RAVE DR6 'n\&o' 
classification, and parallax errors lower than $20\%$. 
The sample is presented in panels corresponding to cuts in effective temperature 
and surface gravity (steps of 500\,K in $\teff$ and 1 dex in $\logg$. For each panel, 
we overplotted a $\teff-\logg$ diagram with the location of the plotted stars 
marked in red.}
\end{figure*}

In \figurename~\ref{kiel_observed_AlphaM}, 
we present abundance patterns for $\alpham$ as a function of the 
overall metallicity $\mh$. We selected $301\,076$ stars with \snr>30 per pixel, 
RAVE DR6 `n\&o' classification ('normal' and 'hot' stars) and 
parallax errors lower than $20\%$. In order to disentangle the different 
stellar classes, we decomposed our sample in bins of 500\,K in $\teff$, 
and 1 dex in $\logg$, and present the $\alpham$ vs. $\mh$ trends 
for different locations in the Kiel diagram (see Appendix~\ref{annex_patterns} 
for similar plots with $\sife$, $\mgfe$, $\alfe$, and $\nife$.)

Dwarf stars exhibit typical low-$\alpham$ sequences, while giants populate 
both the low-$\alpham$ and high-$\alpham$ range up to halo chemistry. 
Red clump stars show a smooth transition from 
the low- to the high-$\alpham$ regime, with a strongly decreasing density. 
On the other hand, in the range of $4000<\teff<4500\,$K and 
$1<\logg<2$, the high-$\alpham$ regime is clearly marked 
by a continuum of stars from solar-$\alpha$ up to 0.25 dex. 
Such behaviour is also observed when plotting $\sife$ and $\mgfe$ 
as a function of $\feh$ (see Appendix~\ref{annex_patterns}).

We note that the low-metallicity high-$\alpham$ plateau shows 
different behaviours in different regions of the Kiel diagram. 
This is mainly driven by the fact that we only have a few stars 
for $\mh<-1\,$dex in the training sample, showing quite different trends. 
For future machine-learning applications, we should put substantial 
efforts into properly mapping the parameter space when creating 
a training sample. The case of $\alfe$ is discussed in 
Appendix~\ref{annex_patterns}.

We have shown that in using a CNN approach and high-resolution 
stellar labels, we are able to provide reliable $\alpham$ values 
for more than $301\,076$ stars, thus extending the scientific 
output of RAVE spectra beyond RAVE DR6. 

In \figurename~\ref{parameters_and_abudances_histo_plus_rave_dr6}, 
we present normalised distributions on CNN chemical abundances in the training, test, 
and observed sample, as well as the corresponding values from RAVE DR6 
($\alphafe, \alfe, \nife$,  \citealt{steinmetz2020b}). 
We first note that both training and test sample distributions show basically the 
same shape. For $\mgfe$, the bi-modality is not well represented 
in the test sample, because of a larger scatter in $\mgfe$ 
at a given $\feh$. As for the atmospheric parameters, the chemical abundances 
in the observed sample track pretty well the training and test sample, 
for this regime of S/N ($\snr>40$). We note that for lower S/N regimes, 
the distributions of the observed sample present larger tails than the training sample. 
Finally the $\alphafe$, $\alfe$, and $\nife$ ratios from RAVE DR6 present broader 
distributions than the present study. Such an effect is already visible in Figure 22 
of \citet{steinmetz2020b}, where RAVE DR6 and APOGEE DR16 are compared. 
The RAVE DR6 abundances show a larger scatter at a given metallicity, mainly 
because of lower spectra resolution. In the present study, besides the intermediate 
resolution of the RAVE spectra, our CNN is able to provide more precise abundances, 
showing narrower distributions. We compare further $\alphafe$ ratios 
between our study and RAVE DR6 in Sect~\ref{discussion_alpha_rave_apogee}.

\section{Determination of uncertainties}\label{errors_section}

\begin{figure}[h]
\centering
\includegraphics[width=1.0\linewidth]{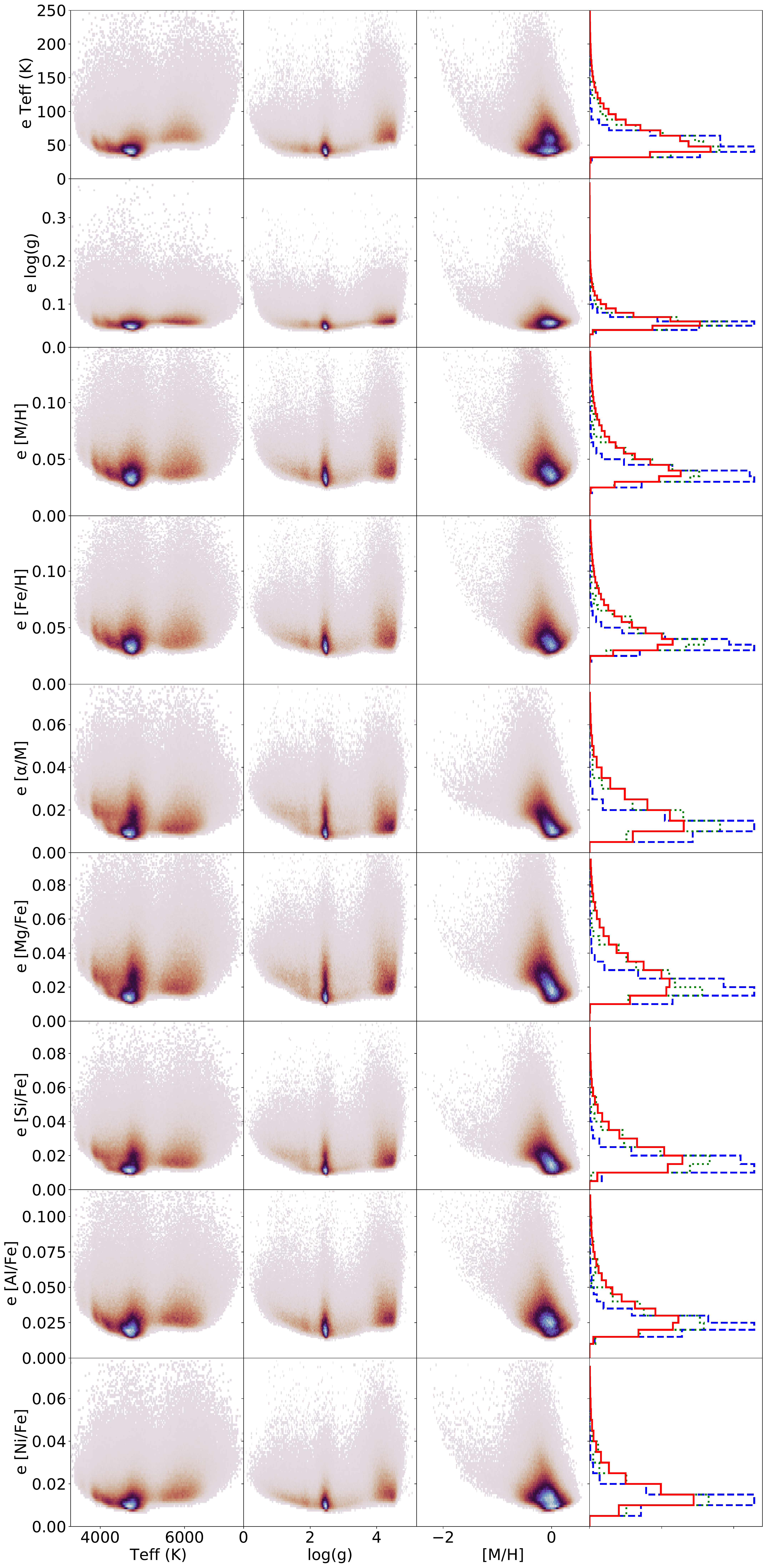}
\caption{\label{errors_histo}Errors of atmospheric 
parameters and chemical abundances plotted as a function of $\teff$, $\logg$ 
and $\mh$ for $391\,005$ stars of the observed sample. We also present 
normalised distribution of errors in the trained labels (blue, dotted), 
the test sample (green, dashed), and the observed sample (red, solid).}
\end{figure}

Despite the fact that we employ the same input labels in every run, 
the CNN does not provide the same trained labels because 
a new set of weights is automatically generated by the CNN during 
each run and the trained labels then change slightly. 
We showed the resulting average trained labels in Sect~\ref{results_training}. 
Here we present the resulting errors (precision), defined as the dispersion of 
each label for the 80 runs. As a result, the errors in both test and 
observed samples are derived in the same fashion. 
In \figurename~\ref{errors_histo}, we present 
the error on our nine atmospheric parameters and abundances as a 
function of $\teff$, $\logg$ and $\mh$, for $391\,035$ stars 
with $\snr>20$ per pixel. The uncertainty for the nine parameters tend 
to increase for both the hot and the cool tails. The same effect is visible 
for the stars with $\logg<2$. On average, the dwarf stars tend to show larger 
errors than the giants. The uncertainties on the nine parameters tend 
to increase with respect to the bulk of errors for the metal-poor tail. 
In the same figure, we present normalised distributions of uncertainties 
for the observed sample, together with the training and test samples. Overall, 
the trained labels show on average smaller errors than the test and 
the observed sample, mostly because the training sample covers a 
higher \snr\ range. The test and observed sample tend to track each 
other well, meaning that we do not over-fit our model.

As a test, we added random offsets to the labels of the training sample, 
drawn from Gaussians with widths given by the quoted uncertainties 
from APOGEE DR16. We observed that the resulting error distributions 
barely change. 

A recent study by \citet{bialek2019} adopted a negative log-likelihood 
criterion instead of a mean squared error loss-function as employed in our 
study. In that way, they were able to derive the individuals error of the 
predicted atmospheric parameters. We explored such a criterion. 
Because of the limited number of stars in our training sample, 
this criterion did not provide improved results. We therefore kept 
a simple mean squared error loss-function and errors derived over 
several CNN runs.

The present uncertainties reflect, in fact, the internal 
dispersion of the CNN. Figure~\ref{errors_histo} shows that the method 
is internally precise and stable if we consider such types of series 
of trainings (Monte-Carlo type). As a consequence, such uncertainties 
could be then underestimated, with respect to typical external errors that 
we would expect at such a resolution. Typical external errors for classical 
pipelines using RAVE spectra report errors of roughly 100K in $\teff$, 
0.15-0.2 dex in $\logg$, and 0.10-0.15 dex in metallicity and chemical abundances 
(see for exemple \citealt{steinmetz2020b}). However, as presented in 
\figurename\ref{repeats}, we note that the dispersion in atmospheric 
parameters and abundances for a star with several RAVE observations 
is very compatible with the uncertainties derived with our method.

Machine-learning methods are, within limits, able to extrapolate and 
provide parametrisations for stars outside the boundaries of the 
training sample parameter space. 
Together with individual uncertainties on the parameters and abundances, 
we provide individual flags for such stars. As an example, a star 
parametrised with an effective temperature inside the training sample 
space will have {\tt flag\_teff=0}, while the flag will be equal to 1 
if $\teff$ is outside that range. Stars with flags equal to 1 may suffer 
from systematics caused by extrapolation outside the training sample 
parameter space.


\section{Validation of atmospheric parameters and abundances}\label{validation}

In this section, we proceed to several comparisons 
with respect to external datasets in order to 
validate our atmospheric parameters and chemical abundances.
We refer the reader to Appendix~\ref{section_clusters} 
for a comparison with stellar clusters and to
Appendix~\ref{section_cnn_vs_high_res} for a comparison of our CNN results 
with a sample of HR data.

\subsection{Validation of surface gravities with asteroseismic data}

The asteroseismology of stars with solar-like oscillations is now 
widely used in large spectroscopic surveys as an additional 
constraint since it ultimately calibrates the $\logg$ 
measured from spectra (RAVE: \citealt{Valentini2017}; GES: 
\citealt{Pancino2012}; APOGEE: \citealt{Pinsonneault2018}; 
LAMOST: \citealt{Wang2016}; GALAH: \citealt{Kos2017}). For 
stars with solar-like oscillations, as well as red giants, 
$\Delta\nu$, the frequency at maximum oscillation power, 
is used for determining $\logg_{\rm seismo}$ using only 
the additional parameter, $\teff$. 
The $\logg_{\rm seismo}$ value depends very weakly
\footnote{According to \citealt{Morel2012}, a shift of 100 K in $\teff$ 
changes $\logg_{\rm seismo}$ only by 0.005 dex.}
on $\teff$, making this quantity reliable even for surveys affected by 
degeneracies such as RAVE \citep{kordopatis2011a, kordopatis2013}.

The RAVE survey has some overlap with the fields observed by the 
K2 mission, the re-purposed 
Kepler satellite \citep{VanCleve2016}. In \citet{Valentini2017}, a first 
comparison (and consequent calibration) of the RAVE spectroscopic $\logg$ 
with the seismic value was performed using 89 targets in K2-Campaign 1. 
Information on the RAVE-K2 sample, the reduction of the seismic data, and 
the calculation of the seismic $\logg$ can be found in \cite{Valentini2017}. 
In the first six Campaigns of K2, solar-like oscillations were detected 
for 462 red giants \citep[][Valentini et al, in prep.]{steinmetz2020b} and 
the seismic $\logg$ was derived. Here, we compare these seismic $\logg$ 
values with the values determined using our CNN.  

\begin{figure*}[h]
\centering
\includegraphics[width=1.0\linewidth]{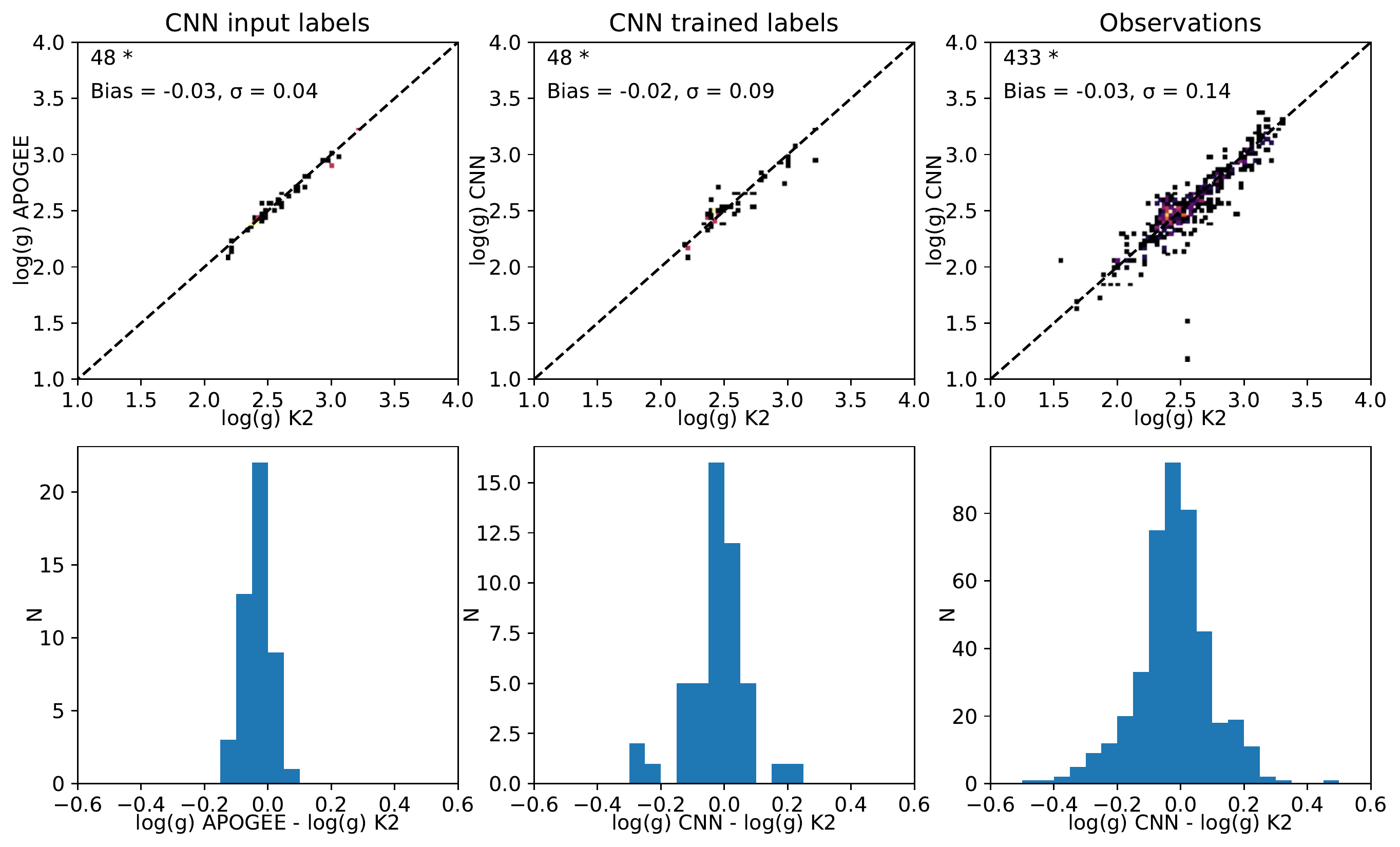}
\caption{\label{cnn_vs_seismo}Comparison of surface gravities from the 
present study with K2 asteroseismic data. Left: 
Comparison with the $\logg$ labels from APOGEE DR16 used as input 
by our CNN. Middle: Comparison with averaged labels trained 
by the CNN. Right: Comparison with averaged $\logg$ predicted for 
common stars in the observed sample. Mean difference and scatter 
are indicated in the top left corner of each panel.}
\end{figure*}

\figurename~\ref{cnn_vs_seismo} shows that the labels (APOGEE DR16) 
and the K2 $\logg$ values exhibit a tight and un-biased 
1-to-1 relation (left panel, $\rm bias=-0.03$~dex and dispersion $\sigma=0.04$~dex).
The K2 $\logg$ values also agree well with the labels trained by 
the CNN (middle panel), with a slightly 
higher scatter ($\sigma=0.09\,$dex). Finally, in the right panel of \figurename~\ref{cnn_vs_seismo}, 
we compare the predicted surface gravity 
for 433 common stars of our observed sample with K2 data, 
finding an very good agreement with a very small bias and a dispersion 
of 0.14 dex. We note that the $\logg$ values from RAVE DR6 show a 
larger scatter with respect to K2 data than our CNN $\logg$ values 
(see Figure 23 of \citealt{steinmetz2020a}).

Keeping in mind that we are limited by the narrow spectral range of 
the RAVE spectra, those comparisons illustrate all the potential of 
a method based on CNN.
A more detailed discussion on the impact of the use of photometry can be found in 
Sect.~\ref{section_phot_vs_no_phot}.

\subsection{Comparison with RAVE DR6 BDASP $\logg$}

In the latest data release of RAVE (DR6, \citealt{steinmetz2020b}), 
improved $\logg$ estimates based on Gaia DR2 parallaxes and 
Bayesian isochrone fitting are provided, thanks to 
the BDASP pipeline \citep{mcmillan2018}. This section is dedicated 
to comparing RAVE/BDASP surface gravities to those derived by our 
CNN in the present study. 

\begin{figure*}[h]
\centering
\includegraphics[width=1.0\linewidth]{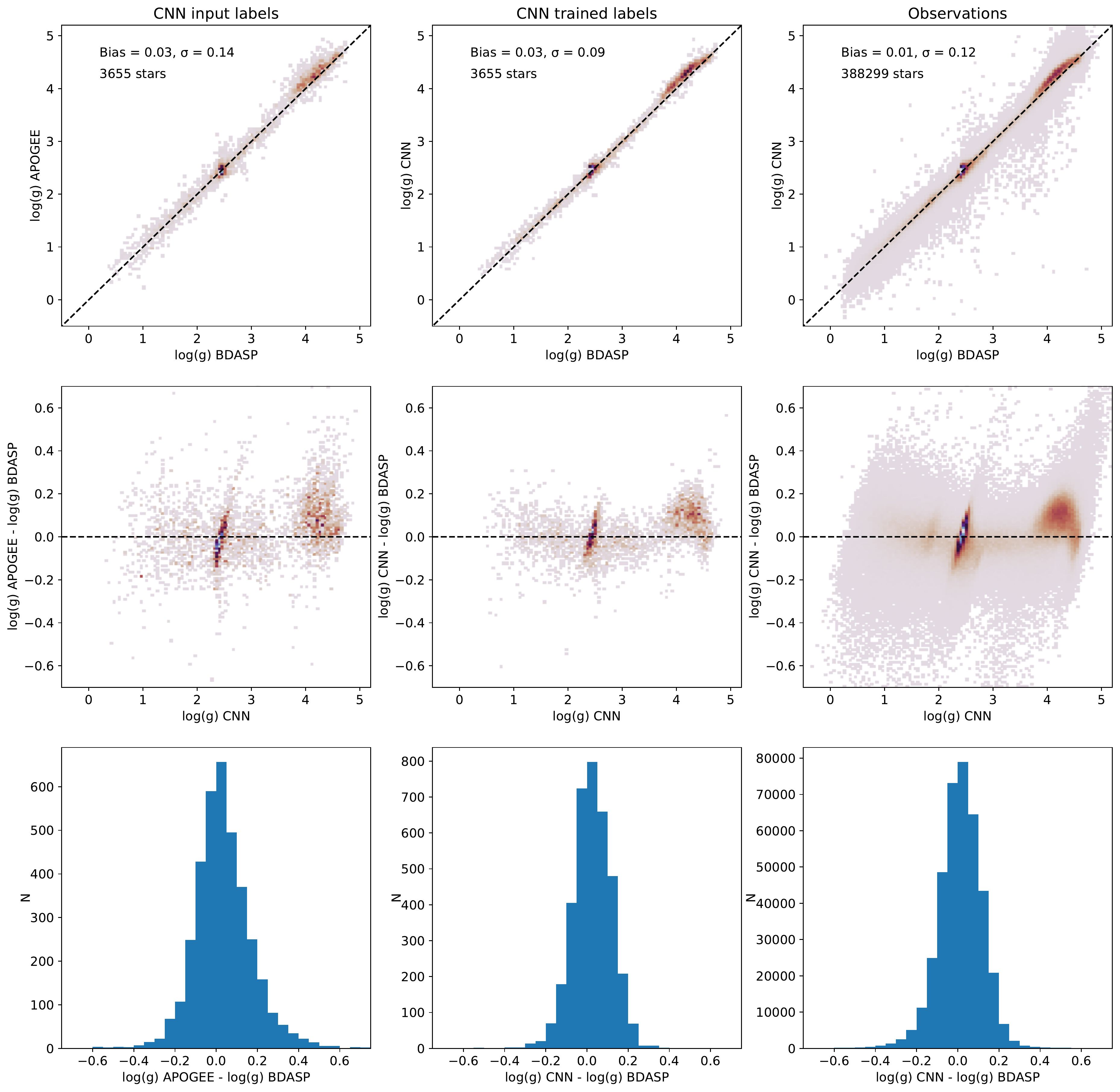}
\caption{\label{cnn_vs_bdasp}Left: Comparisons of the $\logg$ values 
used as input labels of our CNN (APOGEE DR16 $\logg$) with respect to 
$\logg$ values of \citet{steinmetz2020b}. We also show a residual plot and an 
histogram of the difference. Mean difference and scatter are indicated 
in the top-left corner. Middle: Comparison of $\logg$ values trained by 
our CNN with respect to $\logg$ values of \citet{steinmetz2020b}. 
Right: Comparison of the $\logg$ values derived by our CNN for 
$388\,299$ stars of our observed sample with respect to the $\logg$ values 
of \citet{steinmetz2020b}.}
\end{figure*}

The left panel of \figurename~\ref{cnn_vs_bdasp} 
compares the input APOGEE DR16 $\logg$ with those of BDASP . 
The dwarfs ($\logg>3.5$) show a  shift of about +0.1 dex, while the 
giants do not show any bias with respect 
to RAVE DR6. The typical dispersion is 0.14 dex for both types 
of stars with a bias of 0.05 dex. We notice that the surface 
gravities provided by APOGEE DR16 show a smaller dispersion 
around the red clump as compared to RAVE DR6, hence, the presence 
of a diagonal line at $\logg\sim2.5$. 

Concerning the labels trained by our CNN, 
the bias decreases slightly (+0.04 dex), while the scatter 
drops  to 0.09 dex. This decrease in the scatter is directly due 
to the fact that we use absolute magnitudes 
during the training process, leading to more precise $\logg$ values 
(see Sect.~\ref{section_phot_vs_no_phot} for more details). 
If no absolute magnitudes are used during the training phase, 
the scatter doubles to 0.17.

Finally, in the right column of \figurename~\ref{cnn_vs_bdasp} 
we compare the surface gravities predicted for $388\,299$ 
stars of the observed 
sample (\snr>20) with respect to RAVE DR6. Again, the biases 
for giants and dwarfs keep the same shape as in the previous comparisons, 
and the scatter tends to still be quite low (0.12 dex). We notice that 
the scatter $\sigma$ increases to 0.37 dex when no photometry is 
used in the training phase. A discussion on the impact of the use of 
photometry can be found in Sect.~\ref{section_phot_vs_no_phot}.

As a final note on this topic, we recall that the input $\teff$ of the 
BDASP pipeline is the InfraRed Flux Method $\teff$ (see 
\citet{steinmetz2020b} for more details). The BDASP $\teff$ 
tends to be very similar to this input. We explicitly 
compare our $\teff$ to $\teff$ IRFM in the next section.

\subsection{Validation of effective temperatures with IRFM temperatures}

A data product of the sixth data release of RAVE is the 
effective temperature derived via to the Infrared Flux Method 
(IRFM, \citealt{casagrande2006,casagrande2010}, see \citealt{steinmetz2020b} 
for more details). In this section, we compare our effective temperatures 
to those provided by RAVE DR6. We compared the $\teff$ used in the training sample 
(APOGEE DR16 $\teff$), those learned by the network, and those derived for 
the observed sample (for \snr>20). 

\begin{figure*}[h]
\centering
\includegraphics[width=1.0\linewidth]{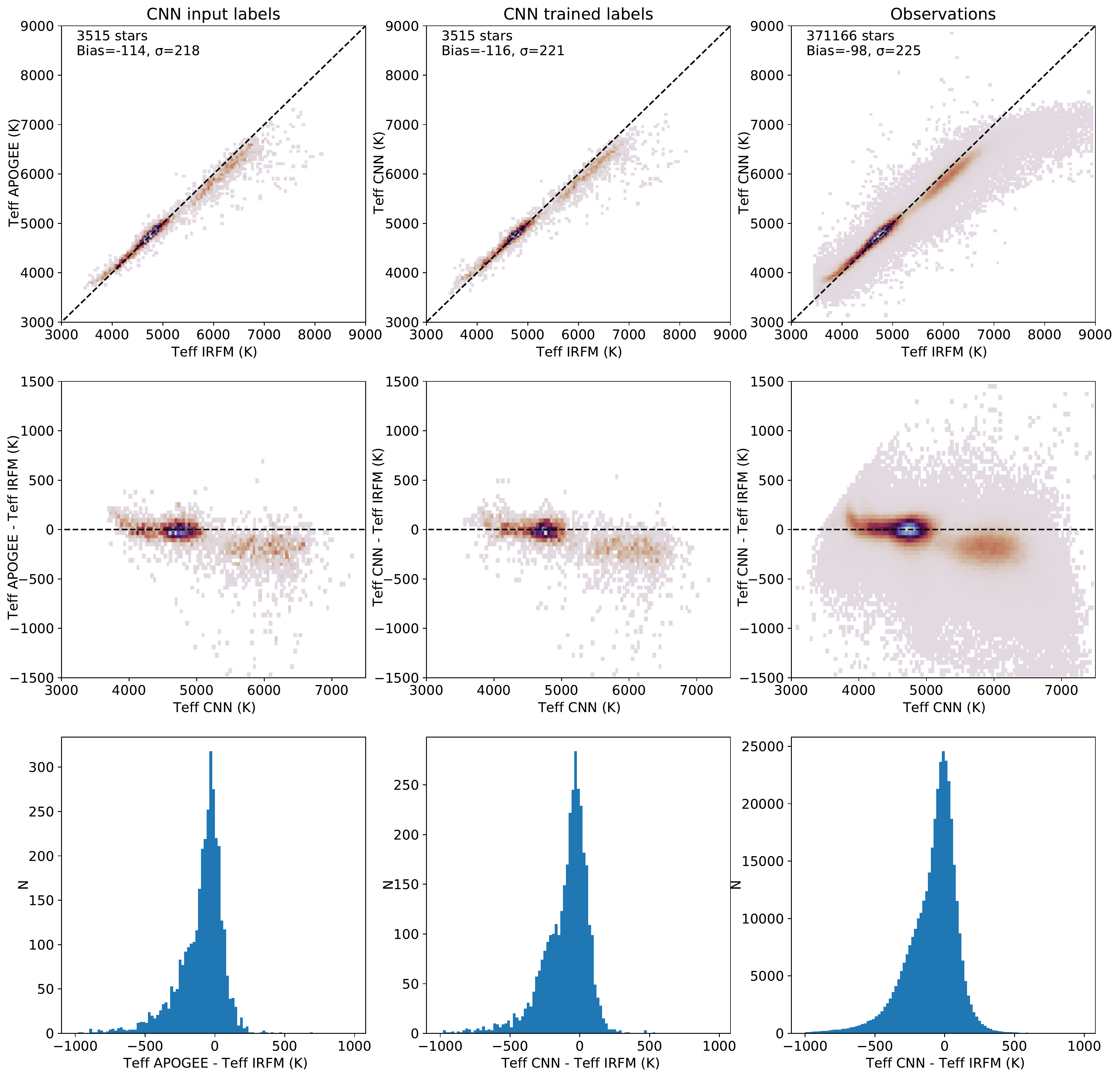}
\caption{\label{cnn_vs_irfm}Left: Comparisons of the 
input label $\teff$ for our CNN (APOGEE DR16 labels) with the 
IRFM temperatures of \citet{steinmetz2020b}. Mean difference 
and scatter are indicated in the top-left corner. We also 
show a residual plot and an histogram of the difference. 
Middle: Comparison of the labels $\teff$ trained by the CNN 
with the IRFM temperatures of \citet{steinmetz2020b}. 
Right: Comparison of the $\teff$ values derived for our 
whole observed data-set (for \snr>20) with the IRFM temperatures 
of \citet{steinmetz2020b}.}
\end{figure*}

The results are presented in \figurename~\ref{cnn_vs_irfm}. We first 
see that there is a shift between the effective temperatures used as 
labels in our study and those of \citet{steinmetz2020b} for hot stars 
($\teff>5\,200\,$K) which are offset by -250\,K (constant with temperature, 
with 260\,K scatter). Those stars are mainly dwarfs. On the other hand, 
the cool stars of the training sample ($\teff<5\,200\,$K, mostly giants) 
show a tight and unbiased one-to-one relation with respect to the IRFM temperatures 
(mean difference of -20\,K and dispersion of 90\,K). Overall, the dispersion 
is about 220\,K for the 3651 stars of the training sample.

We note that stars with $\teff>5\,200\,$K tend to be cooler by 250\,K 
with respect to the IRFM $\teff$. The $\logg$ of such stars will be then 
systematically higher. This could serve as an explanation for the higher $\logg$ 
measured by our CNN with respect to BDASP $\logg$ (see previous section, 
\figurename~\ref{cnn_vs_bdasp}).
Once the CNN is trained, the effective temperatures still show 
the same behaviour with respect to the IRFM $\teff$.

Finally, we can see that the measured $\teff$ in $371\,166$ stars 
of the observed sample match in the same way the RAVE IRFM $\teff$, 
with a larger scatter than the training sample mainly due to the 
presence of stars with lower \snr.
Overall, the effective temperatures used in the training sample 
(from APOGEE DR16), those trained, and those predicted agree rather 
well with the $\teff$ IRFM from \citet{steinmetz2020b}. Finally, we note 
that this comparison only provides an assessment of the biases 
and scatter with respect to APOGEE DR16.

\subsection{Validation with repeat observations}

\begin{figure*}[h]
\centering
\includegraphics[width=1.0\linewidth]{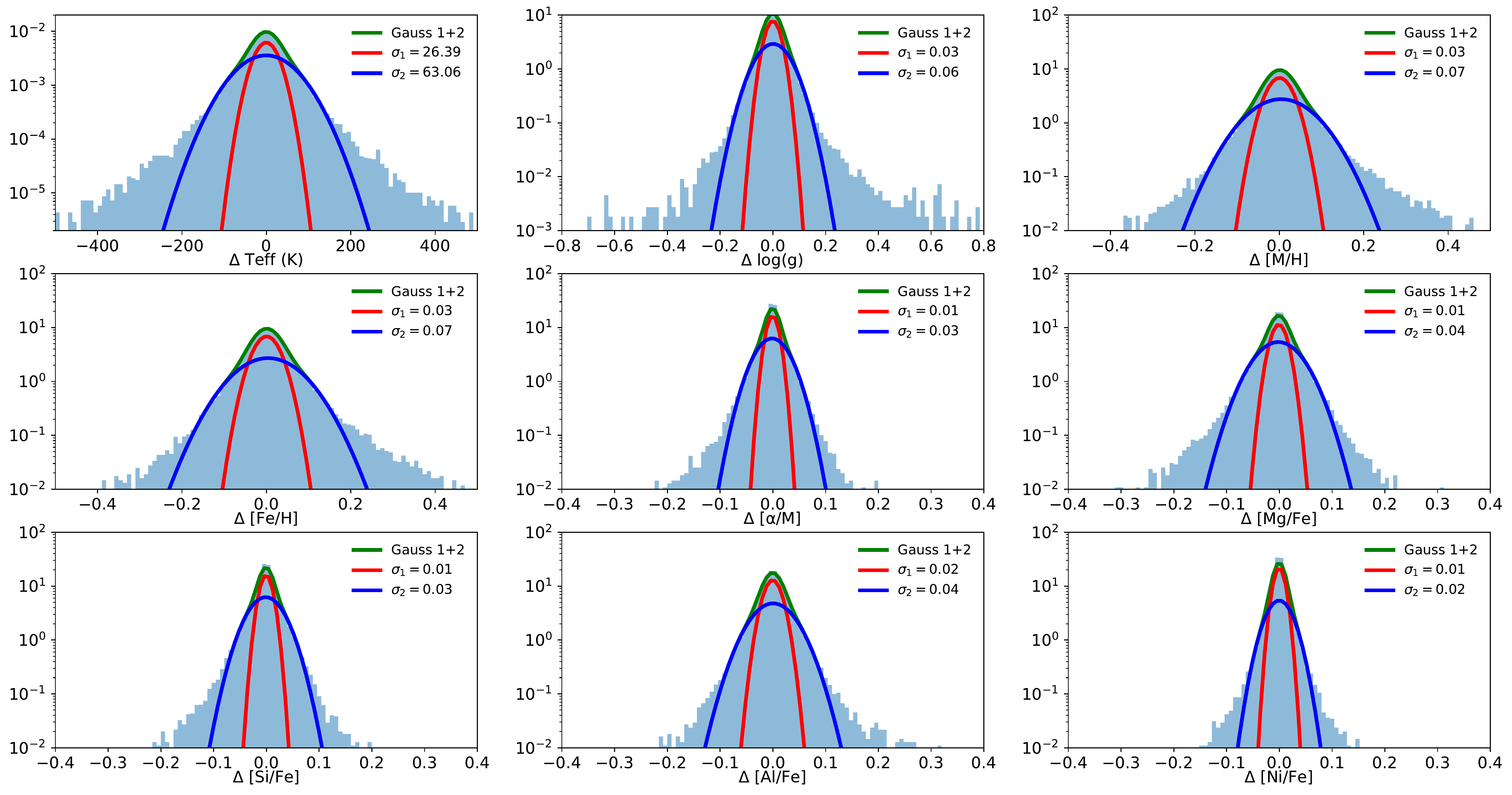}
\caption{\label{repeats}Differences in atmospheric parameters and chemical 
abundances for $80\,342$ stars based on several observations and $\snr>20$. }
\end{figure*}

Another way to show the reliability of our atmospheric 
parameters and chemical abundances is to investigate stars 
with repeated observations. We follow the same procedure 
as in \citet{steinmetz2020a,steinmetz2020b}. Briefly, for a given 
star with several observations, we computed the differences in 
atmospheric parameters and chemical abundances. For all stars with multiple 
repeats, we analyzed the distribution of those differences. We 
approximated the distribution function by a combination 
of two Gaussians using a least-squares fit. The results are presented 
in \figurename~\ref{repeats}, for all repeats ($80\,342$ stars, $\snr>20$). 
Firstly, we can see that the distributions are roughly similar in 
shape for $\teff$, $\logg$, $\mh$, and $\feh$. 
On the other hand, the chemical abundances of $\alpham$, $\mgfe$, 
$\sife$, $\alfe$, and $\nife$ present asymmetric tails. 
The typical dispersion of the distribution for the effective temperature 
is about $\sim50\,$K, while for the surface gravity, the dispersion 
is below 0.05 dex. The dispersion increases to $80\,$K for $\teff$ 
and 0.14 dex for $\logg$ if we do not use photometry to introduce additional information. 
For $\mh$ and $\feh$, the typical dispersion over all 
repeats is of the order of 0.05 dex. Finally, for $\alpham$, 
$\mgfe$, $\sife$, $\alfe$, and $\nife$, a dispersion of 0.02-0.03 dex 
is measured over all repeats. These results imply that the CNN 
is precise (low dispersion within repeats) and accurate 
(overall difference distributions centered on zero) 
in determining atmospheric parameters and chemical abundances 
of RAVE spectra. We note that such dispersion 
within repeats in consistent with the typical uncertainties reported 
in Sect.~\ref{errors_section} for both atmospheric parameters 
and chemical abundances.

\subsection{Comparison with RAVE DR6 $\alpham$ ratios}\label{discussion_alpha_rave_apogee}

The RAVE spectra cover the near-infrared CaII triplet, 
which is a key spectral feature in the process of placing constraints 
on the overall $\alpha$ enrichment of stars. 
In this section, we compare the $\alpham$ derived 
in the present study by our CNN to the 
$\alphafe$ derived in \citet{steinmetz2020b} by a more 
classical approach (synthetic spectra grid + optimisation 
method). Both quantities were derived using the same 
observed spectra.

\begin{figure}[h]
\centering
\includegraphics[width=1.0\linewidth]{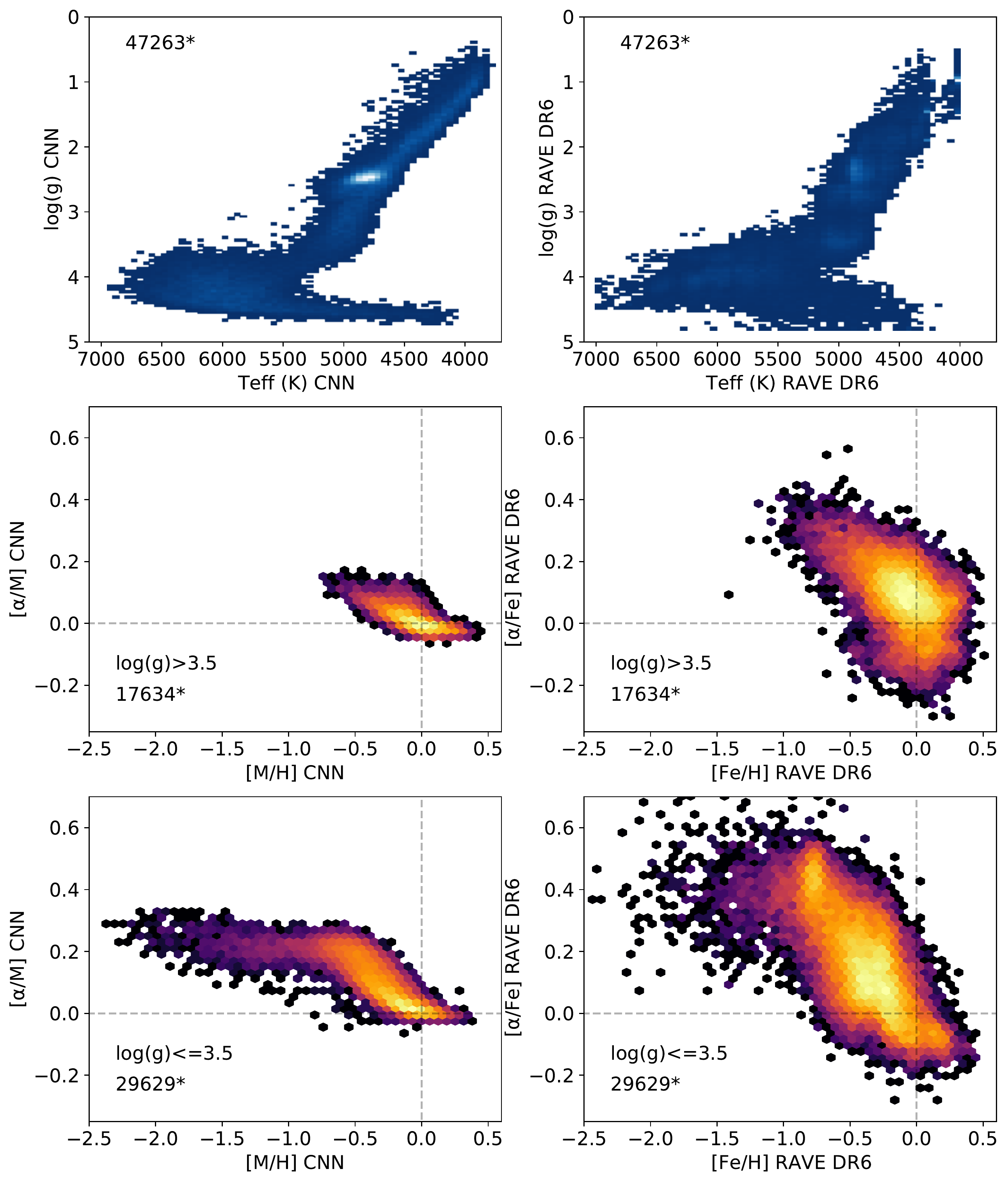}
\caption{\label{comparison_dr6_cnn_alpha}Top: 
$\teff$ vs. $\logg$ for 69959 stars, derived with our CNN (left) 
and derived by RAVE DR6 (right). Middle: 
Abundance pattern for $30\,988$ dwarfs, derived 
by our CNN (left) and RAVE DR6 (right). 
Bottom: Same plots for $38\,671$ giants.}
\end{figure}

In \figurename~\ref{comparison_dr6_cnn_alpha}, we 
present an abundance pattern comparison between 
the present study ($\alpham$ vs. $\mh$) and RAVE DR6 
($\alphafe$ vs. $\feh$), for $69\,659$ dwarfs and giants ($\snr>20$).
We adopt the same quality criteria presented in 
\citet{steinmetz2020b} to select the best RAVE DR6 $\alphafe$ ratios. 

We first show a typical Kiel diagram for each sample 
(CNN top-left, RAVE DR6 top-right). Using our CNN approach 
with combined spectroscopy, photometry and astrometry, we are 
able to tackle the degeneracy caused by RAVE's narrow wavelength 
range, especially in the very cool regime. 

The abundances derived by RAVE DR6 show a larger scatter 
at a given metallicity. In the metal-poor regime, the CNN 
results show a tight $\alpham$ sequence. Overall, both 
studies show the same main chemical features, for both giants 
and dwarfs. They also cover the same metallicity range. 
We note that in the metallicity range of $-1<\mh<+0\,$dex, the CNN $\mh$ 
present a shift of +0.14 dex with respect to RAVE DR6 $\feh$, 
while for both metal-poor and metal-rich tails, the bias is basically null. 
The differences in trends and zero-points originate from a different 
calibration between the two studies, one based on APOGEE data, 
while the RAVE DR6 is based on synthetic spectra grid.

\subsection{Exotic star detection capabilities}\label{section_outliers}

\begin{figure}[h]
\centering
\includegraphics[width=1.0\linewidth]{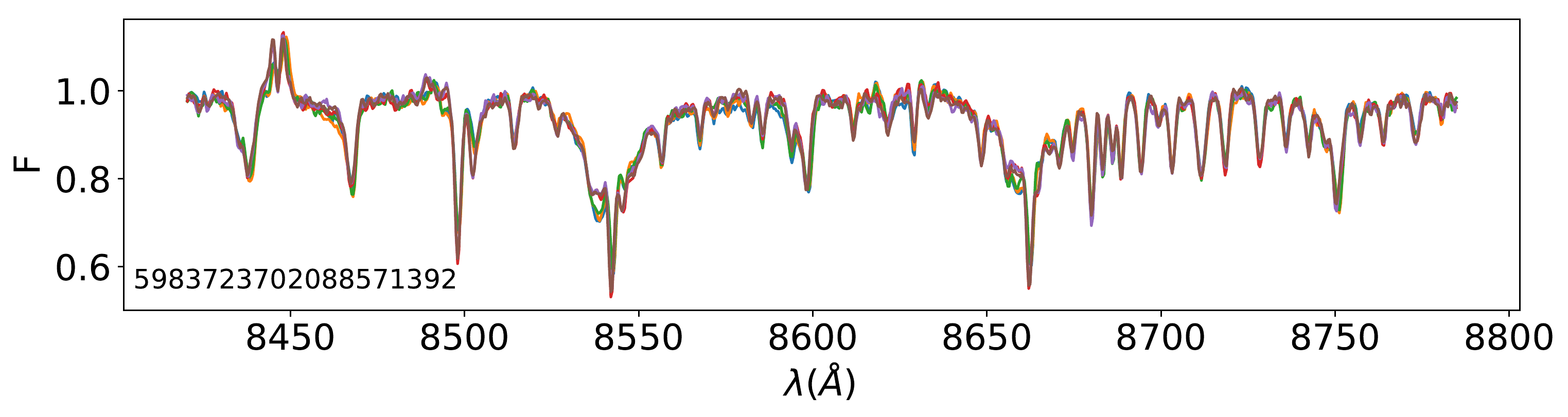}
\caption{\label{yellow_super_giant}Normalised RAVE DR6 spectra of the 
target Gaia '5983723702088571392'. The six spectra are plotted in different 
colours.}
\end{figure}

Neural networks are particularly efficient with regard to classifying objects. 
In addition, peculiar 
stars are expected to be detected by such a machine-learning pipeline; 
by peculiar, we mean that the CNN is able to parameterise stars in 
regions where the training sample parameter space is poorly covered. 
We illustrate this point by the example of the known yellow supergiant 
(spectral type F3I, \citealt{houk1978}, Gaia\_sourceid='5983723702088571392'), 
which has been observed six times by the RAVE survey. 
The normalised RAVE DR6 spectra are presented in \figurename~\ref{yellow_super_giant}. 
This star has been characterised as 'normal' by RAVE DR6. 
Its Gaia DR2 parallax error is 10\%.
The mean atmospheric parameters and errors derived by our CNN from 
the six repeats are the following: $\teff=5423\pm355\,$K, $\logg=1.02\pm0.53$, 
$\mh=-0.36\pm0.20\,$dex. The average RAVE DR6 parameters derived 
with the BDASP pipeline (using Gaia DR2 and isochrone fitting) are the 
following: $\teff=5047\pm213\,$K, $\logg=1.39\pm0.08$, $\mh=+0.28\pm0.15\,$dex. 
In spite of the differences in the approach, the CNN and BDASP 
methods tend to put this star in the same region of the Kiel diagram, within 
1-$\sigma$ errors. The overall metallicity shows the largest scatter, with CNN 
and BDASP consistent within 2-$\sigma$.

On the other hand, the RAVE DR6 parameters by the MADERA pipeline (pure spectroscopy) 
are the following: $\teff=5986\pm95\,$K, $\logg=3.63\pm0.15$, $\mh=+0.51\pm0.09\,$dex. 
Those parameters are consistent to those derived by our CNN, only using 
spectroscopic data (no photometry or parallaxes), within 2-$\sigma$ in $\teff$ and 
1-$\sigma$ in $\logg$ and $\mh$: $\teff=6401\pm150\,$K, $\logg=3.90\pm0.20$, $\mh=+0.50\pm0.11\,$dex.

\section{Including  versus excluding photometry }\label{section_phot_vs_no_phot}

\begin{figure*}[h]
\centering
\includegraphics[width=0.95\linewidth]{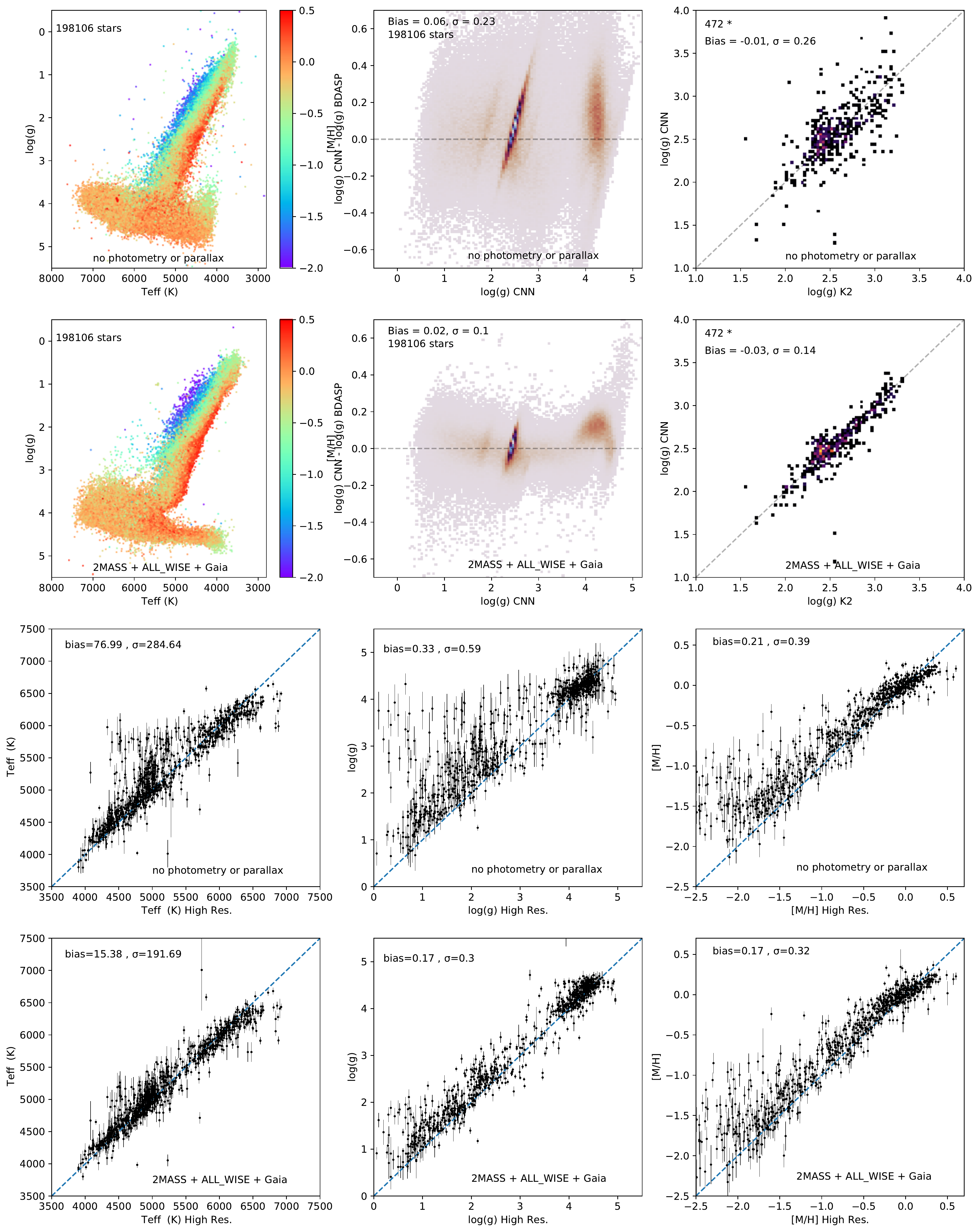}
\caption{\label{phot_vs_no_phot}Systematic comparisons 
of parameters from our CNN with or without photometry (2MASS+ALL\_WISE+GaiaDR2) 
and astrometry (Gaia DR2). Top-left: The Kiel diagrams are colour-codded in $\mh$ 
for $198\,106$ stars ('n\&o' classification) with $\snr>40$ and parallax 
errors lower than $20\%$; Top middle: Comparison of $\logg$ with respect to 
RAVE DR6 $\logg$ for the same stars; Top-right: Comparison of CNN $\logg$ values 
with respect to K2 $\logg$ values; Bottom: Comparisons of CNN $\teff$, $\logg$, and 
$\mh$ values with respect to the high-resolution sample (one-to-one relations).}
\end{figure*}

We show here that adding absolute photometric magnitudes during the training 
phase of the CNN significantly improves the quality of 
the derived effective temperature and surface gravity, and, to a 
lesser extent, the overall metallicity. We recall that colours are 
key indicators of effective temperatures and that colours and 
absolute magnitudes help to constrain surface gravities.

To do so, we simply re-trained our CNN a hundred times, 
with the same overall architecture but removing the photometric neurons, 
meaning that we only use pure spectroscopic data from RAVE. 
We kept the same training sample. We simultaneously 
predicted $\teff$, $\logg$, $\mh$, $\feh$, plus individual abundances 
for the observed data.

In \figurename~\ref{phot_vs_no_phot}, we present the 
resulting Kiel diagram of $\teff$ and $\logg$, colour-coded in $\mh$. 
We only show data with $\snr>40$, that is, stars with good observational data. 
Compared to the Kiel diagram derived including absolute magnitudes, 
the pure spectroscopic results still have all the typical features, 
like the cool dwarf sequence, the turn-off, or the giant branch. 
On the other hand, the cool dwarfs sequence suffers from large scatter, 
while degeneracies appear for very cool giants (large $\logg$ scatter for a given 
$\teff$). The red giant branch appears as a straight sequence. 
Finally, the metallicity sequence in the giant branch is not as 
well-defined as when absolute magnitudes are used. 
The wavelength range around the CaII triplet is known to suffer from degeneracies 
when deriving atmospheric parameters \citep{kordopatis2011a}. We note that 
including absolute magnitudes helps us to break these degeneracies, without applying 
any prior or restraining the parameter space of the training sample. 
The mean error in $\teff$ is increased by $\sim20\,$K when no 
absolute magnitudes are used.

We then compare our surface gravities with those from RAVE BDASP $\logg$. 
When using 2MASS+ALL\_WISE+Gaia, we can see that the average difference 
between both studies is one quarter of the one based purely on 
spectroscopy, while the dispersion drops from 0.23 to 0.09 dex.

Next, we compare our purely spectroscopic $\logg$ values to 
those provided by K2. Without photometric input, the scatter is 
much larger (0.26 dex) with a tiny bias. We note that the purely 
spectroscopic $\teff$ values show a slightly higher 
dispersion with respect to those derived including absolute 
magnitudes during the training phase. 

Finally, we compare $\teff$, $\logg$, and $\mh$ derived 
from purely spectroscopic data by our CNN to those of 
the high-resolution sample presented in Appendix~\ref{section_cnn_vs_high_res}
(only stars with $\snr>20$). Without absolute magnitudes, 
we observed a significantly larger dispersion in $\logg$ (0.58 dex) 
and bias (+0.26 dex), as compared to the high-resolution sample. 
This is also the case for the effective temperature, with a 
slightly larger bias (55\,K instead of no bias) and a dispersion larger 
by 80\,K. Finally, the metallicity derived purely by spectroscopic data 
suffers from a slightly higher bias and dispersion with respect to 
the literature sample. The main improvement is actually notable 
for $\mh$<-1.5 dex, consistent with previous remarks on the Kiel diagram.

With these comparisons, we demonstrate that purely spectroscopic data 
can still provide quite satisfying outputs, 
however, adding photometry as well as astrometric parallaxes provides a major gain 
with a strong increase 
in precision and accuracy, mainly for effective temperature 
and surface gravity. We are able to efficiently break the degeneracies 
in the $\teff-\logg$ space, caused by limited 
spectral range of RAVE spectra, particularly in the cool regime.


\section{Science verification}\label{science_verification}

\subsection{Abundance-kinematical properties of the Milky Way components}

Here, we investigate some implications for the chemical and 
kinematical properties of the Milky Way. We adopted the kinematics 
from RAVE DR6 
\citep{steinmetz2020b} and followed the same approach as 
\citet{gratton2003} 
and \citet{boeche2013a}. We first kinematically selected a thin disc component
with low eccentricity stars ($e<0.25$) and low maximum altitude 
($\text{Z}_{\text{max}}<0.8\,$kpc). 
We identified a dissipative collapse component, mainly composed of 
thick disc and halo stars with $e>0.25$, $\text{Z}_{\text{max}}>0.8\,$kpc, 
and $V_{\phi}>40\,$\kms. Finally, 
we characterised an accretion component, composed of halo and 
accreted stars ($V_{\phi}<40\,$\kms).

\begin{figure*}[h]
\centering
\includegraphics[width=1.0\linewidth]{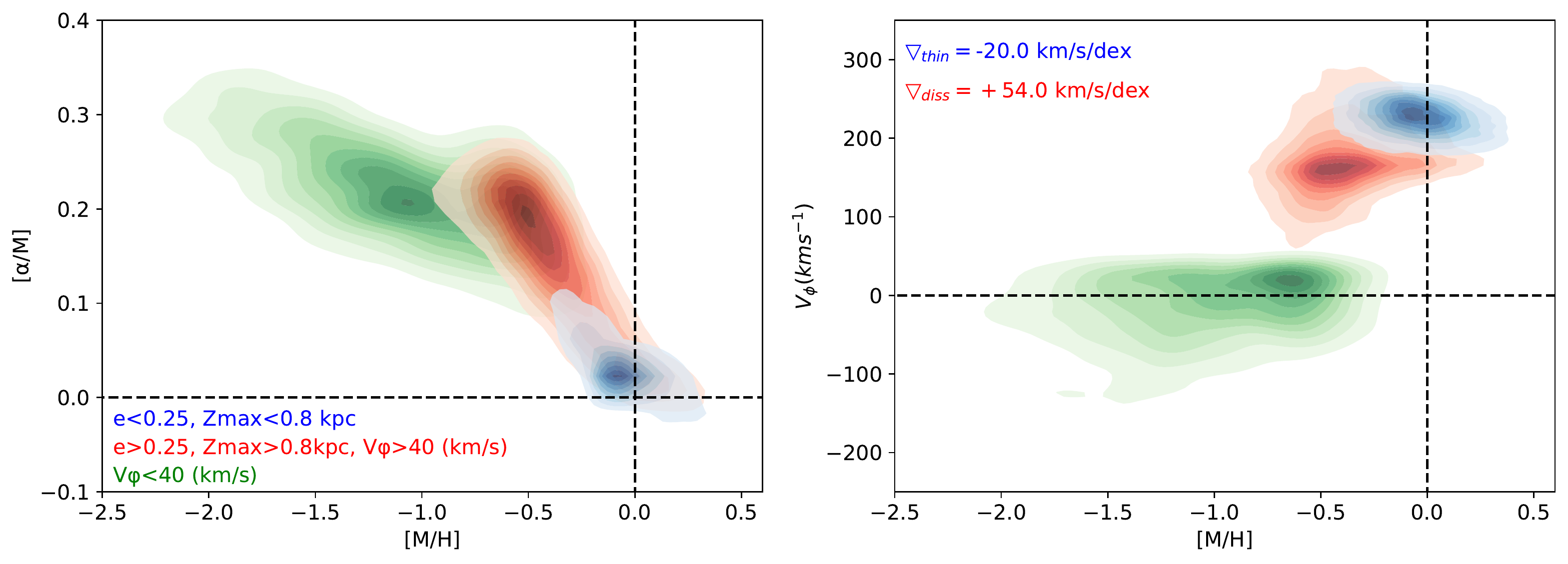}
\caption{\label{thin_thick_halo}Left: 
$\alpham$ vs. $\mh$ contour plots of the thin disc component (75\,642 stars, blue), 
the dissipative collapse component (15\,433 stars, red), and the accretion component 
(1\,400 stars, green). Right: Galactocentric rotational 
velocities $V_{\phi}$ as a function of $\mh$. We only show stars with 
parallax errors lower than $20\%$, $\snr>40$ and 'n\&o' RAVE classification. 
We estimated the gradients of $V_{\phi}$ vs. $\feh$ in the thin disc component 
and in the dissipative collapse component and find good agreement  with 
literature values, depsite our different selection criteria (see for 
example \citet{kordopatis2011b} and \citet{lee2011}).}
\end{figure*}

In \figurename~\ref{thin_thick_halo}, we present the $\alpham$ pattern 
for these three components for giant stars ($\logg<3.5$). The thin disc is 
mainly confined to $\mh>-1\,$dex, while the dissipative collapse component 
shows a large metallicity range, a few metal-rich stars, 
including halo stars with metallicities higher than $-2\,$dex, 
and a narrow $\alpham$ sequence. The accretion component is only 
composed of metal-poor stars, in the range $-2.0<\mh<-0.5$.
We note that the mean error on $\mh$ and $\alpham$ increases with 
decreasing metallicity for the three components. These findings
are in good agreement with \citet{boeche2013a}.

We measured the gradients of $V_{\phi}$ vs. $\mh$ 
in both the thin disc and dissipative collapse components. 
The thin disc component shows an anti-correlation ($\nabla=-20$\kms/dex), while a 
strong correlation is visible in the dissipative collapse component 
($\nabla=+54$\kms/dex). Such gradients are consistent with previous works, 
like for example \citet{lee2011} with SEGUE data or \citet{kordopatis2011b}, 
despite different selection functions. We note, however, that the positive 
gradient in the dissipative collapse components results from the superposition 
of mono-$\alpham$ sub-populations with negative slopes, as was recently shown 
using RAVE DR5 data \citep{wojno2018, minchev2019}. These simple science 
applications show the potential of the CNN abundances.

\subsection{Chemical cartography of $\alpham$ ratio in the galactic discs}

\begin{figure*}[h]
\centering
\includegraphics[width=1.0\linewidth]{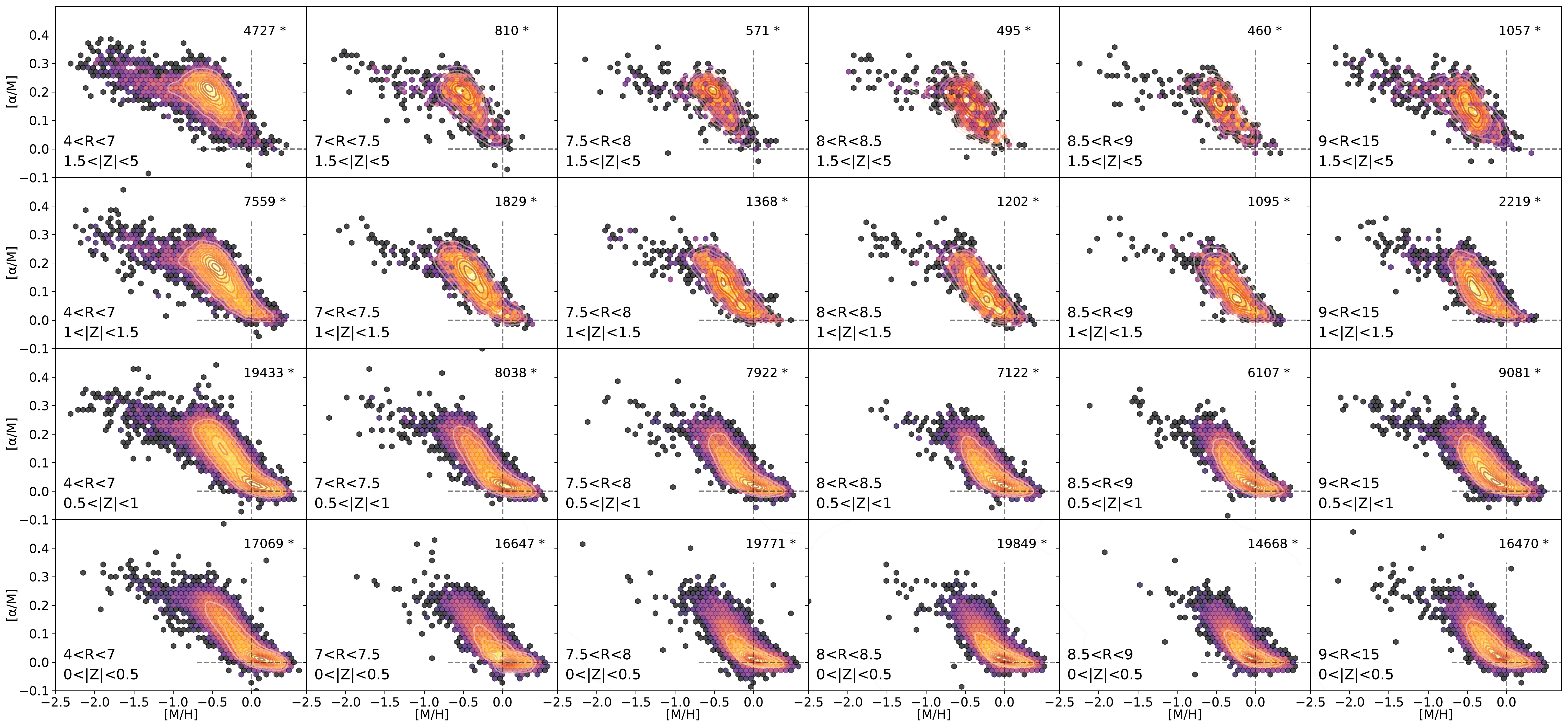}
\caption{\label{hayden_plot}$\alpham$ ratio as a function of 
$\feh$ for several bins of $R$ and $|Z|$. $\alpham$ and $\feh$ were 
derived through our CNN while the $R$ and $|Z|$ come 
from the DR6 of RAVE \citep{steinmetz2020b}. Hexagonal bins and contour plots 
of the data are presented together. In total, we 
present trends for $185\,569$ stars with $\snr>30$, parallax errors 
lower than $20\%$, and RAVE 'n\&o' classification.}
\end{figure*}

In this section, we investigate the spatial transition 
between the $\alpham$-rich and $\alpham$-poor populations of the Milky Way.
We once again take advantage of the orbital parameters provided by 
the sixth data release of RAVE \citep{steinmetz2020b}. 
We present, in \figurename~\ref{hayden_plot}, the behaviour of 
the $\alpham$ ratio as a function of $\feh$ for different bins of mean 
Galactocentric radii ($R$) and heights above the Galactic plane ($|Z|$). 
The figure shows hexagonal density maps and contour plots 
for a total of $185\,569$ giant stars with $\snr>30$, parallax errors 
lower than $20\%$, and RAVE 'n\&o' classification. We observe 
that the $\alpham$-poor population dominates at low Galactic 
heights ($|Z|<0.5$kpc), while $\alpham-$rich stars are mostly 
located at larger height above the plane ($|Z|>0.5$kpc). In between, 
there is a very smooth transition. We note that such 
observations are also valid for the $\mgfe$ and $\sife$ ratios, with 
slightly larger scatter. We find consistent results with the study 
of \citet{hayden2015} based on APOGEE DR12. For the same Galactic volume, 
our results are a good match with the recent study by 
\citet{queiroz2019} based on APOGEE DR16.
We show that we are able to complement RAVE DR6 and ultimately provide chemical abundance 
trends for a larger sample of stars with improved precision.

\section{Caveats}

The present project relies entirely on the cross-match 
between a few thousand RAVE and APOGEE targets, which, together
with the limitations of the two respective surveys, results in a 
number of possible caveats.

Firstly, the spectral range of RAVE spectra, $[8\,410-8\,795]$\AA, 
contains plenty of features with which to derive $\alpham$ ratios, such 
as Ca, Ti, Mg, Si, and O spectral lines. The $\alpham$ labels 
adopted here come from the DR16 of APOGEE. This survey uses a 
different wavelength range (1.51–1.70$\mu$m), nonetheless, its 
wavelength coverage contains similar elements as RAVE 
contributing to the $\alpham$ mixture, 
apart from Ne and S. On the other hand, it is known that the most 
significant contributors of the spectral features are Ca, Ti, Si, O, and Mg.
In this context, using the RAVE spectral range to constrain 
$\alpham$ is reasonable.
    
Secondly, we clearly have a lack of stars at low metallicity ($\mh\lesssim-1$) 
in the training sample, which us mainly due to the fact that 
we have few metal-poor stars in RAVE \citep{matijevic2017} and 
in the cross-match with APOGEE DR16. The mapping of the parameters space 
for those stars is quite limited. For future studies, 
it is important to carefully build a training sample with good mapping 
of the parameters in the metal-poor regime. More and more metal-poor 
stars are being observed, for example, in the Pristine Survey
\citep{starkenburg2017, youakim2017} and they are key stars for 
obtaining a more homogeneous mapping of the parameters space.
    
Finally, out of the approximately $400\,000$ stars of the APOGEE survey 
DR16, our training sample contains roughly $4\,000$ stars 
in common with the RAVE survey. It is clear 
that in the present study the performances of our CNN approach is limited 
by the small size of the training sample. We have seen that the training 
and test samples can then suffer from slightly different coverage 
in the parameter space. The APOGEE and 
RAVE surveys are characterised by different selection functions. 
The selection function of the training sample is then characterised 
by traits common to both surveys. This is a caveat in our study, 
but the goal for the moment is not to characterise the selection 
function in full as this will 
be the object of a future study. Our message here to the 
community is that we call for everyone to make a special effort 
to creating unbiased training samples, especially for the next 
generation of spectroscopic surveys, such as 4MOST, GAIA and WEAVE.


\section{Database and public code}

Here, we present our catalogue of atmospheric parameters ($\teff$, $\logg$ and $\mh$), 
along with chemical abundances ($\feh$, $\alpham$, $\sife$, $\mgfe$, $\alfe$, and $\nife$) 
for $420\,165$ stars (summarised in \tablename~\ref{catalogue}). The 
data table is available at: doi://10.17876/rave/dr.6/19.

\begin{table*}
\centering
\begin{tabular}[c]{l l l l l}
\hline
\hline
Col  & Format & Units & Label  &  Explanations \\
\hline
1   & char  & -    & rave\_obs\_id & RAVE Obs ID                \\
2   & char  & -    & sourceid     & Gaia Source ID              \\
3   & float & K    & teff         & Effective temperature       \\
4   & float & K    & eteff        & Error of $\teff$            \\
5   & int   & -    & flag\_teff   & Boundary flag for $\teff$   \\
6   & float & \cms & logg         & Surface gravity             \\
7   & float & \cms & elogg        & Error on $\logg$            \\
8   & int   & -    & flag\_logg   & Boundary flag for $\logg$   \\
9   & float & dex  & mh           & Overall metallicity         \\
10  & float & dex  & emh          & Error on $\mh$              \\
11  & int   & -    & flag\_mh     & Boundary flag for $\mh$     \\
12  & float & dex  & feh          & $\feh$ ratio                \\
13  & float & dex  & efeh         & Error on $\feh$             \\
14  & int   & -    & flag\_feh    & Boundary flag for $\feh$    \\
15  & float & dex  & alpham       & $\alpham$ ratio             \\
16  & float & dex  & ealpham      & Error on $\alpham$          \\
17  & int   & -    & flag\_alpham & Boundary flag for $\alpham$ \\
18  & float & dex  & sife         & $\sife$ ratio               \\
19  & float & dex  & esife        & Error on $\sife$            \\
20  & int   & -    & flag\_sife   & Boundary flag for $\sife$   \\
21  & float & dex  & mgfe         & $\mgfe$ ratio               \\
22  & float & dex  & emgfe        & Error on $\mgfe$            \\
23  & int   & -    & flag\_mgfe   & Boundary flag for $\mgfe$   \\
24  & float & dex  & alfe         & $\alfe$ ratio               \\
25  & float & dex  & ealfe        & Error on $\alfe$            \\
26  & int   & -    & flag\_alfe   & Boundary flag for $\alfe$   \\
27  & float & dex  & nife         & $\nife$ ratio               \\
28  & float & dex  & enife        & Error on $\nife$            \\
29  & int   & -    & flag\_nife   & Boundary flag for $\nife$   \\
20  & float & /pix & snr          & Signal-to-noise ratio       \\
\hline 
\hline
\end{tabular}
\caption{\label{catalogue}Atmospheric parameters, chemical abundances, 
and boundary flags of the publicly available online catalogue for $420\,165$ stars.}
\end{table*}

The CNN architecture, stellar labels, stellar photometry and 
spectra used in this paper are accessible via 
"https://github.com/gguiglion/CNN\_Guiglion\_et\_al\_2020". The CNN can be easily 
applied to any current spectroscopic archive or survey to derive atmospheric parameters, 
chemical abundances, and also other extra parameters such as rotational velocity.


\section{Conclusion}\label{conclusiooooonnnn}

Here, we list here the main results of our study.

Based on APOGEE DR16, we built a training 
sample composed of $3\,904$ stars in common with RAVE DR6. 
These stars have high quality atmospheric parameters 
and chemical abundances for $\feh$, $\alpham$, $\sife$, 
$\mgfe$, $\alfe$, and $\nife$, which we use as labels.
We built a CNN using the Keras libraries 
in Python to train the labels defined above. 
Using these trained labels, we predicted 
atmospheric parameters and chemical abundances 
for $420\,165$ RAVE spectra, with our results 
available online. Our catalogue covers a larger range 
of \snr\ than RAVE DR6, and extends the 
scientific output of the RAVE spectra.
    
Next, we used ALL\_WISE W1\&2,  2MASS $JHK_s$ and Gaia DR2 
$G$, $G_{BP}$, and $G_{RP}$ apparent magnitudes, and 
extinction estimates to derive absolute magnitudes.
We included them in the training process and showed that CNNs 
are efficient in combining spectroscopic and photometric 
data. We gain a dramatic advantage in precision and accuracy, 
especially in $\teff$ and $\logg$, where spectral features 
are overly degenerate (cool main sequence stars, metal-poor giants, 
and very cool giants). We demonstrated that such a comprehensive 
combination of spectra, photometry, and parallaxes allows us to efficiently 
break degeneracies when the spectral range is too narrow to 
provide strong constraints on surface gravity.

In performing a hundred training phases, 
we derived errors of the atmospheric parameters, 
which typically amount to 60\,K in $\teff$, 0.06 in $\logg$, and 
0.02-0.04 dex for individual chemical abundances. 
Such high precision is realistic because the network is 
able to learn the low-  and high-$\alpha$ sequences in the 
Milky Way disc.
We show that for stars with several observations, the network 
is able to provide precise atmospheric parameters and 
abundances among the repeats that typically precise to 50\,K in $\teff$ and 
0.03-0.05 dex in abundances.

We show that the surface gravities 
match nicely with more than 430 asteroseismic gravities 
from the K2 space mission within 0.14 dex dispersion and no bias.
We compared our effective temperature and surface 
gravities with respect to both the IRFM $\teff$ and $\logg$ from 
the DR6 of RAVE and we were able to characterise the systematics 
between the two studies.

It is important to note that different trends 
and zero-point offsets between this work and external 
studies primarily reflect the different calibrations applied to these surveys. 
A systematic comparison between different surveys 
is therefore crucial.
Furthermore, the CNN architecture and weights will be publicly available.

Despite quite a low number statistics in the training sample 
with respect to the number of free parameters to fit, 
we show that such an approach can provide solid 
scientific output. Of course, the performance 
would improve a lot if the size of the training sample was three to four times 
larger, but this pilot study is limited by the current overlap 
with APOGEE DR16. This study allowed us to highlight possible bias 
and systematics induced by using a limited-size training sample with 
a CNN machine-learning method. For the next generation of surveys, 
the community will have to put strong efforts into producing 
large and un-biased training samples.

Our study shows that CNNs are particularly efficient 
in transferring knowledge from one survey at high resolution, such as APOGEE, 
to another at lower resolution, such as RAVE. This study 
gives good insights for ongoing and future spectroscopic surveys, 
such as Gaia-RVS and 4MOST. The Gaia-RVS spectra are expected to be very similar 
to those of RAVE (R$\sim$11400) and we show that adding 
photometry breaks spectral degeneracies; photometry will 
be available for all RVS targets. Efficient training of Gaia-RVS 
data based on higher-resolution surveys could deliver atmospheric 
parameters and abundances for a larger number of RVS stars, 
as it is the case for RAVE in the present paper. The low-resolution 4MOST 
spectra will cover a much larger spectral range ($4\,000-9\,000$ \AA) 
at a slightly lower resolution than Gaia for the 4MIDABLE-LR 
low-resolution survey \citep{chiappini2019}, and Gaia 
photometry will also be available for all targets. Additional constraints could 
then be put on the derivation of $\teff$ and $\logg$ by coupling 
spectroscopy, photometry, and astrometry. Such surveys will deliver 
millions of spectra that can be analysed in only a few 
minutes on a single graphics processing unit once the labels are trained.\\


\begin{acknowledgements}
The authors acknowledge the anonymous 
referee for the comments 
and suggestions that improved the readability of the paper. 
Funding for RAVE  has been provided by: the Leibniz-Institut 
f\"{u}r Astrophysik Potsdam (AIP); the Australian Astronomical 
Observatory;  the Australian National University; the Australian 
Research Council; the French National Research Agency (Programme 
National Cosmology et Galaxies (PNCG) of CNRS/INSU with INP and 
IN2P3, co-funded by CEA and CNES); the German Research Foundation 
(SPP 1177 and SFB 881); the European Research Council (ERC-StG 
240271 Galactica); the Istituto Nazionale di Astrofisica at Padova; 
The Johns Hopkins University; the National Science Foundation of 
the USA (AST-0908326); the W. M. Keck foundation; the Macquarie 
University; the Netherlands Research School for Astronomy; the 
Natural Sciences and Engineering Research Council of Canada; the 
Slovenian Research Agency (research core funding no. P1-0188); the 
Swiss National Science Foundation; the Science \& Technology 
Facilities Council of the UK; Opticon; Strasbourg Observatory; and 
the Universities of Basel, Groningen, Heidelberg, and Sydney. 
TZ acknowledges financial support of the Slovenian 
Research Agency (research core funding No. P1-0188) 
and of the ESA project PHOTO2CHEM (C4000127986). 
FA is grateful for funding from the European Union’s 
Horizon 2020 research and innovation program under the 
Marie Skłodowska-Curie grant agreement No. 800502. 
This work has made use of data from the European
Space Agency (ESA) mission Gaia (http://www.cosmos.esa.int/gaia),
processed by the Gaia Data Processing and Analysis Consortium 
(DPAC, http://www.cosmos.esa.int/web/gaia/dpac/consortium). 
Funding for the DPAC has been provided by national institutions, 
in particular the institutions participating in the Gaia Multilateral 
Agreement. This publication makes use 
of data products from the Wide-field 
Infrared Survey Explorer, which is a joint project of the 
University of California, Los Angeles, and the Jet Propulsion 
Laboratory/California Institute of Technology, funded by 
the National Aeronautics and Space Administration.
This publication makes use of data products from the 
Two Micron All Sky Survey, which is a joint project of 
the University of Massachusetts and the Infrared Processing 
and Analysis Center/California Institute of Technology, 
funded by the National Aeronautics and Space Administration 
and the National Science Foundation.
\end{acknowledgements}


\bibliographystyle{aa}
\bibliography{cite_r_s}

\begin{thebibliography}{94}
\expandafter\ifx\csname natexlab\endcsname\relax\def\natexlab#1{#1}\fi

\bibitem[{{Adibekyan} {et~al.}(2012){Adibekyan}, {Sousa}, {Santos}, {Delgado
  Mena}, {Gonz{\'a}lez Hern{\'a}ndez}, {Israelian}, {Mayor}, \&
  {Khachatryan}}]{adibekyan2012}
{Adibekyan}, V.~Z., {Sousa}, S.~G., {Santos}, N.~C., {et~al.} 2012, \aap, 545,
  A32

\bibitem[{{Ahumada} {et~al.}(2020){Ahumada}, {Allende Prieto}, {Almeida},
  {Anders}, {Anderson}, {Andrews}, {Anguiano}, {Arcodia}, {Armengaud},
  {Aubert}, {Avila}, {Avila-Reese}, {Badenes}, {Balland }, {Barger},
  {Barrera-Ballesteros}, {Basu}, {Bautista}, {Beaton}, {Beers}, {Benavides},
  {Bender}, {Bernardi}, {Bershady}, {Beutler}, {Bidin}, {Bird}, {Bizyaev},
  {Blanc}, {Blanton}, {Boquien}, {Borissova}, {Bovy}, {Brand t}, {Brinkmann},
  {Brownstein}, {Bundy}, {Bureau}, {Burgasser}, {Burtin}, {Cano-D{\'\i}az},
  {Capasso}, {Cappellari}, {Carrera}, {Chabanier}, {Chaplin}, {Chapman},
  {Cherinka}, {Chiappini}, {Doohyun Choi}, {Chojnowski}, {Chung}, {Clerc},
  {Coffey}, {Comerford}, {Comparat}, {da Costa}, {Cousinou}, {Covey}, {Crane},
  {Cunha}, {da Silva Ilha}, {Dai}, {Damsted}, {Darling}, {Davidson}, {Davies},
  {Dawson}, {De}, {de la Macorra}, {De Lee}, {de Andrade Queiroz}, {Deconto
  Machado}, {de la Torre}, {Dell'Agli}, {du Mas des Bourboux},
  {Diamond-Stanic}, {Dillon}, {Donor}, {Drory}, {Duckworth}, {Dwelly},
  {Ebelke}, {Eftekharzadeh}, {Eigenbrot}, {Elsworth}, {Eracleous},
  {Erfanianfar}, {Escoffier}, {Fan}, {Farr}, {Fern{\'a}ndez-Trincado},
  {Feuillet}, {Finoguenov}, {Fofie}, {Fraser-McKelvie}, {Frinchaboy},
  {Fromenteau}, {Fu}, {Galbany}, {Garcia}, {Garc{\'\i}a-Hern{\'a}ndez}, {Garma
  Oehmichen}, {Ge}, {Geimba Maia}, {Geisler}, {Gelfand }, {Goddy},
  {Gonzalez-Perez}, {Grabowski}, {Green}, {Grier}, {Guo}, {Guy}, {Harding},
  {Hasselquist}, {Hawken}, {Hayes}, {Hearty}, {Hekker}, {Hogg}, {Holtzman},
  {Horta}, {Hou}, {Hsieh}, {Huber}, {Hunt}, {Ider Chitham}, {Imig}, {Jaber},
  {Jimenez Angel}, {Johnson}, {Jones}, {J{\"o}nsson}, {Jullo}, {Kim},
  {Kinemuchi}, {Kirkpatrick}, {Kite}, {Klaene}, {Kneib}, {Kollmeier}, {Kong},
  {Kounkel}, {Krishnarao}, {Lacerna}, {Lan}, {Lane}, {Law}, {Le Goff}, {Leung},
  {Lewis}, {Li}, {Lian}, {Lin}, {Long}, {Longa-Pe{\~n}a}, {Lundgren}, {Lyke},
  {Ted Mackereth}, {MacLeod}, {Majewski}, {Manchado}, {Maraston}, {Martini},
  {Masseron}, {Masters}, {Mathur}, {McDermid}, {Merloni}, {Merrifield},
  {M{\'e}sz{\'a}ros}, {Miglio}, {Minniti}, {Minsley}, {Miyaji}, {Mohammad},
  {Mosser}, {Mueller}, {Muna}, {Mu{\~n}oz-Guti{\'e}rrez}, {Myers}, {Nadathur},
  {Nair}, {Nandra}, {do Nascimento}, {Nevin}, {Newman}, {Nidever}, {Nitschelm},
  {Noterdaeme}, {O'Connell}, {Olmstead}, {Oravetz}, {Oravetz}, {Osorio},
  {Pace}, {Padilla}, {Palanque-Delabrouille}, {Palicio}, {Pan}, {Pan},
  {Parker}, {Paviot}, {Peirani}, {Pe{\~n}a Ram{\'r}ez}, {Penny}, {Percival},
  {Perez-Fournon}, {P{\'e}rez-R{\`a}fols}, {Petitjean}, {Pieri},
  {Pinsonneault}, {Poovelil}, {Povick}, {Prakash}, {Price-Whelan}, {Raddick},
  {Raichoor}, {Ray}, {Rembold}, {Rezaie}, {Riffel}, {Riffel}, {Rix}, {Robin},
  {Roman-Lopes}, {Rom{\'a}n-Z{\'u}{\~n}iga}, {Rose}, {Ross}, {Rossi}, {Rowland
  s}, {Rubin}, {Salvato}, {S{\'a}nchez}, {S{\'a}nchez-Menguiano},
  {S{\'a}nchez-Gallego}, {Sayres}, {Schaefer}, {Schiavon}, {Schimoia},
  {Schlafly}, {Schlegel}, {Schneider}, {Schultheis}, {Schwope}, {Seo},
  {Serenelli}, {Shafieloo}, {Shamsi}, {Shao}, {Shen}, {Shetrone}, {Shirley},
  {Silva Aguirre}, {Simon}, {Skrutskie}, {Slosar}, {Smethurst}, {Sobeck},
  {Sodi}, {Souto}, {Stark}, {Stassun}, {Steinmetz}, {Stello}, {Stermer},
  {Storchi-Bergmann}, {Streblyanska}, {Stringfellow}, {Stutz}, {Su{\'a}rez},
  {Sun}, {Taghizadeh-Popp}, {Talbot}, {Tayar}, {Thakar}, {Theriault}, {Thomas},
  {Thomas}, {Tinker}, {Tojeiro}, {Toledo}, {Tremonti}, {Troup}, {Tuttle},
  {Unda-Sanzana}, {Valentini}, {Vargas-Gonz{\'a}lez}, {Vargas-Maga{\~n}a},
  {V{\'a}zquez-Mata}, {Vivek}, {Wake}, {Wang}, {Weaver}, {Weijmans}, {Wild},
  {Wilson}, {Wilson}, {Wolthuis}, {Wood-Vasey}, {Yan}, {Yang}, {Y{\`e}che},
  {Zamora}, {Zarrouk}, {Zasowski}, {Zhang}, {Zhao}, {Zhao}, {Zheng}, {Zheng},
  {Zhu}, \& {Zou}}]{apogeedr16}
{Ahumada}, R., {Allende Prieto}, C., {Almeida}, A., {et~al.} 2020, \apjs, 249,
  3

\bibitem[{{Allende Prieto} {et~al.}(2006){Allende Prieto}, {Beers}, {Wilhelm},
  {Newberg}, {Rockosi}, {Yanny}, \& {Lee}}]{allendeprieto206}
{Allende Prieto}, C., {Beers}, T.~C., {Wilhelm}, R., {et~al.} 2006, \apj, 636,
  804

\bibitem[{{Anders} {et~al.}(2019){Anders}, {Khalatyan}, {Chiappini}, {Queiroz},
  {Santiago}, {Jordi}, {Girardi}, {Brown}, {Matijevi{\v{c}}}, {Monari},
  {Cantat-Gaudin}, {Weiler}, {Khan}, {Miglio}, {Carrillo}, {Romero-G{\'o}mez},
  {Minchev}, {de Jong}, {Antoja}, {Ramos}, {Steinmetz}, \& {Enke}}]{anders2019}
{Anders}, F., {Khalatyan}, A., {Chiappini}, C., {et~al.} 2019, \aap, 628, A94

\bibitem[{{Antoja} {et~al.}(2017){Antoja}, {Kordopatis}, {Helmi}, {Monari},
  {Famaey}, {Wyse}, {Grebel}, {Steinmetz}, {Bland-Hawthorn}, {Gibson},
  {Bienaym{\'e}}, {Navarro}, {Parker}, {Reid}, {Seabroke}, {Siebert},
  {Siviero}, \& {Zwitter}}]{antoja2017}
{Antoja}, T., {Kordopatis}, G., {Helmi}, A., {et~al.} 2017, \aap, 601, A59

\bibitem[{{Arenou} {et~al.}(2018){Arenou}, {Luri}, {Babusiaux}, {Fabricius},
  {Helmi}, {Muraveva}, {Robin}, {Spoto}, {Vallenari}, {Antoja},
  {Cantat-Gaudin}, {Jordi}, {Leclerc}, {Reyl{\'e}}, {Romero-G{\'o}mez}, {Shih},
  {Soria}, {Barache}, {Bossini}, {Bragaglia}, {Breddels}, {Fabrizio},
  {Lambert}, {Marrese}, {Massari}, {Moitinho}, {Robichon}, {Ruiz-Dern},
  {Sordo}, {Veljanoski}, {Eyer}, {Jasniewicz}, {Pancino}, {Soubiran}, {Spagna},
  {Tanga}, {Turon}, \& {Zurbach}}]{arenou2018}
{Arenou}, F., {Luri}, X., {Babusiaux}, C., {et~al.} 2018, \aap, 616, A17

\bibitem[{{Bensby} {et~al.}(2019){Bensby}, {Bergemann}, {Rybizki}, {Lemasle},
  {Howes}, {Kovalev}, {Agertz}, {Asplund}, {Barklem}, {Battistini},
  {Casagrande}, {Chiappini}, {Church}, {Feltzing}, {Ford}, {Gerhard},
  {Kushniruk}, {Kordopatis}, {Lind}, {Minchev}, {McMillan}, {Rix}, {Ryde}, \&
  {Traven}}]{bensby2019}
{Bensby}, T., {Bergemann}, M., {Rybizki}, J., {et~al.} 2019, The Messenger,
  175, 35

\bibitem[{{Bensby} {et~al.}(2014){Bensby}, {Feltzing}, \& {Oey}}]{bensby2014}
{Bensby}, T., {Feltzing}, S., \& {Oey}, M.~S. 2014, \aap, 562, A71

\bibitem[{{Bialek} {et~al.}(2020){Bialek}, {Fabbro}, {Venn}, {Kumar},
  {O'Briain}, \& {Yi}}]{bialek2019}
{Bialek}, S., {Fabbro}, S., {Venn}, K.~A., {et~al.} 2020, \mnras

\bibitem[{Bijaoui {et~al.}(2012)Bijaoui, Recio-Blanco, De~Laverny, \&
  Ordenovic}]{GAUGUIN12}
Bijaoui, A., Recio-Blanco, A., De~Laverny, P., \& Ordenovic, C. 2012,
  {Statistical Methodology}, 9, 55

\bibitem[{{Boeche} {et~al.}(2013{\natexlab{a}}){Boeche}, {Chiappini},
  {Minchev}, {Williams}, {Steinmetz}, {Sharma}, {Kordopatis}, {Bland-Hawthorn},
  {Bienaym{\'e}}, {Gibson}, {Gilmore}, {Grebel}, {Helmi}, {Munari}, {Navarro},
  {Parker}, {Reid}, {Seabroke}, {Siebert}, {Siviero}, {Watson}, {Wyse}, \&
  {Zwitter}}]{boeche2013a}
{Boeche}, C., {Chiappini}, C., {Minchev}, I., {et~al.} 2013{\natexlab{a}},
  \aap, 553, A19

\bibitem[{{Boeche} \& {Grebel}(2018)}]{boeche2018}
{Boeche}, C. \& {Grebel}, E.~K. 2018, {SP\_Ace: Stellar Parameters And Chemical
  abundances Estimator}

\bibitem[{{Boeche} {et~al.}(2014){Boeche}, {Siebert}, {Piffl}, {Just},
  {Steinmetz}, {Grebel}, {Sharma}, {Kordopatis}, {Gilmore}, {Chiappini},
  {Freeman}, {Gibson}, {Munari}, {Siviero}, {Bienaym{\'e}}, {Navarro},
  {Parker}, {Reid}, {Seabroke}, {Watson}, {Wyse}, \& {Zwitter}}]{boeche2014}
{Boeche}, C., {Siebert}, A., {Piffl}, T., {et~al.} 2014, \aap, 568, A71

\bibitem[{{Boeche} {et~al.}(2013{\natexlab{b}}){Boeche}, {Siebert}, {Piffl},
  {Just}, {Steinmetz}, {Sharma}, {Kordopatis}, {Gilmore}, {Chiappini},
  {Williams}, {Grebel}, {Bland-Hawthorn}, {Gibson}, {Munari}, {Siviero},
  {Bienaym{\'e}}, {Navarro}, {Parker}, {Reid}, {Seabroke}, {Watson}, {Wyse}, \&
  {Zwitter}}]{boeche2013b}
{Boeche}, C., {Siebert}, A., {Piffl}, T., {et~al.} 2013{\natexlab{b}}, \aap,
  559, A59

\bibitem[{{Boeche} {et~al.}(2011){Boeche}, {Siebert}, {Williams}, {de Jong},
  {Steinmetz}, {Fulbright}, {Ruchti}, {Bienaym{\'e}}, {Bland-Hawthorn},
  {Campbell}, {Freeman}, {Gibson}, {Gilmore}, {Grebel}, {Helmi}, {Munari},
  {Navarro}, {Parker}, {Reid}, {Seabroke}, {Siviero}, {Watson}, {Wyse}, \&
  {Zwitter}}]{boeche2011}
{Boeche}, C., {Siebert}, A., {Williams}, M., {et~al.} 2011, \aj, 142, 193

\bibitem[{{Buder} {et~al.}(2018){Buder}, {Asplund}, {Duong}, {Kos}, {Lind},
  {Ness}, {Sharma}, {Bland -Hawthorn}, {Casey}, {de Silva}, {D'Orazi},
  {Freeman}, {Lewis}, {Lin}, {Martell}, {Schlesinger}, {Simpson}, {Zucker},
  {Zwitter}, {Amarsi}, {Anguiano}, {Carollo}, {Casagrande}, {{\v{C}}otar},
  {Cottrell}, {da Costa}, {Gao}, {Hayden}, {Horner}, {Ireland}, {Kafle},
  {Munari}, {Nataf}, {Nordlander}, {Stello}, {Ting}, {Traven}, {Watson},
  {Wittenmyer}, {Wyse}, {Yong}, {Zinn}, {{\v{Z}}erjal}, \& {Galah
  Collaboration}}]{buder2018}
{Buder}, S., {Asplund}, M., {Duong}, L., {et~al.} 2018, \mnras, 478, 4513

\bibitem[{{Buder} {et~al.}(2019){Buder}, {Lind}, {Ness}, {Asplund}, {Duong},
  {Lin}, {Kos}, {Casagrande}, {Casey}, {Bland-Hawthorn}, {de Silva}, {D'Orazi},
  {Freeman}, {Martell}, {Schlesinger}, {Sharma}, {Simpson}, {Zucker},
  {Zwitter}, {{\v{C}}otar}, {Dotter}, {Hayden}, {Hyde}, {Kafle}, {Lewis},
  {Nataf}, {Nordlander}, {Reid}, {Rix}, {Sk{\'u}lad{\'o}ttir}, {Stello},
  {Ting}, {Traven}, {Wyse}, \& {Galah Collaboration}}]{buder2019}
{Buder}, S., {Lind}, K., {Ness}, M.~K., {et~al.} 2019, \aap, 624, A19

\bibitem[{{Carretta} {et~al.}(2009){Carretta}, {Bragaglia}, {Gratton}, \&
  {Lucatello}}]{carretta2009}
{Carretta}, E., {Bragaglia}, A., {Gratton}, R., \& {Lucatello}, S. 2009, \aap,
  505, 139

\bibitem[{{Casagrande} {et~al.}(2006){Casagrande}, {Portinari}, \&
  {Flynn}}]{casagrande2006}
{Casagrande}, L., {Portinari}, L., \& {Flynn}, C. 2006, \mnras, 373, 13

\bibitem[{{Casagrande} {et~al.}(2010){Casagrande}, {Ram{\'{\i}}rez},
  {Mel{\'e}ndez}, {Bessell}, \& {Asplund}}]{casagrande2010}
{Casagrande}, L., {Ram{\'{\i}}rez}, I., {Mel{\'e}ndez}, J., {Bessell}, M., \&
  {Asplund}, M. 2010, \aap, 512, A54

\bibitem[{{Casey} {et~al.}(2017){Casey}, {Hawkins}, {Hogg}, {Ness}, {Rix},
  {Kordopatis}, {Kunder}, {Steinmetz}, {Koposov}, {Enke}, {Sanders}, {Gilmore},
  {Zwitter}, {Freeman}, {Casagrand e}, {Matijevi{\v{c}}}, {Seabroke},
  {Bienaym{\'e}}, {Bland-Hawthorn}, {Gibson}, {Grebel}, {Helmi}, {Munari},
  {Navarro}, {Reid}, {Siebert}, \& {Wyse}}]{casey2017}
{Casey}, A.~R., {Hawkins}, K., {Hogg}, D.~W., {et~al.} 2017, \apj, 840, 59

\bibitem[{{Casey} {et~al.}(2016){Casey}, {Hogg}, {Ness}, {Rix}, {Ho}, \&
  {Gilmore}}]{casey2016}
{Casey}, A.~R., {Hogg}, D.~W., {Ness}, M., {et~al.} 2016, arXiv e-prints,
  arXiv:1603.03040

\bibitem[{{Chiappini} {et~al.}(2019){Chiappini}, {Minchev}, {Starkenburg},
  {Anders}, {Fusillo}, {Gerhard}, {Guiglion}, {Khalatyan}, {Kordopatis},
  {Lemasle}, {Matijevic}, {Queiroz}, {Schwope}, {Steinmetz}, {Storm}, {Traven},
  {Tremblay}, {Valentini}, {Andrae}, {Arentsen}, {Asplund}, {Bensby},
  {Bergemann}, {Casagrande}, {Church}, {Cescutti}, {Feltzing}, {Fouesneau},
  {Grebel}, {Kovalev}, {McMillan}, {Monari}, {Rybizki}, {Ryde}, {Rix},
  {Walton}, {Xiang}, {Zucker}, \& {4MIDABLE-Lr Team}}]{chiappini2019}
{Chiappini}, C., {Minchev}, I., {Starkenburg}, E., {et~al.} 2019, The
  Messenger, 175, 30

\bibitem[{Chollet {et~al.}(2015)}]{chollet2015keras}
Chollet, F. {et~al.} 2015, Keras, \url{https://github.com/fchollet/keras}

\bibitem[{{Cire{\textcommabelow s}an} {et~al.}(2011){Cire{\textcommabelow
  s}an}, {Meier}, {Masci}, {Gambardella}, \& {Schmidhuber}}]{ciresan2011}
{Cire{\textcommabelow s}an}, D.~C., {Meier}, U., {Masci}, J., {Gambardella},
  L.~M., \& {Schmidhuber}, J. 2011, arXiv e-prints, arXiv:1102.0183

\bibitem[{{Dalton} {et~al.}(2018){Dalton}, {Trager}, {Abrams}, {Bonifacio},
  {Aguerri}, {Vallenari}, {Middleton}, {Benn}, {Dee}, {Say{\`e}de}, {Lewis},
  {Pragt}, {Pic{\'o}}, {Walton}, {Rey}, {Allende}, {Lhom{\'e}}, {Terrett},
  {Brock}, {Gilbert}, {Ridings}, {Verheijen}, {Tosh}, {Steele}, {Stuik},
  {Kroes}, {Tromp}, {Kragt}, {Lesman}, {Mottram}, {Bates}, {Gribbin}, {Burgal},
  {Herreros}, {Delgado}, {Martin}, {Cano}, {Navarro}, {Irwin}, {Lewis},
  {Gonzales Solares}, {O'Mahony}, {Bianco}, {Zurita}, {ter Horst}, {Molinari},
  {Lodi}, {Guerra}, {Baruffolo}, {Carrasco}, {Farkas}, {Schallig}, {Hill},
  {Smith}, {Drew}, {Poggianti}, {Pieri}, {Jin}, {Dominquez Palmero},
  {Fari{\~n}a}, {Martin}, {Worley}, {Murphy}, {Hidalgo}, {Mignot}, {Bishop},
  {Guest}, {Elswijk}, {de Haan}, {Hanenburg}, {Salasnich}, {Mayya},
  {Izazaga-P{\'e}rez}, \& {Peralta de Arriba}}]{WEAVE}
{Dalton}, G., {Trager}, S., {Abrams}, D.~C., {et~al.} 2018, in Society of
  Photo-Optical Instrumentation Engineers (SPIE) Conference Series, Vol. 10702,
  \procspie, 107021B

\bibitem[{{de Jong} {et~al.}(2019){de Jong}, {Agertz}, {Berbel}, {Aird},
  {Alexander}, {Amarsi}, {Anders}, {Andrae}, {Ansarinejad}, {Ansorge},
  {Antilogus}, {Anwand -Heerwart}, {Arentsen}, {Arnadottir}, {Asplund},
  {Auger}, {Azais}, {Baade}, {Baker}, {Baker}, {Balbinot}, {Baldry}, {Banerji},
  {Barden}, {Barklem}, {Barth{\'e}l{\'e}my-Mazot}, {Battistini}, {Bauer},
  {Bell}, {Bellido-Tirado}, {Bellstedt}, {Belokurov}, {Bensby}, {Bergemann},
  {Bestenlehner}, {Bielby}, {Bilicki}, {Blake}, {Bland-Hawthorn}, {Boeche},
  {Boland}, {Boller}, {Bongard}, {Bongiorno}, {Bonifacio}, {Boudon}, {Brooks},
  {Brown}, {Brown}, {Br{\"u}ggen}, {Brynnel}, {Brzeski}, {Buchert},
  {Buschkamp}, {Caffau}, {Caillier}, {Carrick}, {Casagrande}, {Case}, {Casey},
  {Cesarini}, {Cescutti}, {Chapuis}, {Chiappini}, {Childress}, {Christlieb},
  {Church}, {Cioni}, {Cluver}, {Colless}, {Collett}, {Comparat}, {Cooper},
  {Couch}, {Courbin}, {Croom}, {Croton}, {Daguis{\'e}}, {Dalton}, {Davies},
  {Davis}, {de Laverny}, {Deason}, {Dionies}, {Disseau}, {Doel}, {D{\"o}scher},
  {Driver}, {Dwelly}, {Eckert}, {Edge}, {Edvardsson}, {Youssoufi}, {Elhaddad},
  {Enke}, {Erfanianfar}, {Farrell}, {Fechner}, {Feiz}, {Feltzing}, {Ferreras},
  {Feuerstein}, {Feuillet}, {Finoguenov}, {Ford}, {Fotopoulou}, {Fouesneau},
  {Frenk}, {Frey}, {Gaessler}, {Geier}, {Fusillo}, {Gerhard}, {Giannantonio},
  {Giannone}, {Gibson}, {Gillingham}, {Gonz{\'a}lez-Fern{\'a}ndez},
  {Gonzalez-Solares}, {Gottloeber}, {Gould}, {Grebel}, {Gueguen}, {Guiglion},
  {Haehnelt}, {Hahn}, {Hansen}, {Hartman}, {Hauptner}, {Hawkins}, {Haynes},
  {Haynes}, {Heiter}, {Helmi}, {Aguayo}, {Hewett}, {Hinton}, {Hobbs}, {Hoenig},
  {Hofman}, {Hook}, {Hopgood}, {Hopkins}, {Hourihane}, {Howes}, {Howlett},
  {Huet}, {Irwin}, {Iwert}, {Jablonka}, {Jahn}, {Jahnke}, {Jarno}, {Jin},
  {Jofre}, {Johl}, {Jones}, {J{\"o}nsson}, {Jordan}, {Karovicova}, {Khalatyan},
  {Kelz}, {Kennicutt}, {King}, {Kitaura}, {Klar}, {Klauser}, {Kneib}, {Koch},
  {Koposov}, {Kordopatis}, {Korn}, {Kosmalski}, {Kotak}, {Kovalev}, {Kreckel},
  {Kripak}, {Krumpe}, {Kuijken}, {Kunder}, {Kushniruk}, {Lam}, {Lamer},
  {Laurent}, {Lawrence}, {Lehmitz}, {Lemasle}, {Lewis}, {Li}, {Lidman}, {Lind},
  {Liske}, {Lizon}, {Loveday}, {Ludwig}, {McDermid}, {Maguire}, {Mainieri},
  {Mali}, {Mandel}, {Mandel}, {Mannering}, {Martell}, {Martinez Delgado},
  {Matijevic}, {McGregor}, {McMahon}, {McMillan}, {Mena}, {Merloni}, {Meyer},
  {Michel}, {Micheva}, {Migniau}, {Minchev}, {Monari}, {Muller}, {Murphy},
  {Muthukrishna}, {Nandra}, {Navarro}, {Ness}, {Nichani}, {Nichol}, {Nicklas},
  {Niederhofer}, {Norberg}, {Obreschkow}, {Oliver}, {Owers}, {Pai},
  {Pankratow}, {Parkinson}, {Paschke}, {Paterson}, {Pecontal}, {Parry},
  {Phillips}, {Pillepich}, {Pinard}, {Pirard}, {Piskunov}, {Plank},
  {Pl{\"u}schke}, {Pons}, {Popesso}, {Power}, {Pragt}, {Pramskiy}, {Pryer},
  {Quattri}, {Queiroz}, {Quirrenbach}, {Rahurkar}, {Raichoor}, {Ramstedt},
  {Rau}, {Recio-Blanco}, {Reiss}, {Renaud}, {Revaz}, {Rhode}, {Richard},
  {Richter}, {Rix}, {Robotham}, {Roelfsema}, {Romaniello}, {Rosario},
  {Rothmaier}, {Roukema}, {Ruchti}, {Rupprecht}, {Rybizki}, {Ryde}, {Saar},
  {Sadler}, {Sahl{\'e}n}, {Salvato}, {Sassolas}, {Saunders}, {Saviauk},
  {Sbordone}, {Schmidt}, {Schnurr}, {Scholz}, {Schwope}, {Seifert}, {Shanks},
  {Sheinis}, {Sivov}, {Sk{\'u}lad{\'o}ttir}, {Smartt}, {Smedley}, {Smith},
  {Smith}, {Sorce}, {Spitler}, {Starkenburg}, {Steinmetz}, {Stilz}, {Storm},
  {Sullivan}, {Sutherland}, {Swann}, {Tamone}, {Taylor}, {Teillon}, {Tempel},
  {ter Horst}, {Thi}, {Tolstoy}, {Trager}, {Traven}, {Tremblay}, {Tresse},
  {Valentini}, {van de Weygaert}, {van den Ancker}, {Veljanoski}, {Venkatesan},
  {Wagner}, {Wagner}, {Walcher}, {Waller}, {Walton}, {Wang}, {Winkler},
  {Wisotzki}, {Worley}, {Worseck}, {Xiang}, {Xu}, {Yong}, {Zhao}, {Zheng},
  {Zscheyge}, \& {Zucker}}]{4MOST}
{de Jong}, R.~S., {Agertz}, O., {Berbel}, A.~A., {et~al.} 2019, The Messenger,
  175, 3

\bibitem[{{Fabbro} {et~al.}(2018){Fabbro}, {Venn}, {O'Briain}, {Bialek},
  {Kielty}, {Jahandar}, \& {Monty}}]{fabbro2018}
{Fabbro}, S., {Venn}, K.~A., {O'Briain}, T., {et~al.} 2018, \mnras, 475, 2978

\bibitem[{{Ford} {et~al.}(2005){Ford}, {Jeffries}, \& {Smalley}}]{ford2005}
{Ford}, A., {Jeffries}, R.~D., \& {Smalley}, B. 2005, \mnras, 364, 272

\bibitem[{{Freeman} \& {Bland-Hawthorn}(2002)}]{freeman2002}
{Freeman}, K. \& {Bland-Hawthorn}, J. 2002, \araa, 40, 487

\bibitem[{{Funayama} {et~al.}(2009){Funayama}, {Itoh}, {Oasa}, {Toyota},
  {Hashimoto}, \& {Mukai}}]{funayama2009}
{Funayama}, H., {Itoh}, Y., {Oasa}, Y., {et~al.} 2009, \pasj, 61, 931

\bibitem[{{Gaia Collaboration} {et~al.}(2018{\natexlab{a}}){Gaia
  Collaboration}, {Babusiaux}, {van Leeuwen}, {Barstow}, {Jordi}, {Vallenari},
  {Bossini}, {Bressan}, {Cantat-Gaudin}, {van Leeuwen}, {Brown}, {Prusti}, {de
  Bruijne}, {Bailer-Jones}, {Biermann}, {Evans}, {Eyer}, {Jansen}, {Klioner},
  {Lammers}, {Lindegren}, {Luri}, {Mignard}, {Panem}, {Pourbaix}, {Randich},
  {Sartoretti}, {Siddiqui}, {Soubiran}, {Walton}, {Arenou}, {Bastian},
  {Cropper}, {Drimmel}, {Katz}, {Lattanzi}, {Bakker}, {Cacciari},
  {Casta{\~n}eda}, {Chaoul}, {Cheek}, {De Angeli}, {Fabricius}, {Guerra},
  {Holl}, {Masana}, {Messineo}, {Mowlavi}, {Nienartowicz}, {Panuzzo},
  {Portell}, {Riello}, {Seabroke}, {Tanga}, {Th{\'e}venin}, {Gracia-Abril},
  {Comoretto}, {Garcia-Reinaldos}, {Teyssier}, {Altmann}, {Andrae}, {Audard},
  {Bellas-Velidis}, {Benson}, {Berthier}, {Blomme}, {Burgess}, {Busso},
  {Carry}, {Cellino}, {Clementini}, {Clotet}, {Creevey}, {Davidson}, {De
  Ridder}, {Delchambre}, {Dell'Oro}, {Ducourant},
  {Fern{\'a}ndez-Hern{\'a}ndez}, {Fouesneau}, {Fr{\'e}mat}, {Galluccio},
  {Garc{\'\i}a-Torres}, {Gonz{\'a}lez-N{\'u}{\~n}ez}, {Gonz{\'a}lez-Vidal},
  {Gosset}, {Guy}, {Halbwachs}, {Hambly}, {Harrison}, {Hern{\'a}ndez},
  {Hestroffer}, {Hodgkin}, {Hutton}, {Jasniewicz}, {Jean-Antoine-Piccolo},
  {Jordan}, {Korn}, {Krone-Martins}, {Lanzafame}, {Lebzelter}, {L{\"o}ffler},
  {Manteiga}, {Marrese}, {Mart{\'\i}n-Fleitas}, {Moitinho}, {Mora}, {Muinonen},
  {Osinde}, {Pancino}, {Pauwels}, {Petit}, {Recio-Blanco}, {Richards},
  {Rimoldini}, {Robin}, {Sarro}, {Siopis}, {Smith}, {Sozzetti}, {S{\"u}veges},
  {Torra}, {van Reeven}, {Abbas}, {Abreu Aramburu}, {Accart}, {Aerts},
  {Altavilla}, {{\'A}lvarez}, {Alvarez}, {Alves}, {Anderson}, {Andrei},
  {Anglada Varela}, {Antiche}, {Antoja}, {Arcay}, {Astraatmadja}, {Bach},
  {Baker}, {Balaguer-N{\'u}{\~n}ez}, {Balm}, {Barache}, {Barata}, {Barbato},
  {Barblan}, {Barklem}, {Barrado}, {Barros}, {Bartholom{\'e} Mu{\~n}oz},
  {Bassilana}, {Becciani}, {Bellazzini}, {Berihuete}, {Bertone}, {Bianchi},
  {Bienaym{\'e}}, {Blanco-Cuaresma}, {Boch}, {Boeche}, {Bombrun}, {Borrachero},
  {Bouquillon}, {Bourda}, {Bragaglia}, {Bramante}, {Breddels}, {Brouillet},
  {Br{\"u}semeister}, {Brugaletta}, {Bucciarelli}, {Burlacu}, {Busonero},
  {Butkevich}, {Buzzi}, {Caffau}, {Cancelliere}, {Cannizzaro}, {Carballo},
  {Carlucci}, {Carrasco}, {Casamiquela}, {Castellani}, {Castro-Ginard},
  {Charlot}, {Chemin}, {Chiavassa}, {Cocozza}, {Costigan}, {Cowell}, {Crifo},
  {Crosta}, {Crowley}, {Cuypers}, {Dafonte}, {Damerdji}, {Dapergolas}, {David},
  {David}, {de Laverny}, {De Luise}, {De March}, {de Martino}, {de Souza}, {de
  Torres}, {Debosscher}, {del Pozo}, {Delbo}, {Delgado}, {Delgado}, {Diakite},
  {Diener}, {Distefano}, {Dolding}, {Drazinos}, {Dur{\'a}n}, {Edvardsson},
  {Enke}, {Eriksson}, {Esquej}, {Eynard Bontemps}, {Fabre}, {Fabrizio},
  {Faigler}, {Falc{\~a}o}, {Farr{\`a}s Casas}, {Federici}, {Fedorets},
  {Fernique}, {Figueras}, {Filippi}, {Findeisen}, {Fonti}, {Fraile}, {Fraser},
  {Fr{\'e}zouls}, {Gai}, {Galleti}, {Garabato}, {Garc{\'\i}a-Sedano},
  {Garofalo}, {Garralda}, {Gavel}, {Gavras}, {Gerssen}, {Geyer}, {Giacobbe},
  {Gilmore}, {Girona}, {Giuffrida}, {Glass}, {Gomes}, {Granvik}, {Gueguen},
  {Guerrier}, {Guiraud}, {Guti{\'e}}, {Haigron}, {Hatzidimitriou}, {Hauser},
  {Haywood}, {Heiter}, {Helmi}, {Heu}, {Hilger}, {Hobbs}, {Hofmann}, {Holland},
  {Huckle}, {Hypki}, {Icardi}, {Jan{\ss}en}, {Jevardat de Fombelle}, {Jonker},
  {Juh{\'a}sz}, {Julbe}, {Karampelas}, {Kewley}, {Klar}, {Kochoska}, {Kohley},
  {Kolenberg}, {Kontizas}, {Kontizas}, {Koposov}, {Kordopatis},
  {Kostrzewa-Rutkowska}, {Koubsky}, {Lambert}, {Lanza}, {Lasne}, {Lavigne}, {Le
  Fustec}, {Le Poncin-Lafitte}, {Lebreton}, {Leccia}, {Leclerc},
  {Lecoeur-Taibi}, {Lenhardt}, {Leroux}, {Liao}, {Licata}, {Lindstr{\o}m},
  {Lister}, {Livanou}, {Lobel}, {L{\'o}pez}, {Managau}, {Mann}, {Mantelet},
  {Marchal}, {Marchant}, {Marconi}, {Marinoni}, {Marschalk{\'o}}, {Marshall},
  {Martino}, {Marton}, {Mary}, {Massari}, {Matijevi{\v{c}}}, {Mazeh},
  {McMillan}, {Messina}, {Michalik}, {Millar}, {Molina}, {Molinaro},
  {Moln{\'a}r}, {Montegriffo}, {Mor}, {Morbidelli}, {Morel}, {Morris},
  {Mulone}, {Muraveva}, {Musella}, {Nelemans}, {Nicastro}, {Noval},
  {O'Mullane}, {Ord{\'e}novic}, {Ord{\'o}{\~n}ez-Blanco}, {Osborne}, {Pagani},
  {Pagano}, {Pailler}, {Palacin}, {Palaversa}, {Panahi}, {Pawlak},
  {Piersimoni}, {Pineau}, {Plachy}, {Plum}, {Poggio}, {Poujoulet},
  {Pr{\v{s}}a}, {Pulone}, {Racero}, {Ragaini}, {Rambaux}, {Ramos-Lerate},
  {Regibo}, {Reyl{\'e}}, {Riclet}, {Ripepi}, {Riva}, {Rivard}, {Rixon},
  {Roegiers}, {Roelens}, {Romero-G{\'o}mez}, {Rowell}, {Royer}, {Ruiz-Dern},
  {Sadowski}, {Sagrist{\`a} Sell{\'e}s}, {Sahlmann}, {Salgado}, {Salguero},
  {Sanna}, {Santana-Ros}, {Sarasso}, {Savietto}, {Schultheis}, {Sciacca},
  {Segol}, {Segovia}, {S{\'e}gransan}, {Shih}, {Siltala}, {Silva}, {Smart},
  {Smith}, {Solano}, {Solitro}, {Sordo}, {Soria Nieto}, {Souchay}, {Spagna},
  {Spoto}, {Stampa}, {Steele}, {Steidelm{\"u}ller}, {Stephenson}, {Stoev},
  {Suess}, {Surdej}, {Szabados}, {Szegedi-Elek}, {Tapiador}, {Taris}, {Tauran},
  {Taylor}, {Teixeira}, {Terrett}, {Teyssand ier}, {Thuillot}, {Titarenko},
  {Torra Clotet}, {Turon}, {Ulla}, {Utrilla}, {Uzzi}, {Vaillant}, {Valentini},
  {Valette}, {van Elteren}, {Van Hemelryck}, {Vaschetto}, {Vecchiato},
  {Veljanoski}, {Viala}, {Vicente}, {Vogt}, {von Essen}, {Voss}, {Votruba},
  {Voutsinas}, {Walmsley}, {Weiler}, {Wertz}, {Wevers}, {Wyrzykowski},
  {Yoldas}, {{\v{Z}}erjal}, {Ziaeepour}, {Zorec}, {Zschocke}, {Zucker},
  {Zurbach}, \& {Zwitter}}]{babusiaux2018}
{Gaia Collaboration}, {Babusiaux}, C., {van Leeuwen}, F., {et~al.}
  2018{\natexlab{a}}, \aap, 616, A10

\bibitem[{{Gaia Collaboration} {et~al.}(2018{\natexlab{b}}){Gaia
  Collaboration}, {Brown}, {Vallenari}, {Prusti}, {de Bruijne}, {Babusiaux},
  {Bailer-Jones}, {Biermann}, {Evans}, {Eyer}, \& et~al.}]{GaiaDR2}
{Gaia Collaboration}, {Brown}, A.~G.~A., {Vallenari}, A., {et~al.}
  2018{\natexlab{b}}, \aap, 616, A1

\bibitem[{{Garc{\'\i}a P{\'e}rez} {et~al.}(2016){Garc{\'\i}a P{\'e}rez},
  {Allende Prieto}, {Holtzman}, {Shetrone}, {M{\'e}sz{\'a}ros}, {Bizyaev},
  {Carrera}, {Cunha}, {Garc{\'\i}a-Hern{\'a}ndez}, {Johnson}, {Majewski},
  {Nidever}, {Schiavon}, {Shane}, {Smith}, {Sobeck}, {Troup}, {Zamora},
  {Weinberg}, {Bovy}, {Eisenstein}, {Feuillet}, {Frinchaboy}, {Hayden},
  {Hearty}, {Nguyen}, {O'Connell}, {Pinsonneault}, {Wilson}, \&
  {Zasowski}}]{garciaperez2016}
{Garc{\'\i}a P{\'e}rez}, A.~E., {Allende Prieto}, C., {Holtzman}, J.~A.,
  {et~al.} 2016, \aj, 151, 144

\bibitem[{{Gilmore} {et~al.}(2012){Gilmore}, {Randich}, {Asplund}, {Binney},
  {Bonifacio}, {Drew}, {Feltzing}, {Ferguson}, {Jeffries}, {Micela}, \&
  et~al.}]{GES}
{Gilmore}, G., {Randich}, S., {Asplund}, M., {et~al.} 2012, The Messenger, 147,
  25

\bibitem[{{Gratton} {et~al.}(2003){Gratton}, {Carretta}, {Desidera},
  {Lucatello}, {Mazzei}, \& {Barbieri}}]{gratton2003}
{Gratton}, R.~G., {Carretta}, E., {Desidera}, S., {et~al.} 2003, \aap, 406, 131

\bibitem[{{Guiglion} {et~al.}(2016){Guiglion}, {de Laverny}, {Recio-Blanco},
  {Worley}, {De Pascale}, {Masseron}, {Prantzos}, \&
  {Mikolaitis}}]{guiglion2016}
{Guiglion}, G., {de Laverny}, P., {Recio-Blanco}, A., {et~al.} 2016, \aap, 595,
  A18

\bibitem[{{Hanke} {et~al.}(2018){Hanke}, {Hansen}, {Koch}, \&
  {Grebel}}]{hanke2018}
{Hanke}, M., {Hansen}, C.~J., {Koch}, A., \& {Grebel}, E.~K. 2018, \aap, 619,
  A134

\bibitem[{{Hayden} {et~al.}(2015){Hayden}, {Bovy}, {Holtzman}, {Nidever},
  {Bird}, {Weinberg}, {Andrews}, {Majewski}, {Allende Prieto}, {Anders},
  {Beers}, {Bizyaev}, {Chiappini}, {Cunha}, {Frinchaboy},
  {Garc{\'\i}a-Her{\'n}and ez}, {Garc{\'\i}a P{\'e}rez}, {Girardi}, {Harding},
  {Hearty}, {Johnson}, {M{\'e}sz{\'a}ros}, {Minchev}, {O'Connell}, {Pan},
  {Robin}, {Schiavon}, {Schneider}, {Schultheis}, {Shetrone}, {Skrutskie},
  {Steinmetz}, {Smith}, {Wilson}, {Zamora}, \& {Zasowski}}]{hayden2015}
{Hayden}, M.~R., {Bovy}, J., {Holtzman}, J.~A., {et~al.} 2015, \apj, 808, 132

\bibitem[{{Houk}(1978)}]{houk1978}
{Houk}, N. 1978, {Michigan catalogue of two-dimensional spectral types for the
  HD stars}

\bibitem[{{Jofr{\'e}} {et~al.}(2019){Jofr{\'e}}, {Heiter}, \&
  {Soubiran}}]{jofre2019}
{Jofr{\'e}}, P., {Heiter}, U., \& {Soubiran}, C. 2019, \araa, 57, 571

\bibitem[{{Johnson} \& {Pilachowski}(2010)}]{johnson2010}
{Johnson}, C.~I. \& {Pilachowski}, C.~A. 2010, \apj, 722, 1373

\bibitem[{{J{\"o}nsson} {et~al.}(2020){J{\"o}nsson}, {Holtzman}, {Prieto},
  {Cunha}, {Garc{\'\i}a-Hern{\'a}ndez}, {Hasselquist}, {Masseron}, {Osorio},
  {Shetrone}, {Smith}, {Stringfellow}, {Bizyaev}, {Edvardsson}, {Majewski},
  {M{\'e}sz{\'a}ros}, {Souto}, {Zamora}, {Beaton}, {Bovy}, {Donor},
  {Pinsonneault}, {Poovelil}, \& {Sobeck}}]{jonsson2020}
{J{\"o}nsson}, H., {Holtzman}, J.~A., {Prieto}, C.~A., {et~al.} 2020, \aj, 160,
  120

\bibitem[{{Kordopatis} {et~al.}(2015){Kordopatis}, {Binney}, {Gilmore}, {Wyse},
  {Belokurov}, {McMillan}, {Hatfield}, {Grebel}, {Steinmetz}, {Navarro},
  {Seabroke}, {Minchev}, {Chiappini}, {Bienaym{\'e}}, {Bland-Hawthorn},
  {Freeman}, {Gibson}, {Helmi}, {Munari}, {Parker}, {Reid}, {Siebert},
  {Siviero}, \& {Zwitter}}]{kordopatis2015}
{Kordopatis}, G., {Binney}, J., {Gilmore}, G., {et~al.} 2015, \mnras, 447, 3526

\bibitem[{{Kordopatis} {et~al.}(2013){Kordopatis}, {Gilmore}, {Steinmetz},
  {Boeche}, {Seabroke}, {Siebert}, {Zwitter}, {Binney}, {de Laverny},
  {Recio-Blanco}, {Williams}, {Piffl}, {Enke}, {Roeser}, {Bijaoui}, {Wyse},
  {Freeman}, {Munari}, {Carrillo}, {Anguiano}, {Burton}, {Campbell}, {Cass},
  {Fiegert}, {Hartley}, {Parker}, {Reid}, {Ritter}, {Russell}, {Stupar},
  {Watson}, {Bienaym{\'e}}, {Bland -Hawthorn}, {Gerhard}, {Gibson}, {Grebel},
  {Helmi}, {Navarro}, {Conrad}, {Famaey}, {Faure}, {Just}, {Kos},
  {Matijevi{\v{c}}}, {McMillan}, {Minchev}, {Scholz}, {Sharma}, {Siviero}, {de
  Boer}, \& {{\v{Z}}erjal}}]{kordopatis2013}
{Kordopatis}, G., {Gilmore}, G., {Steinmetz}, M., {et~al.} 2013, \aj, 146, 134

\bibitem[{{Kordopatis} {et~al.}(2011{\natexlab{a}}){Kordopatis},
  {Recio-Blanco}, {de Laverny}, {Bijaoui}, {Hill}, {Gilmore}, {Wyse}, \&
  {Ordenovic}}]{kordopatis2011a}
{Kordopatis}, G., {Recio-Blanco}, A., {de Laverny}, P., {et~al.}
  2011{\natexlab{a}}, \aap, 535, A106

\bibitem[{{Kordopatis} {et~al.}(2011{\natexlab{b}}){Kordopatis},
  {Recio-Blanco}, {de Laverny}, {Gilmore}, {Hill}, {Wyse}, {Helmi}, {Bijaoui},
  {Zoccali}, \& {Bienaym{\'e}}}]{kordopatis2011b}
{Kordopatis}, G., {Recio-Blanco}, A., {de Laverny}, P., {et~al.}
  2011{\natexlab{b}}, \aap, 535, A107

\bibitem[{{Kos} {et~al.}(2017){Kos}, {Lin}, {Zwitter}, {{\v{Z}}erjal},
  {Sharma}, {Bland -Hawthorn}, {Asplund}, {Casey}, {De Silva}, {Freeman},
  {Martell}, {Simpson}, {Schlesinger}, {Zucker}, {Anguiano}, {Bacigalupo},
  {Bedding}, {Betters}, {Da Costa}, {Duong}, {Hyde}, {Ireland}, {Kafle},
  {Leon-Saval}, {Lewis}, {Munari}, {Nataf}, {Stello}, {Tinney}, {Traven},
  {Watson}, \& {Wittenmyer}}]{Kos2017}
{Kos}, J., {Lin}, J., {Zwitter}, T., {et~al.} 2017, \mnras, 464, 1259

\bibitem[{{Kunder} {et~al.}(2017){Kunder}, {Kordopatis}, {Steinmetz},
  {Zwitter}, {McMillan}, {Casagrande}, {Enke}, {Wojno}, {Valentini},
  {Chiappini}, {Matijevi{\v{c}}}, {Siviero}, {de Laverny}, {Recio-Blanco},
  {Bijaoui}, {Wyse}, {Binney}, {Grebel}, {Helmi}, {Jofre}, {Antoja}, {Gilmore},
  {Siebert}, {Famaey}, {Bienaym{\'e}}, {Gibson}, {Freeman}, {Navarro},
  {Munari}, {Seabroke}, {Anguiano}, {{\v{Z}}erjal}, {Minchev}, {Reid},
  {Bland-Hawthorn}, {Kos}, {Sharma}, {Watson}, {Parker}, {Scholz}, {Burton},
  {Cass}, {Hartley}, {Fiegert}, {Stupar}, {Ritter}, {Hawkins}, {Gerhard},
  {Chaplin}, {Davies}, {Elsworth}, {Lund}, {Miglio}, \& {Mosser}}]{kunder2017}
{Kunder}, A., {Kordopatis}, G., {Steinmetz}, M., {et~al.} 2017, \aj, 153, 75

\bibitem[{{Lee} {et~al.}(2011){Lee}, {Beers}, {An}, {Ivezi{\'c}}, {Just},
  {Rockosi}, {Morrison}, {Johnson}, {Sch{\"o}nrich}, {Bird}, {Yanny},
  {Harding}, \& {Rocha-Pinto}}]{lee2011}
{Lee}, Y.~S., {Beers}, T.~C., {An}, D., {et~al.} 2011, \apj, 738, 187

\bibitem[{{Leung} \& {Bovy}(2019)}]{leung2019a}
{Leung}, H.~W. \& {Bovy}, J. 2019, \mnras, 483, 3255

\bibitem[{{Lindegren} {et~al.}(2018){Lindegren}, {Hern{\'a}ndez}, {Bombrun},
  {Klioner}, {Bastian}, {Ramos-Lerate}, {de Torres}, {Steidelm{\"u}ller},
  {Stephenson}, {Hobbs}, {Lammers}, {Biermann}, {Geyer}, {Hilger}, {Michalik},
  {Stampa}, {McMillan}, {Casta{\~n}eda}, {Clotet}, {Comoretto}, {Davidson},
  {Fabricius}, {Gracia}, {Hambly}, {Hutton}, {Mora}, {Portell}, {van Leeuwen},
  {Abbas}, {Abreu}, {Altmann}, {Andrei}, {Anglada}, {Balaguer-N{\'u}{\~n}ez},
  {Barache}, {Becciani}, {Bertone}, {Bianchi}, {Bouquillon}, {Bourda},
  {Br{\"u}semeister}, {Bucciarelli}, {Busonero}, {Buzzi}, {Cancelliere},
  {Carlucci}, {Charlot}, {Cheek}, {Crosta}, {Crowley}, {de Bruijne}, {de
  Felice}, {Drimmel}, {Esquej}, {Fienga}, {Fraile}, {Gai}, {Garralda},
  {Gonz{\'a}lez-Vidal}, {Guerra}, {Hauser}, {Hofmann}, {Holl}, {Jordan},
  {Lattanzi}, {Lenhardt}, {Liao}, {Licata}, {Lister}, {L{\"o}ffler},
  {Marchant}, {Martin-Fleitas}, {Messineo}, {Mignard}, {Morbidelli}, {Poggio},
  {Riva}, {Rowell}, {Salguero}, {Sarasso}, {Sciacca}, {Siddiqui}, {Smart},
  {Spagna}, {Steele}, {Taris}, {Torra}, {van Elteren}, {van Reeven}, \&
  {Vecchiato}}]{lindegren2018}
{Lindegren}, L., {Hern{\'a}ndez}, J., {Bombrun}, A., {et~al.} 2018, \aap, 616,
  A2

\bibitem[{{Matijevi{\v{c}}} {et~al.}(2017){Matijevi{\v{c}}}, {Chiappini},
  {Grebel}, {Wyse}, {Zwitter}, {Bienaym{\'e}}, {Bland -Hawthorn}, {Freeman},
  {Gibson}, {Gilmore}, {Helmi}, {Kordopatis}, {Kunder}, {Munari}, {Navarro},
  {Parker}, {Reid}, {Seabroke}, {Siviero}, {Steinmetz}, \&
  {Watson}}]{matijevic2017}
{Matijevi{\v{c}}}, G., {Chiappini}, C., {Grebel}, E.~K., {et~al.} 2017, \aap,
  603, A19

\bibitem[{{Matteucci} \& {Francois}(1989)}]{matteucci1989}
{Matteucci}, F. \& {Francois}, P. 1989, \mnras, 239, 885

\bibitem[{{McMillan} {et~al.}(2018){McMillan}, {Kordopatis}, {Kunder},
  {Binney}, {Wojno}, {Zwitter}, {Steinmetz}, {Bland-Hawthorn}, {Gibson},
  {Gilmore}, {Grebel}, {Helmi}, {Munari}, {Navarro}, {Parker}, {Seabroke},
  {Watson}, \& {Wyse}}]{mcmillan2018}
{McMillan}, P.~J., {Kordopatis}, G., {Kunder}, A., {et~al.} 2018, \mnras, 477,
  5279

\bibitem[{{Minchev} {et~al.}(2014){Minchev}, {Chiappini}, {Martig},
  {Steinmetz}, {de Jong}, {Boeche}, {Scannapieco}, {Zwitter}, {Wyse}, {Binney},
  {Bland-Hawthorn}, {Bienaym{\'e}}, {Famaey}, {Freeman}, {Gibson}, {Grebel},
  {Gilmore}, {Helmi}, {Kordopatis}, {Lee}, {Munari}, {Navarro}, {Parker},
  {Quillen}, {Reid}, {Siebert}, {Siviero}, {Seabroke}, {Watson}, \&
  {Williams}}]{minchev_2014}
{Minchev}, I., {Chiappini}, C., {Martig}, M., {et~al.} 2014, \apjl, 781, L20

\bibitem[{{Minchev} {et~al.}(2019){Minchev}, {Matijevic}, {Hogg}, {Guiglion},
  {Steinmetz}, {Anders}, {Chiappini}, {Martig}, {Queiroz}, \&
  {Scannapieco}}]{minchev2019}
{Minchev}, I., {Matijevic}, G., {Hogg}, D.~W., {et~al.} 2019, \mnras, 487, 3946

\bibitem[{{Morel} \& {Miglio}(2012)}]{Morel2012}
{Morel}, T. \& {Miglio}, A. 2012, \mnras, 419, L34

\bibitem[{{Ness} {et~al.}(2015){Ness}, {Hogg}, {Rix}, {Ho}, \&
  {Zasowski}}]{ness2015}
{Ness}, M., {Hogg}, D.~W., {Rix}, H.~W., {Ho}, A. Y.~Q., \& {Zasowski}, G.
  2015, \apj, 808, 16

\bibitem[{{Nordstr{\"o}m} {et~al.}(2004){Nordstr{\"o}m}, {Mayor}, {Andersen},
  {Holmberg}, {Pont}, {J{\o}rgensen}, {Olsen}, {Udry}, \&
  {Mowlavi}}]{nordstrom2004}
{Nordstr{\"o}m}, B., {Mayor}, M., {Andersen}, J., {et~al.} 2004, \aap, 418, 989

\bibitem[{{Pancino} \& {Gaia-ESO Survey consortium}(2012)}]{Pancino2012}
{Pancino}, E. \& {Gaia-ESO Survey consortium}, o.~b.~o.~t. 2012, ArXiv e-prints

\bibitem[{{Pasquini} {et~al.}(2004){Pasquini}, {Randich}, {Zoccali}, {Hill},
  {Charbonnel}, \& {Nordstr{\"o}m}}]{pasquini2004}
{Pasquini}, L., {Randich}, S., {Zoccali}, M., {et~al.} 2004, \aap, 424, 951

\bibitem[{{Pinsonneault} {et~al.}(2018){Pinsonneault}, {Elsworth}, {Tayar},
  {Serenelli}, {Stello}, {Zinn}, {Mathur}, {Garc{\'\i}a}, {Johnson}, {Hekker},
  {Huber}, {Kallinger}, {M{\'e}sz{\'a}ros}, {Mosser}, {Stassun}, {Girardi},
  {Rodrigues}, {Silva Aguirre}, {An}, {Basu}, {Chaplin}, {Corsaro}, {Cunha},
  {Garc{\'\i}a-Hern{\'a}ndez}, {Holtzman}, {J{\"o}nsson}, {Shetrone}, {Smith},
  {Sobeck}, {Stringfellow}, {Zamora}, {Beers}, {Fern{\'a}ndez- Trincado},
  {Frinchaboy}, {Hearty}, \& {Nitschelm}}]{Pinsonneault2018}
{Pinsonneault}, M.~H., {Elsworth}, Y.~P., {Tayar}, J., {et~al.} 2018, The
  Astrophysical Journal Supplement Series, 239, 32

\bibitem[{{Queiroz} {et~al.}(2020){Queiroz}, {Anders}, {Chiappini},
  {Khalatyan}, {Santiago}, {Steinmetz}, {Valentini}, {Miglio}, {Bossini},
  {Barbuy}, {Minchev}, {Minniti}, {Garc{\'\i}a Hern{\'a}ndez}, {Schultheis},
  {Beaton}, {Beers}, {Bizyaev}, {Brownstein}, {Cunha},
  {Fern{\'a}ndez-Trincado}, {Frinchaboy}, {Lane}, {Majewski}, {Nataf},
  {Nitschelm}, {Pan}, {Roman-Lopes}, {Sobeck}, {Stringfellow}, \&
  {Zamora}}]{queiroz2019}
{Queiroz}, A.~B.~A., {Anders}, F., {Chiappini}, C., {et~al.} 2020, \aap, 638,
  A76

\bibitem[{{Queiroz} {et~al.}(2018){Queiroz}, {Anders}, {Santiago}, {Chiappini},
  {Steinmetz}, {Dal Ponte}, {Stassun}, {da Costa}, {Maia}, {Crestani}, {Beers},
  {Fern{\'a}ndez-Trincado}, {Garc{\'\i}a-Hern{\'a}ndez}, {Roman-Lopes}, \&
  {Zamora}}]{queiroz2018}
{Queiroz}, A.~B.~A., {Anders}, F., {Santiago}, B.~X., {et~al.} 2018, \mnras,
  476, 2556

\bibitem[{{Recio-Blanco} {et~al.}(2006){Recio-Blanco}, {Bijaoui}, \& {de
  Laverny}}]{matisse}
{Recio-Blanco}, A., {Bijaoui}, A., \& {de Laverny}, P. 2006, \mnras, 370, 141

\bibitem[{{Reddy} {et~al.}(2006){Reddy}, {Lambert}, \& {Prieto}}]{reddy2006}
{Reddy}, B.~E., {Lambert}, D.~L., \& {Prieto}, C.~A. 2006, VizieR Online Data
  Catalog, J/MNRAS/367/1329

\bibitem[{{Reddy} {et~al.}(2003){Reddy}, {Tomkin}, {Lambert}, \& {Allende
  Prieto}}]{reddy2003}
{Reddy}, B.~E., {Tomkin}, J., {Lambert}, D.~L., \& {Allende Prieto}, C. 2003,
  VizieR Online Data Catalog, J/MNRAS/340/304

\bibitem[{{Ruchti} {et~al.}(2010){Ruchti}, {Fulbright}, {Wyse}, {Gilmore},
  {Bienaym{\'e}}, {Binney}, {Bland -Hawthorn}, {Campbell}, {Freeman}, {Gibson},
  {Grebel}, {Helmi}, {Munari}, {Navarro}, {Parker}, {Reid}, {Seabroke},
  {Siebert}, {Siviero}, {Steinmetz}, {Watson}, {Williams}, \&
  {Zwitter}}]{ruchti2010}
{Ruchti}, G.~R., {Fulbright}, J.~P., {Wyse}, R.~F.~G., {et~al.} 2010, \apjl,
  721, L92

\bibitem[{{Ruchti} {et~al.}(2011){Ruchti}, {Fulbright}, {Wyse}, {Gilmore},
  {Grebel}, {Bienaym{\'e}}, {Bland-Hawthorn}, {Freeman}, {Gibson}, {Munari},
  {Navarro}, {Parker}, {Reid}, {Seabroke}, {Siebert}, {Siviero}, {Steinmetz},
  {Watson}, {Williams}, \& {Zwitter}}]{ruchti2011}
{Ruchti}, G.~R., {Fulbright}, J.~P., {Wyse}, R.~F.~G., {et~al.} 2011, \apj,
  743, 107

\bibitem[{{Santiago} {et~al.}(2016){Santiago}, {Brauer}, {Anders}, {Chiappini},
  {Queiroz}, {Girardi}, {Rocha-Pinto}, {Balbinot}, {da Costa}, {Maia},
  {Schultheis}, {Steinmetz}, {Miglio}, {Montalb{\'a}n}, {Schneider}, {Beers},
  {Frinchaboy}, {Lee}, \& {Zasowski}}]{santiago2016}
{Santiago}, B.~X., {Brauer}, D.~E., {Anders}, F., {et~al.} 2016, \aap, 585, A42

\bibitem[{{Sch{\"o}nrich} \& {Bergemann}(2014)}]{schonrich2014}
{Sch{\"o}nrich}, R. \& {Bergemann}, M. 2014, \mnras, 443, 698

\bibitem[{{Skrutskie} {et~al.}(2006){Skrutskie}, {Cutri}, {Stiening},
  {Weinberg}, {Schneider}, {Carpenter}, {Beichman}, {Capps}, {Chester},
  {Elias}, {Huchra}, {Liebert}, {Lonsdale}, {Monet}, {Price}, {Seitzer},
  {Jarrett}, {Kirkpatrick}, {Gizis}, {Howard}, {Evans}, {Fowler}, {Fullmer},
  {Hurt}, {Light}, {Kopan}, {Marsh}, {McCallon}, {Tam}, {Van Dyk}, \&
  {Wheelock}}]{2MASS}
{Skrutskie}, M.~F., {Cutri}, R.~M., {Stiening}, R., {et~al.} 2006, \aj, 131,
  1163

\bibitem[{{Smiljanic} {et~al.}(2016){Smiljanic}, {Romano}, {Bragaglia},
  {Donati}, {Magrini}, {Friel}, {Jacobson}, {Randich}, {Ventura}, {Lind},
  {Bergemann}, {Nordlander}, {Morel}, {Pancino}, {Tautvai{\v{s}}ien{\.{e}}},
  {Adibekyan}, {Tosi}, {Vallenari}, {Gilmore}, {Bensby}, {Fran{\c{c}}ois},
  {Koposov}, {Lanzafame}, {Recio-Blanco}, {Bayo}, {Carraro}, {Casey},
  {Costado}, {Franciosini}, {Heiter}, {Hill}, {Hourihane}, {Jofr{\'e}},
  {Lardo}, {de Laverny}, {Lewis}, {Monaco}, {Morbidelli}, {Sacco}, {Sbordone},
  {Sousa}, {Worley}, \& {Zaggia}}]{smiljanic2017}
{Smiljanic}, R., {Romano}, D., {Bragaglia}, A., {et~al.} 2016, \aap, 589, A115

\bibitem[{{Soubiran} \& {Girard}(2005)}]{soubiran_2005}
{Soubiran}, C. \& {Girard}, P. 2005, \aap, 438, 139

\bibitem[{{Starkenburg} {et~al.}(2017){Starkenburg}, {Martin}, {Youakim},
  {Aguado}, {Allende Prieto}, {Arentsen}, {Bernard}, {Bonifacio}, {Caffau},
  {Carlberg}, {C{\^o}t{\'e}}, {Fouesneau}, {Fran{\c{c}}ois}, {Franke},
  {Gonz{\'a}lez Hern{\'a}ndez}, {Gwyn}, {Hill}, {Ibata}, {Jablonka},
  {Longeard}, {McConnachie}, {Navarro}, {S{\'a}nchez-Janssen}, {Tolstoy}, \&
  {Venn}}]{starkenburg2017}
{Starkenburg}, E., {Martin}, N., {Youakim}, K., {et~al.} 2017, \mnras, 471,
  2587

\bibitem[{{Steinmetz}(2003)}]{RAVE}
{Steinmetz}, M. 2003, in Astronomical Society of the Pacific Conference Series,
  Vol. 298, GAIA Spectroscopy: Science and Technology, ed. U.~{Munari}, 381

\bibitem[{{Steinmetz} {et~al.}(2020{\natexlab{a}}){Steinmetz}, {Guiglion},
  {McMillan}, {Matijevi{\v{c}}}, {Enke}, {Kordopatis}, {Zwitter}, {Valentini},
  {Chiappini}, {Casagrande}, {Wojno}, {Anguiano}, {Bienaym{\'e}}, {Bijaoui},
  {Binney}, {Burton}, {Cass}, {de Laverny}, {Fiegert}, {Freeman}, {Fulbright},
  {Gibson}, {Gilmore}, {Grebel}, {Helmi}, {Kunder}, {Munari}, {Navarro},
  {Parker}, {Ruchti}, {Recio-Blanco}, {Reid}, {Seabroke}, {Siviero}, {Siebert},
  {Stupar}, {Watson}, {Williams}, {Wyse}, {Anders}, {Antoja}, {Birko},
  {Bland-Hawthorn}, {Bossini}, {Garc{\'\i}a}, {Carrillo}, {Chaplin},
  {Elsworth}, {Famaey}, {Gerhard}, {Jofre}, {Just}, {Mathur}, {Miglio},
  {Minchev}, {Monari}, {Mosser}, {Ritter}, {Rodrigues}, {Scholz}, {Sharma},
  {Sysoliatina}, \& {RAVE Collaboration}}]{steinmetz2020b}
{Steinmetz}, M., {Guiglion}, G., {McMillan}, P.~J., {et~al.}
  2020{\natexlab{a}}, \aj, 160, 83

\bibitem[{{Steinmetz} {et~al.}(2020{\natexlab{b}}){Steinmetz},
  {Matijevi{\v{c}}}, {Enke}, {Zwitter}, {Guiglion}, {McMillan}, {Kordopatis},
  {Valentini}, {Chiappini}, {Casagrande}, {Wojno}, {Anguiano}, {Bienaym{\'e}},
  {Bijaoui}, {Binney}, {Burton}, {Cass}, {de Laverny}, {Fiegert}, {Freeman},
  {Fulbright}, {Gibson}, {Gilmore}, {Grebel}, {Helmi}, {Kunder}, {Munari},
  {Navarro}, {Parker}, {Ruchti}, {Recio-Blanco}, {Reid}, {Seabroke}, {Siviero},
  {Siebert}, {Stupar}, {Watson}, {Williams}, {Wyse}, {Anders}, {Antoja},
  {Birko}, {Bland-Hawthorn}, {Bossini}, {Garc{\'\i}a}, {Carrillo}, {Chaplin},
  {Elsworth}, {Famaey}, {Gerhard}, {Jofre}, {Just}, {Mathur}, {Miglio},
  {Minchev}, {Monari}, {Mosser}, {Ritter}, {Rodrigues}, {Scholz}, {Sharma},
  {Sysoliatina}, \& {RAVE Collaboration}}]{steinmetz2020a}
{Steinmetz}, M., {Matijevi{\v{c}}}, G., {Enke}, H., {et~al.}
  2020{\natexlab{b}}, \aj, 160, 82

\bibitem[{{Steinmetz} {et~al.}(2006){Steinmetz}, {Zwitter}, {Siebert},
  {Watson}, {Freeman}, {Munari}, {Campbell}, {Williams}, {Seabroke}, {Wyse},
  {Parker}, {Bienaym{\'e}}, {Roeser}, {Gibson}, {Gilmore}, {Grebel}, {Helmi},
  {Navarro}, {Burton}, {Cass}, {Dawe}, {Fiegert}, {Hartley}, {Russell},
  {Saunders}, {Enke}, {Bailin}, {Binney}, {Bland-Hawthorn}, {Boeche}, {Dehnen},
  {Eisenstein}, {Evans}, {Fiorucci}, {Fulbright}, {Gerhard}, {Jauregi}, {Kelz},
  {Mijovi{\'c}}, {Minchev}, {Parmentier}, {Pe{\~n}arrubia}, {Quillen}, {Read},
  {Ruchti}, {Scholz}, {Siviero}, {Smith}, {Sordo}, {Veltz}, {Vidrih}, {von
  Berlepsch}, {Boyle}, \& {Schilbach}}]{steinmetz2006}
{Steinmetz}, M., {Zwitter}, T., {Siebert}, A., {et~al.} 2006, \aj, 132, 1645

\bibitem[{{Ting} {et~al.}(2019){Ting}, {Conroy}, {Rix}, \&
  {Cargile}}]{ting2019}
{Ting}, Y.-S., {Conroy}, C., {Rix}, H.-W., \& {Cargile}, P. 2019, \apj, 879, 69

\bibitem[{{Valenti} \& {Fischer}(2005)}]{valenti2005}
{Valenti}, J.~A. \& {Fischer}, D.~A. 2005, \apjs, 159, 141

\bibitem[{{Valenti} \& {Piskunov}(1996)}]{valenti1996}
{Valenti}, J.~A. \& {Piskunov}, N. 1996, \aaps, 118, 595

\bibitem[{{Valentini} {et~al.}(2017){Valentini}, {Chiappini}, {Davies},
  {Elsworth}, {Mosser}, {Lund}, {Miglio}, {Chaplin}, {Rodrigues}, {Boeche},
  {Steinmetz}, {Matijevi{\v c}}, {Kordopatis}, {Bland-Hawthorn}, {Munari},
  {Bienaym{\'e}}, {Freeman}, {Gibson}, {Gilmore}, {Grebel}, {Helmi}, {Kunder},
  {McMillan}, {Navarro}, {Parker}, {Reid}, {Seabroke}, {Sharma}, {Siviero},
  {Watson}, {Wyse}, {Zwitter}, \& {Mott}}]{Valentini2017}
{Valentini}, M., {Chiappini}, C., {Davies}, G.~R., {et~al.} 2017, \aap, 600,
  A66

\bibitem[{{Van Cleve} {et~al.}(2016){Van Cleve}, {Howell}, {Smith}, {Clarke},
  {Thompson}, {Bryson}, {Lund}, {Handberg}, \& {Chaplin}}]{VanCleve2016}
{Van Cleve}, J.~E., {Howell}, S.~B., {Smith}, J.~C., {et~al.} 2016, \pasp, 128,
  075002

\bibitem[{{Wang} {et~al.}(2016){Wang}, {Wang}, {Wu}, {Zhao}, {Li}, {Luo},
  {Liu}, {Zhang}, {Hou}, {Wang}, \& {Cao}}]{Wang2016}
{Wang}, L., {Wang}, W., {Wu}, Y., {et~al.} 2016, \aj, 152, 6

\bibitem[{{Wojno} {et~al.}(2018){Wojno}, {Kordopatis}, {Steinmetz}, {McMillan},
  {Binney}, {Famaey}, {Monari}, {Minchev}, {Wyse}, {Antoja}, {Siebert},
  {Carrillo}, {Bland -Hawthorn}, {Grebel}, {Zwitter}, {Bienaym{\'e}}, {Gibson},
  {Kunder}, {Munari}, {Navarro}, {Parker}, {Reid}, \& {Seabroke}}]{wojno2018}
{Wojno}, J., {Kordopatis}, G., {Steinmetz}, M., {et~al.} 2018, \mnras, 477,
  5612

\bibitem[{{Wright} {et~al.}(2010){Wright}, {Eisenhardt}, {Mainzer}, {Ressler},
  {Cutri}, {Jarrett}, {Kirkpatrick}, {Padgett}, {McMillan}, {Skrutskie},
  {Stanford}, {Cohen}, {Walker}, {Mather}, {Leisawitz}, {Gautier}, {McLean},
  {Benford}, {Lonsdale}, {Blain}, {Mendez}, {Irace}, {Duval}, {Liu}, {Royer},
  {Heinrichsen}, {Howard}, {Shannon}, {Kendall}, {Walsh}, {Larsen}, {Cardon},
  {Schick}, {Schwalm}, {Abid}, {Fabinsky}, {Naes}, \& {Tsai}}]{ALLWISE}
{Wright}, E.~L., {Eisenhardt}, P. R.~M., {Mainzer}, A.~K., {et~al.} 2010, \aj,
  140, 1868

\bibitem[{{Wyse} \& {Gilmore}(1988)}]{wyse1988}
{Wyse}, R. F.~G. \& {Gilmore}, G. 1988, \aj, 95, 1404

\bibitem[{{Yanny} {et~al.}(2009){Yanny}, {Rockosi}, {Newberg}, {Knapp},
  {Adelman-McCarthy}, {Alcorn}, {Allam}, {Allende Prieto}, {An}, {Anderson},
  {Anderson}, {Bailer-Jones}, {Bastian}, {Beers}, {Bell}, {Belokurov},
  {Bizyaev}, {Blythe}, {Bochanski}, {Boroski}, {Brinchmann}, {Brinkmann},
  {Brewington}, {Carey}, {Cudworth}, {Evans}, {Evans}, {Gates}, {G{\"a}nsicke},
  {Gillespie}, {Gilmore}, {Nebot Gomez-Moran}, {Grebel}, {Greenwell}, {Gunn},
  {Jordan}, {Jordan}, {Harding}, {Harris}, {Hendry}, {Holder}, {Ivans},
  {Ivezi{\v{c}}}, {Jester}, {Johnson}, {Kent}, {Kleinman}, {Kniazev},
  {Krzesinski}, {Kron}, {Kuropatkin}, {Lebedeva}, {Lee}, {French Leger},
  {L{\'e}pine}, {Levine}, {Lin}, {Long}, {Loomis}, {Lupton}, {Malanushenko},
  {Malanushenko}, {Margon}, {Martinez-Delgado}, {McGehee}, {Monet}, {Morrison},
  {Munn}, {Neilsen}, {Nitta}, {Norris}, {Oravetz}, {Owen}, {Padmanabhan},
  {Pan}, {Peterson}, {Pier}, {Platson}, {Re Fiorentin}, {Richards}, {Rix},
  {Schlegel}, {Schneider}, {Schreiber}, {Schwope}, {Sibley}, {Simmons},
  {Snedden}, {Allyn Smith}, {Stark}, {Stauffer}, {Steinmetz}, {Stoughton},
  {SubbaRao}, {Szalay}, {Szkody}, {Thakar}, {Sivarani}, {Tucker}, {Uomoto},
  {Vanden Berk}, {Vidrih}, {Wadadekar}, {Watters}, {Wilhelm}, {Wyse}, {Yarger},
  \& {Zucker}}]{yanny2009}
{Yanny}, B., {Rockosi}, C., {Newberg}, H.~J., {et~al.} 2009, \aj, 137, 4377

\bibitem[{{Yoshii}(1981)}]{yoshii1981}
{Yoshii}, Y. 1981, \aap, 97, 280

\bibitem[{{Youakim} {et~al.}(2017){Youakim}, {Starkenburg}, {Aguado}, {Martin},
  {Fouesneau}, {Gonz{\'a}lez Hern{\'a}ndez}, {Allende Prieto}, {Bonifacio},
  {Gentile}, {Kielty}, {C{\^o}t{\'e}}, {Jablonka}, {McConnachie}, {S{\'a}nchez
  Janssen}, {Tolstoy}, \& {Venn}}]{youakim2017}
{Youakim}, K., {Starkenburg}, E., {Aguado}, D.~S., {et~al.} 2017, \mnras, 472,
  2963

\bibitem[{{Zhang} {et~al.}(2019){Zhang}, {Zhao}, {Yang}, {Wang}, \&
  {Zuo}}]{zhang2019}
{Zhang}, X., {Zhao}, G., {Yang}, C.~Q., {Wang}, Q.~X., \& {Zuo}, W.~B. 2019,
  \pasp, 131, 094202

\bibitem[{{Zwitter} {et~al.}(2008){Zwitter}, {Siebert}, {Munari}, {Freeman},
  {Siviero}, {Watson}, {Fulbright}, {Wyse}, {Campbell}, {Seabroke}, {Williams},
  {Steinmetz}, {Bienaym{\'e}}, {Gilmore}, {Grebel}, {Helmi}, {Navarro},
  {Anguiano}, {Boeche}, {Burton}, {Cass}, {Dawe}, {Fiegert}, {Hartley},
  {Russell}, {Veltz}, {Bailin}, {Binney}, {Bland-Hawthorn}, {Brown}, {Dehnen},
  {Evans}, {Re Fiorentin}, {Fiorucci}, {Gerhard}, {Gibson}, {Kelz}, {Kujken},
  {Matijevi{\v c}}, {Minchev}, {Parker}, {Pe{\~n}arrubia}, {Quillen}, {Read},
  {Reid}, {Roeser}, {Ruchti}, {Scholz}, {Smith}, {Sordo}, {Tolstoi},
  {Tomasella}, {Vidrih}, \& {Wylie-de Boer}}]{zwitter2008}
{Zwitter}, T., {Siebert}, A., {Munari}, U., {et~al.} 2008, \aj, 136, 421

\end{thebibliography}


\appendix


\section{Effect of parallax errors on 
CNN performance}\label{section_parallax_errors}

In the present study, 94\% of the RAVE DR6 targets have good Gaia DR2 parallaxes, 
with an error better than $20\%$. Deriving absolute magnitudes from such parallaxes 
and apparent magnitudes is then safe in the context of the present paper. 
This high success rate is, however, an immediate consequence of the relatively 
bright magnitude limit of $I<13$ for RAVE, with the majority of the stars 
even having $I<12$. The overall Gaia RVS survey will, however, probe considerably 
fainter objects. The low-resolution surveys like Gaia RVS or 4MIDABLE-LR of 
4MOST \citep{chiappini2019} will probe a much larger 
volume than RAVE. There is then a risk that many targets suffer from large 
parallax errors. Here, we discuss the impact of such large parallax errors on 
the determination of atmospheric parameters and abundances.

\begin{figure*}[h]
\centering
\includegraphics[width=1.0\linewidth]{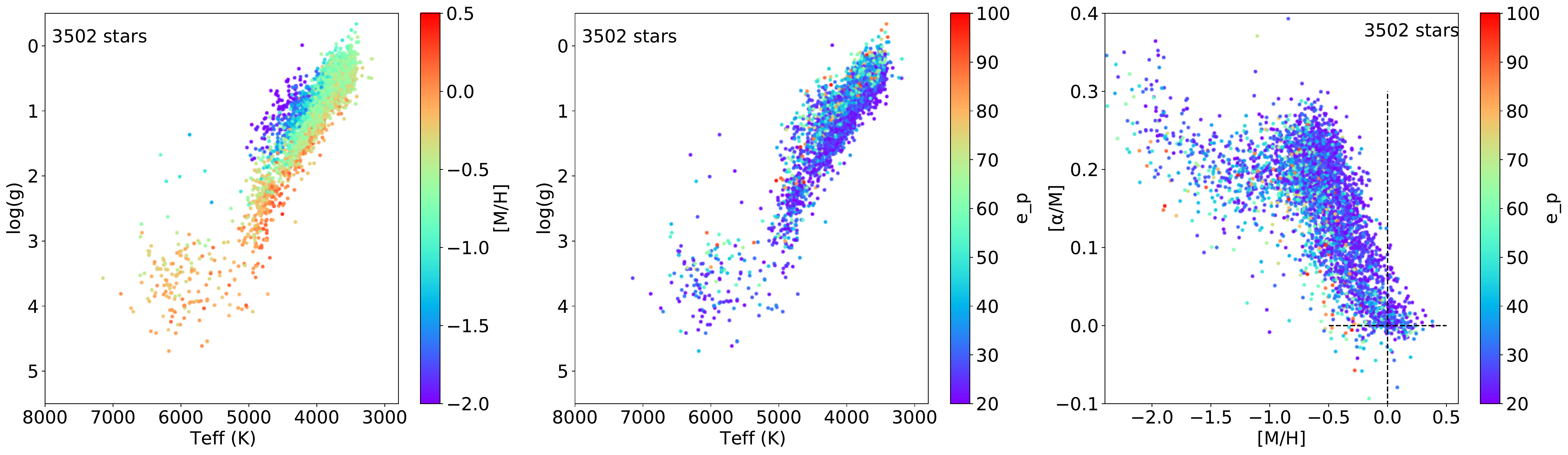}
\caption{\label{observed_kiel_diagram_parallax_error_cuts}Kiel diagram of $3\,502$ 
stars of the observed sample, with $e_p>20\%$, colour-coded by $\mh$ (top panel) 
and $e_p$ (middle panel). Only stars with $\snr>50$ per pixel are plotted. 
The same stars are presented in the bottom panels in the $\alpham $ versus $\mh$ plane, 
colour-coded by parallax errors.}
\end{figure*}

In \figurename~\ref{observed_kiel_diagram_parallax_error_cuts}, we present 
CNN results for $3\,502$ stars of the observed sample with parallax errors, 
$e_p>20\%$, and $\snr>50$ per pixel. Despite the large parallax errors, we can recover a 
proper giant branch with a clear metallicity sequence. Most of the stars 
with $e_p>40\%$ are actually located in the upper part of either the main 
sequence or the cool giant branch. Such stars should 
be thus used with caution. For all stars with $e_p>20\%$, 
the $\alpham $ versus $\mh$ abundance patterns do not show systematics, 
meaning that chemical abundances are less sensitive to less precise parallaxes 
(absolute magnitudes constraining mainly $\teff$ and $\logg$).

To check if the CNN could learn from lower quality data, we added in our 
training sample $\sim150$ more stars with parallax errors higher than 20\%. 
Adding such stars did not improve the training phase or the atmospheric 
parameters of the observed sample stars with parallax errors larger than 20\%. 


\section{Validation of atmospheric parameters 
with stellar clusters}\label{section_clusters}

\begin{figure*}[h]
\centering
\includegraphics[width=1.0\linewidth]{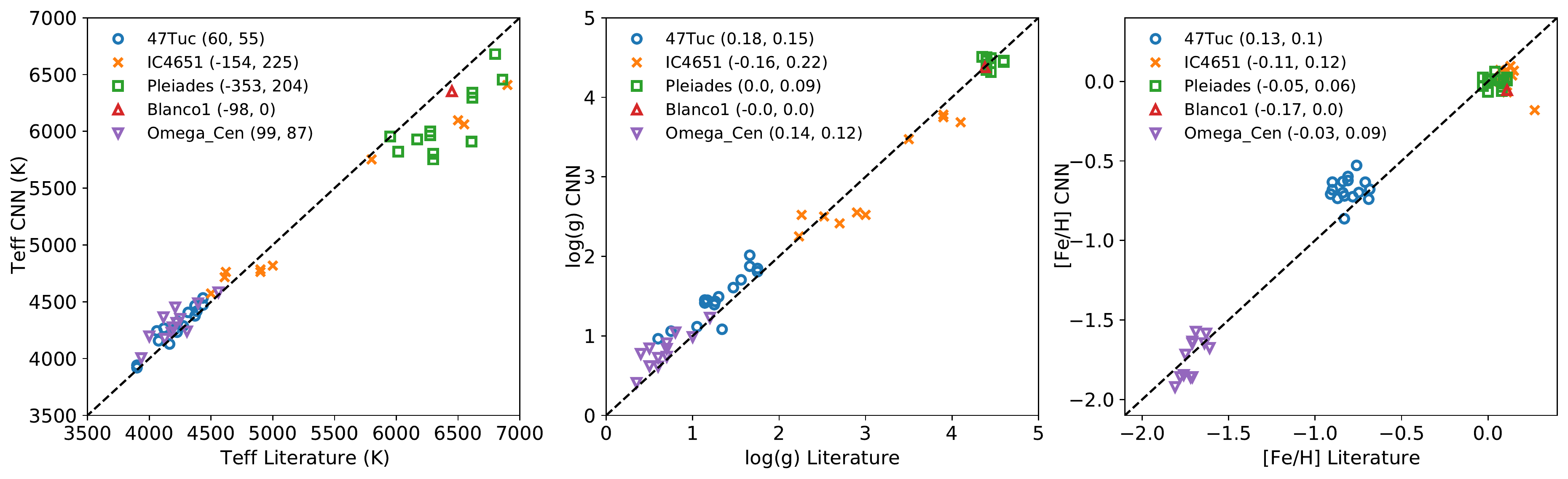}
\caption{\label{cnn_vs_clusters}Comparison of $\teff$, $\logg$, and $\feh$ 
between the present study and compilation of five stellar clusters 
with common stars with RAVE: 47Tuc ($\circ$), 
IC4651 ($\times$), Pleiades ($\square$), Blanco1 ($\triangle$), and 
Omega Centauri ($\triangledown$). 
The mean difference and dispersion are indicated together with the cluster name.}
\end{figure*}

Here we compare the CNN results with 41 stars from four 
clusters used in RAVE DR6 for calibration purposes:
47Tuc \citep{carretta2009}, Pleiades \citep{funayama2009}, 
Blanco1 \citep{ford2005}, IC4651 \citep{pasquini2004}, and
Omega Centauri \citep{johnson2010}. 
The results are presented in \figurename~\ref{cnn_vs_clusters}.

The giants tend to match pretty well between our study and the 
literature, with slight variations from cluster to cluster. The Pleiades 
show no discernible offset in $\logg$ and $\feh,$ while a large mean difference 
is measured for $\teff$ (-353\,K). We have both giant and dwarf stars 
in common with IC4651, and they tend to show a good match with our study. 
The dispersion in $\feh$ drops to 0.03 when only considering stars with 
$\snr>40$. We only have one star in common with Blanco 1, but we find good 
agreement between the literature and our study. Finally, the cluster 
47Tuc presents an offset of +0.13\,dex in $\feh$ with respect to the literature, 
while the dispersion is about 0.1\,dex. We note that we have a total of 13 
stars from 47Tuc and the Pleiades in our training sample. 
We have 12 giants in common with the metal-poor globular cluster 
Omega Centauri. The $\feh$ values of our CNN do not show any bias with 
respect to the literature, and the dispersion is about 0.1\,dex. The 
Omega Centauri stars span lower $\logg$ values that 47Tuc, mainly
$\logg<1$. We show that the CNN is able to provide reliable 
parameterisation of metal-poor super-giant stars. 

The systematics observed in the three parameters  come directly from systematics 
in the APOGEE DR16 labels. Overall, the typical dispersion $\sigma$ 
in $\teff$ and $\feh$ tends to decrease when selecting stars with $\snr>40$, 
but stays constant for $\logg$.

\section{Validation of atmospheric parameters and chemical abundances 
with the HR sample}\label{section_cnn_vs_high_res}

Here, compare  our atmospheric parameters and chemical 
abundances with those from high-resolution (HR) studies 
in the literature. We took a high resolution sample 
compiled and used for validation purposes in RAVE DR6 
\citep{steinmetz2020b}. It includes more than 1700 stars, 
taken from several studies, among them with available 
chemical abundances \citet{reddy2003, valenti2005, soubiran_2005, 
reddy2006, ruchti2011, adibekyan2012, bensby2014} and Gaia-ESO Survey DR5.

\begin{figure*}[h]
\centering
\includegraphics[width=1.0\linewidth]{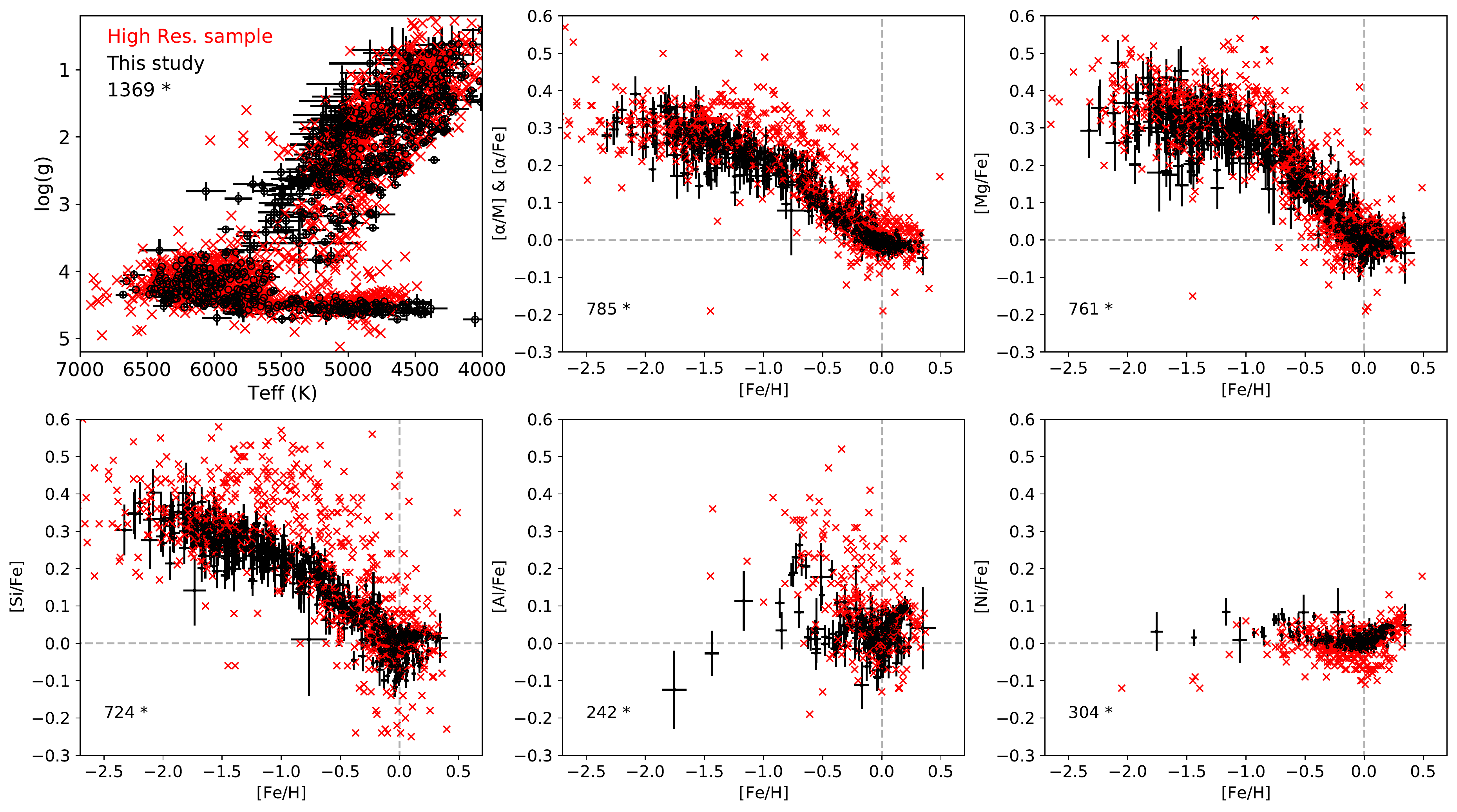}
\caption{\label{cnn_vs_high_res_diagrams}Kiel diagram and chemical abundances 
patterns for stars in common between our study (black circles) and the literature 
(red crosses).}
\end{figure*}

We present a Kiel diagram and abundance patterns 
for stars of the high-resolution sample and from the present 
study in \figurename~\ref{cnn_vs_high_res_diagrams}. 
We only selected stars with $\snr>20$. 
Basically, the main and giant sequences match pretty well. 
The $\alpham$, $\sife$ patterns
tend to match for $\feh>-0.5\,$dex, while at lower metallicity
the CNN abundances tend to be systematically lower. 
This comes from the fact that  $\alpham$ and $\sife$ do 
not reach values higher than +0.30 dex in APOGEE DR16. On the other hand, 
$\mgfe$ matches rather well between our CNN results and the literature. 
The $\alfe$ ratios are reasonably consistent around solar $\feh$, but 
the scatter increases for the metal-poor regime. Finally, $\nife$ 
is rather flat in both samples, as expected for an Fe-peak element. 

\begin{figure*}[h]
\centering
\includegraphics[width=1.0\linewidth]{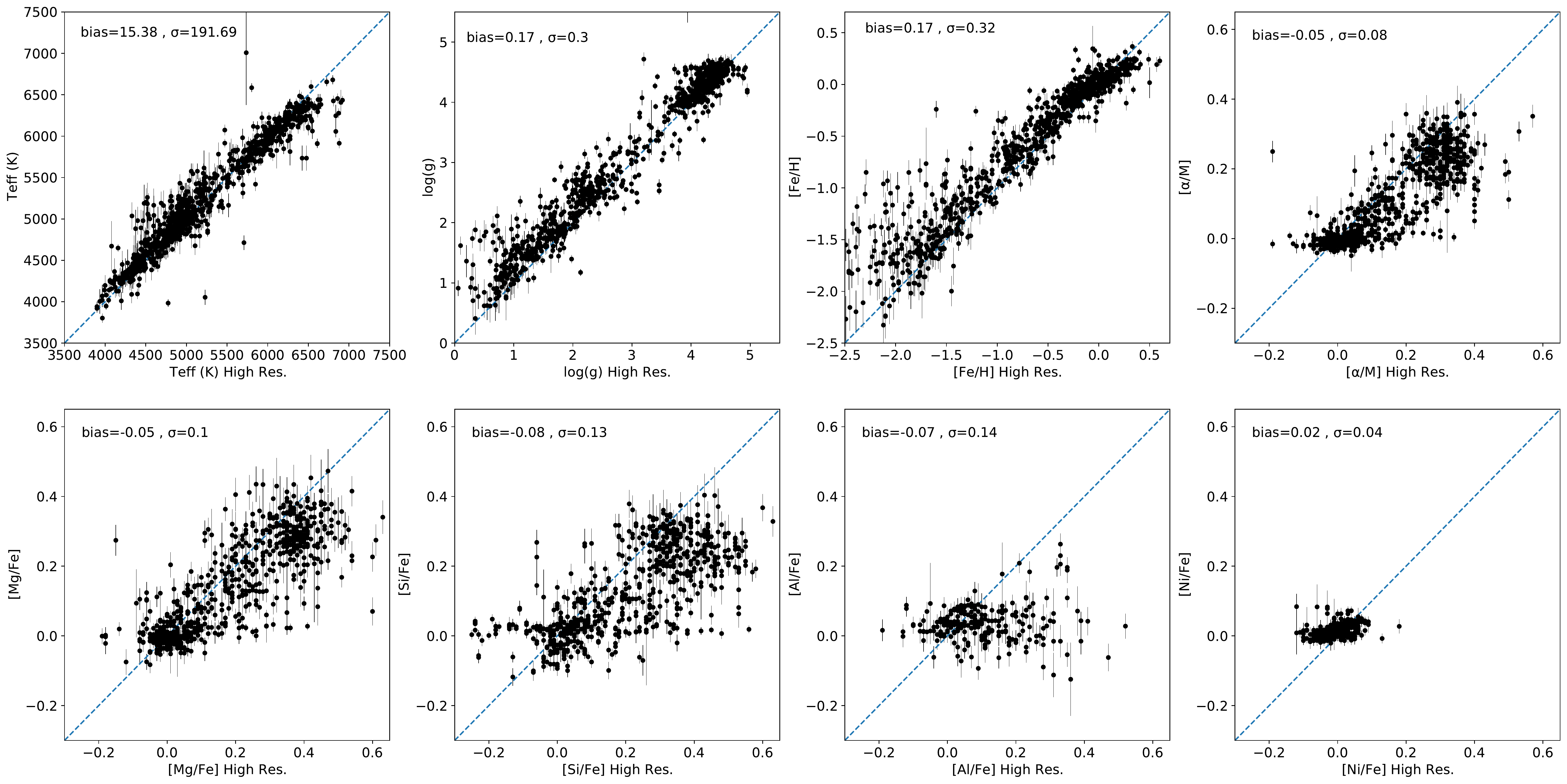}
\caption{\label{cnn_vs_high_res}Atmospheric parameters and 
chemical abundances derived by our CNN, as a function 
of values from the literature. Mean bias and dispersion are 
indicated in the top left corner of each panel.}
\end{figure*}

In \figurename~\ref{cnn_vs_high_res}, we present the 1-to-1 relations 
between the high-resolution sample and the present study. This illustrates 
the differences in the trends and zero-points very well. The typical 
dispersion is about 200\,K in $\teff$ (no bias), while it is around 0.3 for 
$\logg$ (bias of 0.13 dex) and $\feh$ ($\sim0.3$ bias). We observe 
an increase of the scatter with decreasing $\feh$. We note that the overall scatter 
in $\feh$ drops to 0.2 dex if we only select stars with $\snr>50$. All 
other abundances show quite a small dispersion, roughly 0.1 dex. In fact, 
shifts in the trends or in the 
zero-points reflect more  a systematic difference of the calibration between the
APOGEE DR16 surveys and the test sample, rather than an incorrect estimation of 
parameters or abundances. Such differences are to be expected considering the 
differences in instrument specifications, resolution, wavelength range, and 
wavelength coverage. 

\section{Chemical abundance patterns of $\mgfe$, $\sife$, 
$\alfe$ and $\nife$}\label{annex_patterns}

\begin{figure*}
\centering
\includegraphics[width=0.95\linewidth]{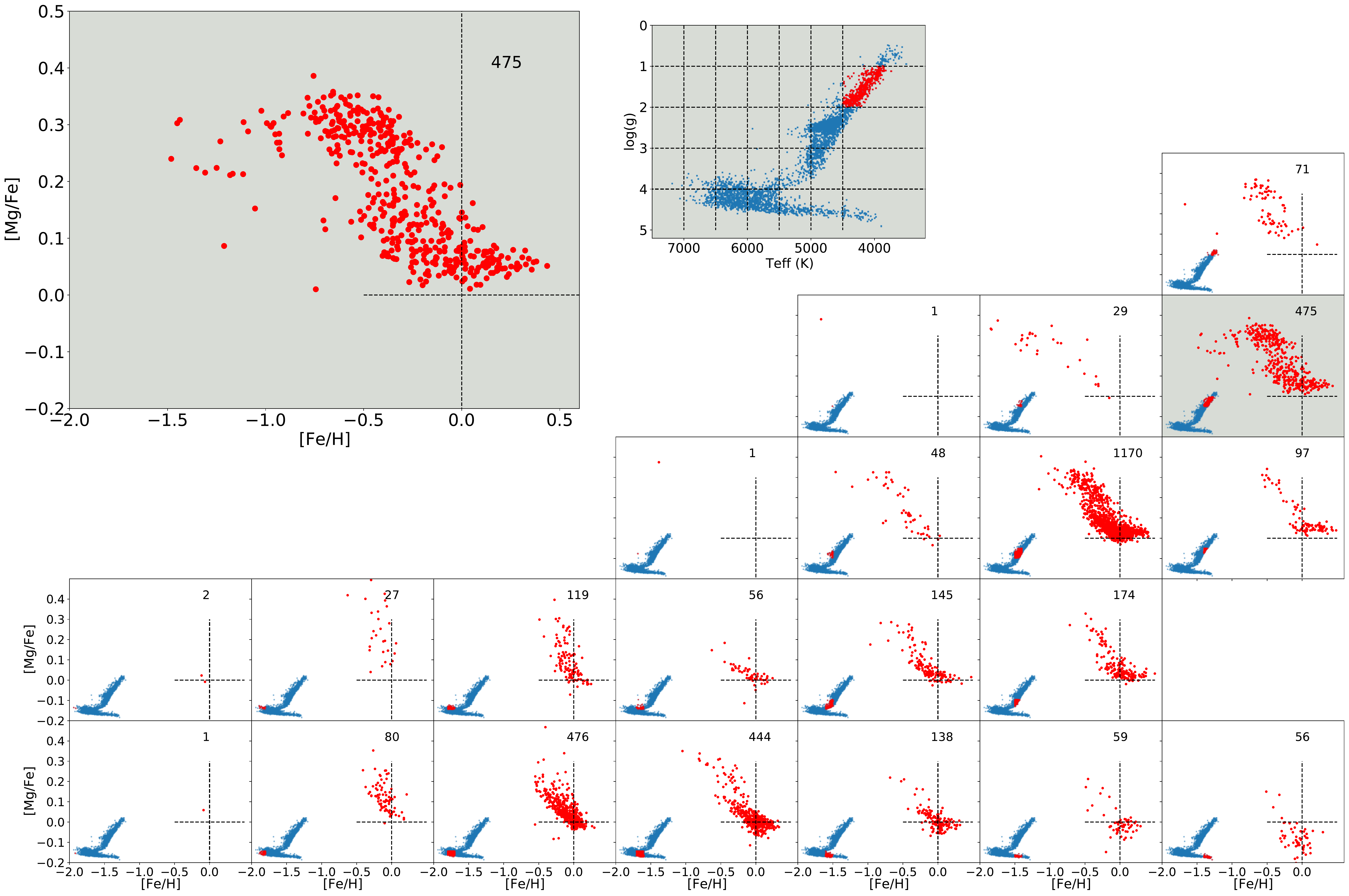}
\includegraphics[width=0.95\linewidth]{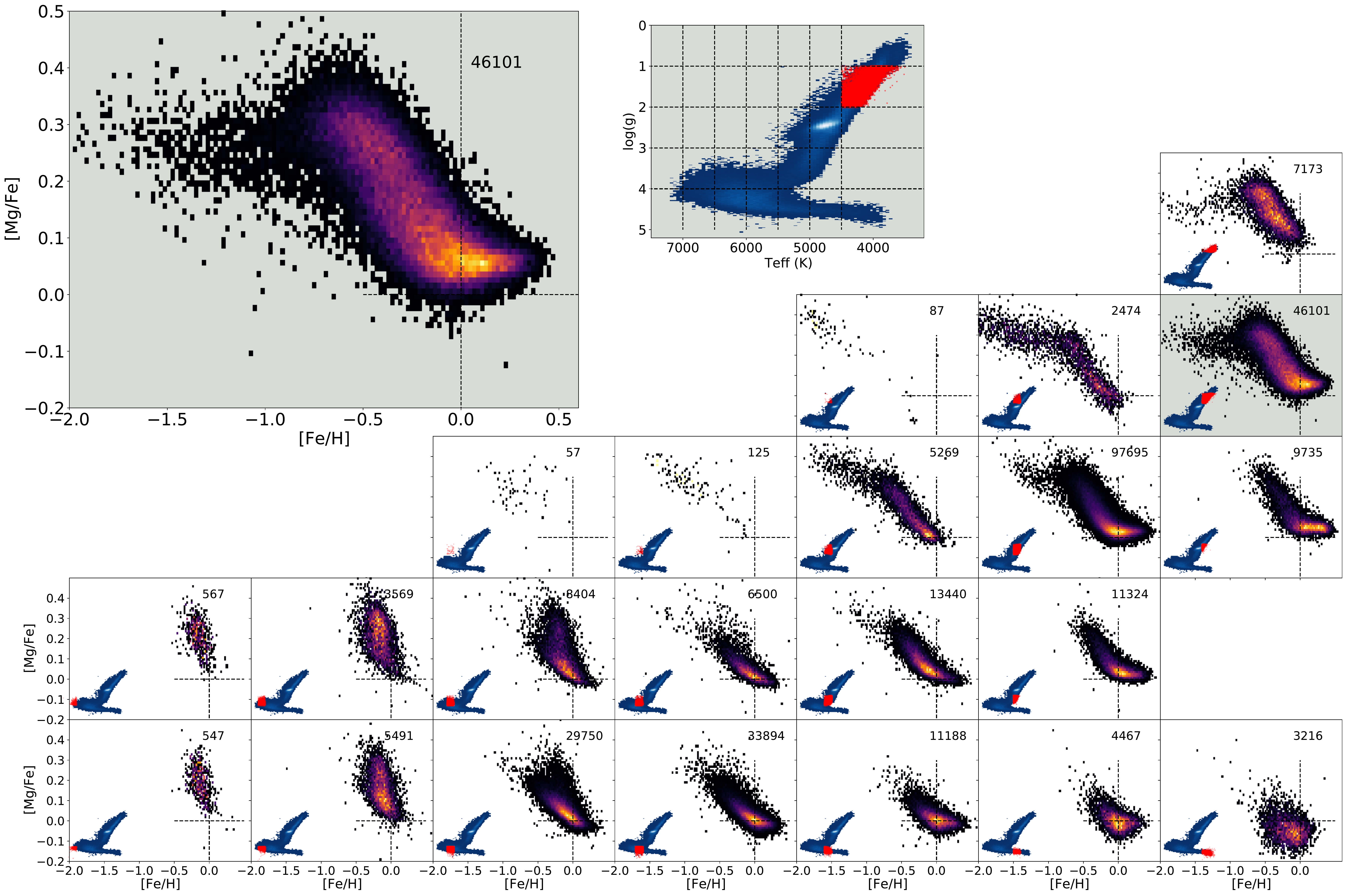}
\caption{\label{kiel_observed_MgFe}Top: $\mgfe$ vs. $\feh$ for 
the training sample. Bottom: $\mgfe$ vs. $\feh$ for $301\,076$ stars 
of the observed sample with \snr>30, RAVE DR6 'n\&o' classification, 
and parallax errors lower than $20\%$. For each panel, 
we overplotted a $\teff-\logg$ diagram with the location of the plotted stars 
marked in red.}
\end{figure*}

\begin{figure*}
\centering
\includegraphics[width=0.95\linewidth]{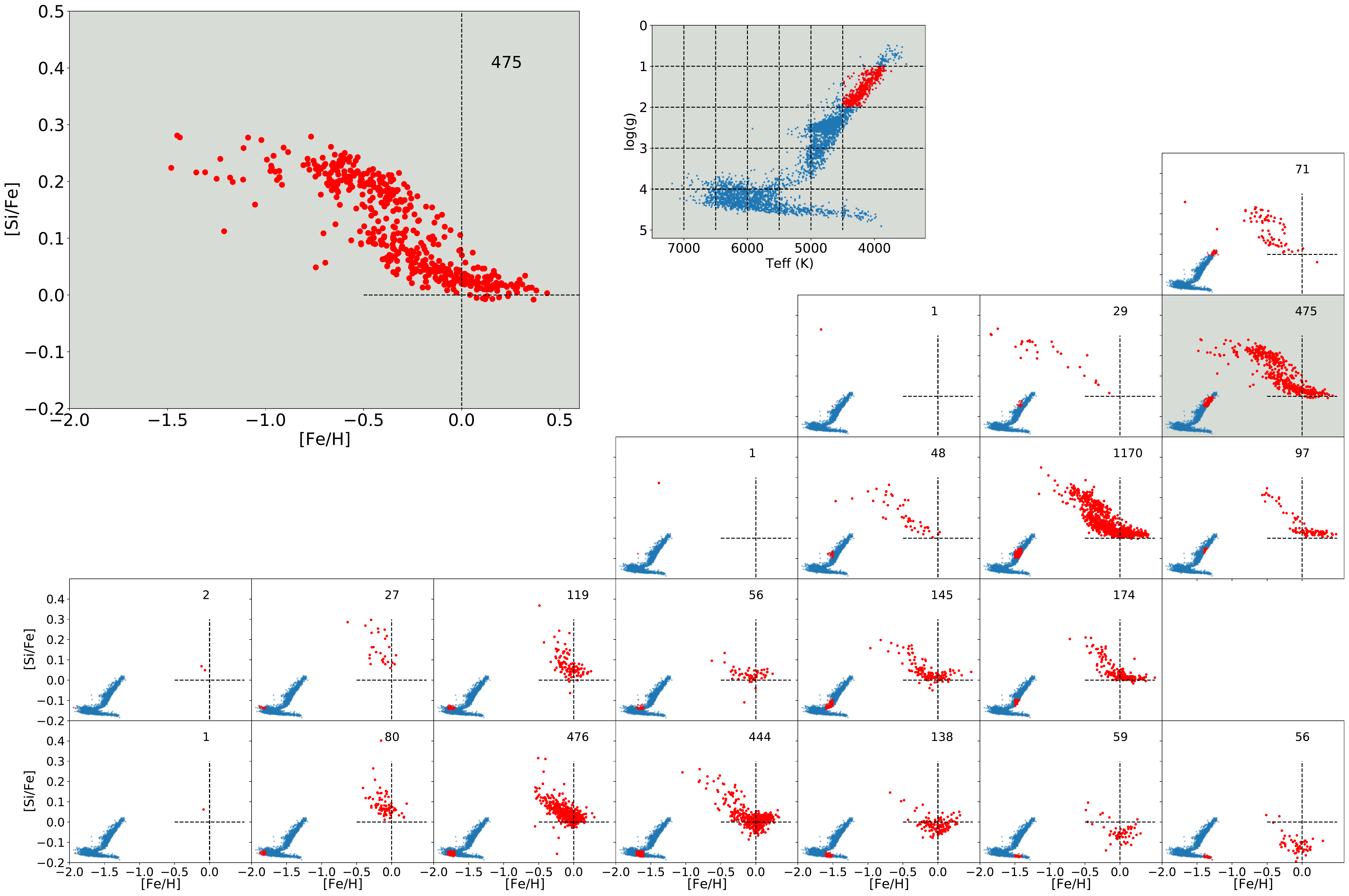}
\includegraphics[width=0.95\linewidth]{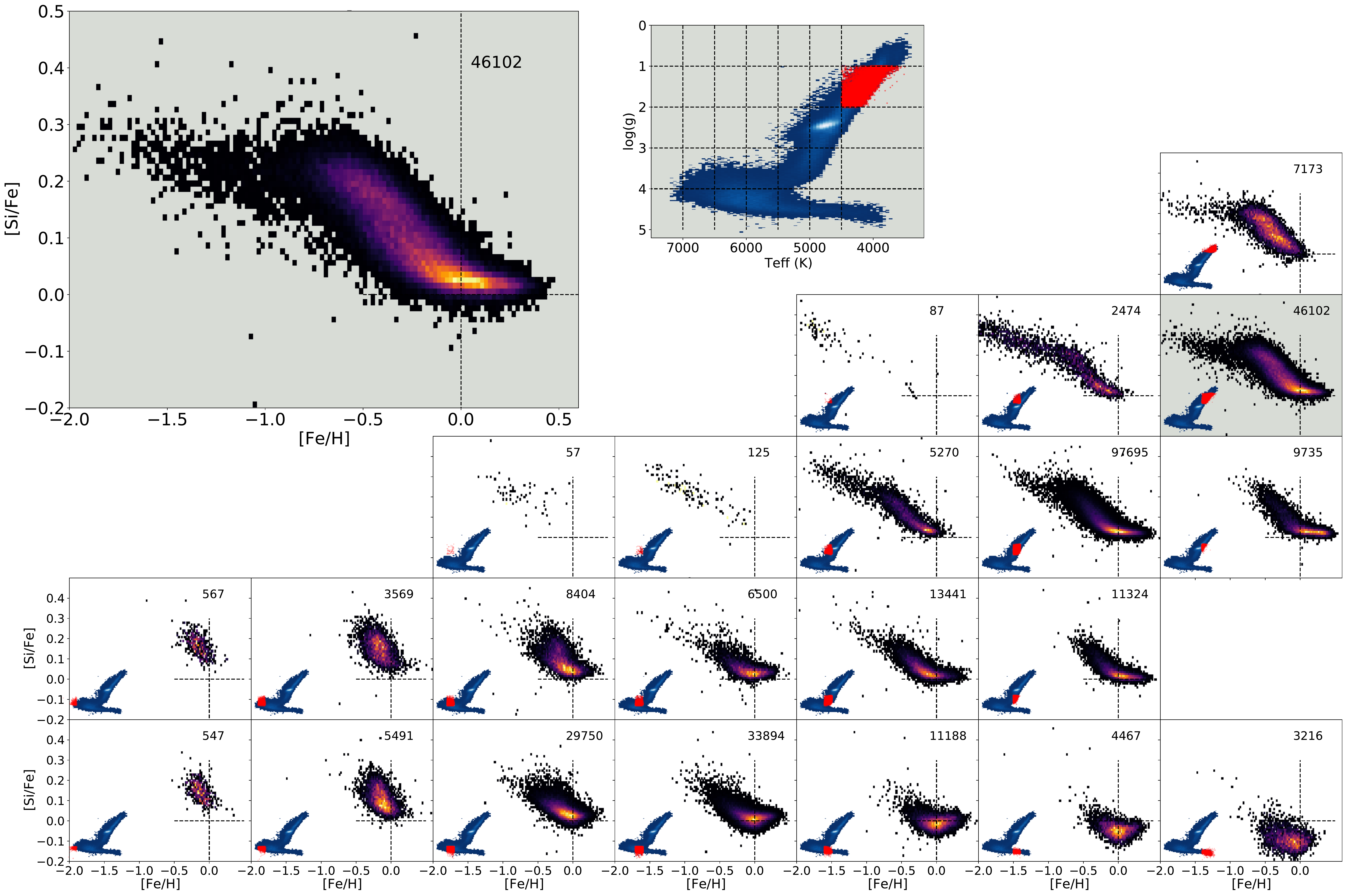}
\caption{\label{kiel_observed_SiFe}Top: $\sife$ vs. $\feh$ for 
the training sample. Bottom: $\sife$ vs. $\feh$ for $301\,076$ stars 
of the observed sample with \snr>30, RAVE DR6 'n\&o' classification, 
and parallax errors lower than $20\%$. For each panel, 
we overplotted a $\teff-\logg$ diagram with the location of the plotted stars 
marked in red.}
\end{figure*}

\begin{figure*}
\centering
\includegraphics[width=0.95\linewidth]{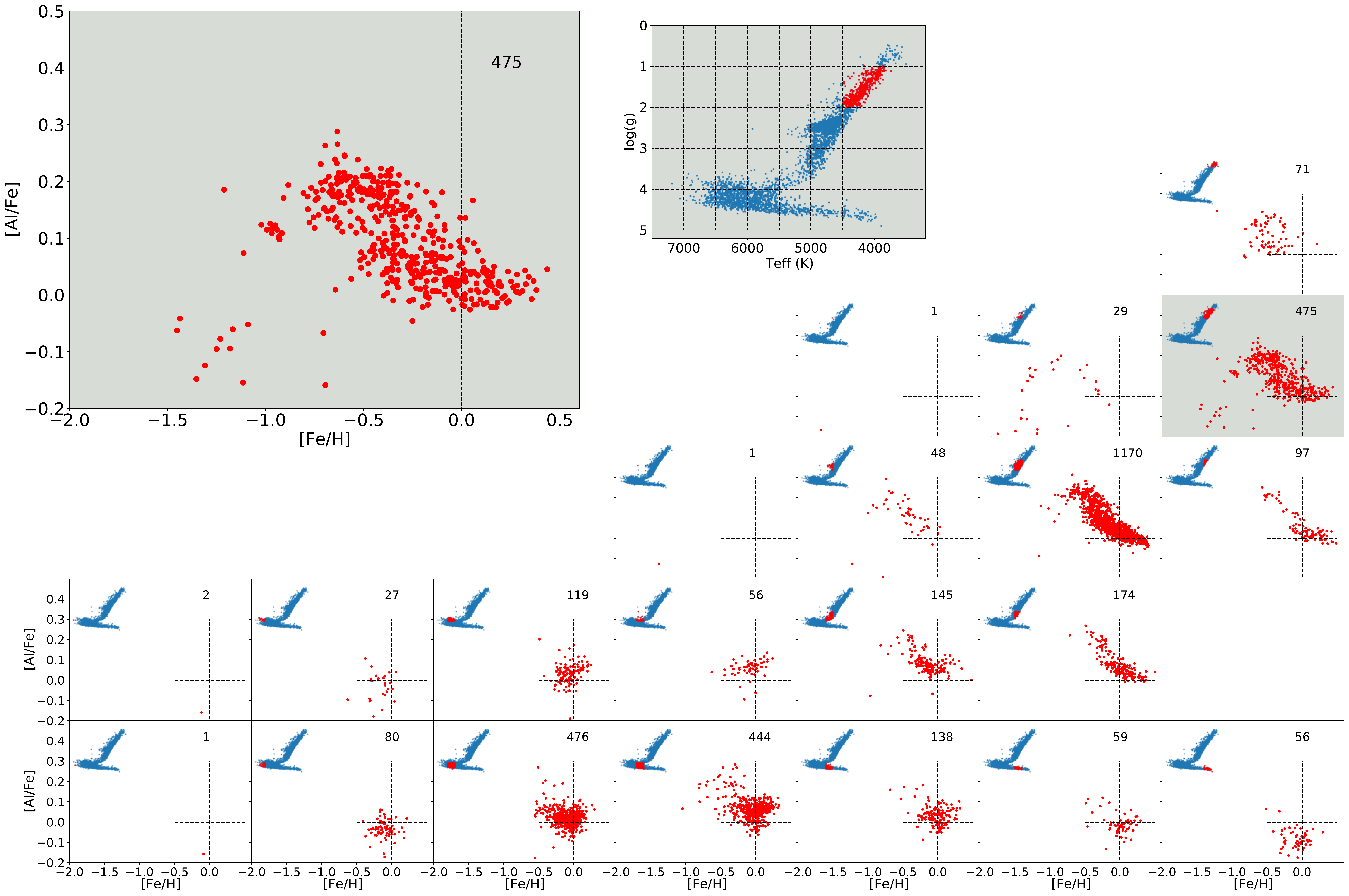}
\includegraphics[width=0.95\linewidth]{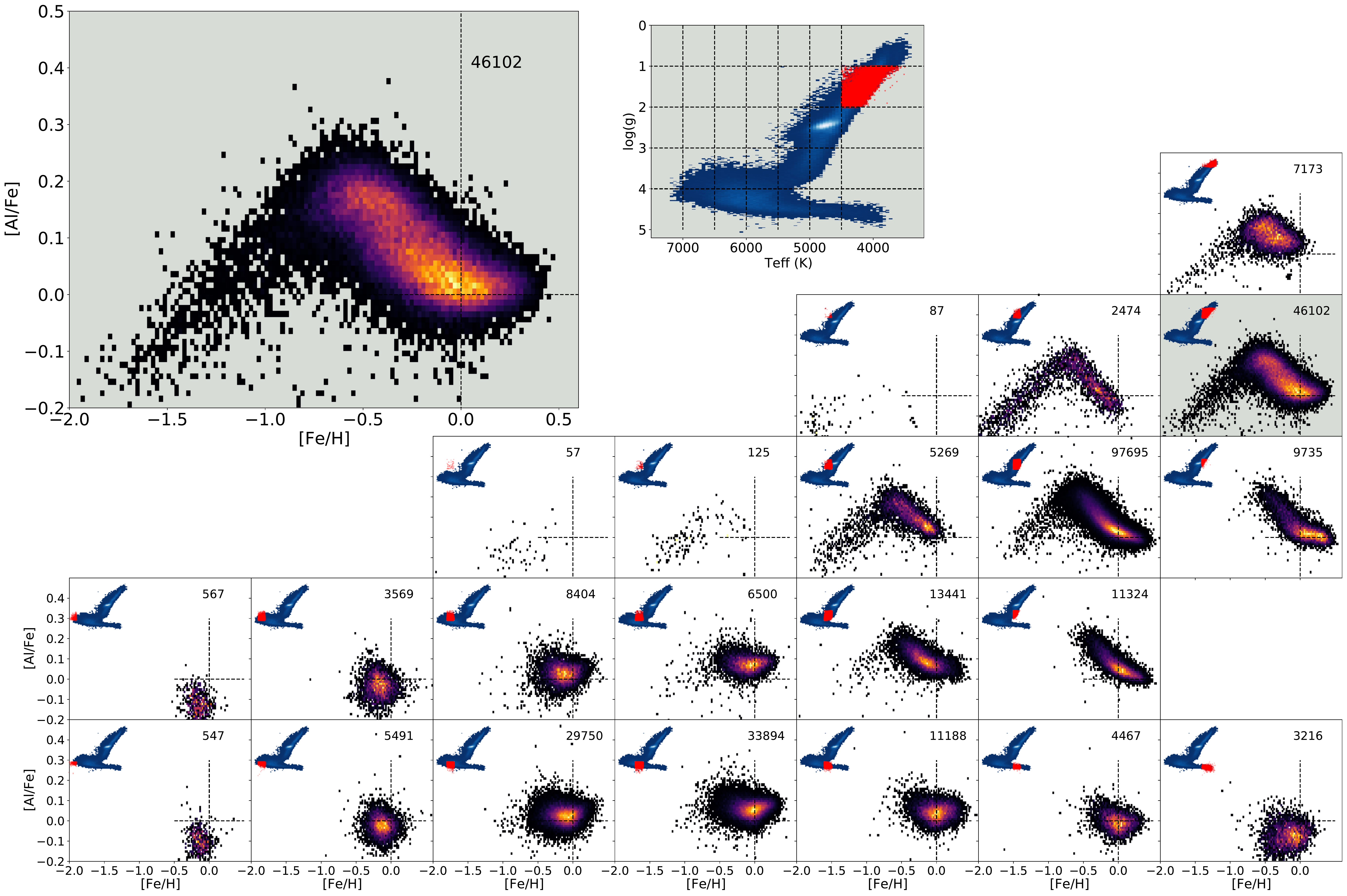}
\caption{\label{kiel_observed_AlFe}Top: $\alfe$ vs. $\feh$ for 
the training sample. Bottom: $\alfe$ vs. $\feh$ for $301\,076$ stars 
of the observed sample with \snr>30, RAVE DR6 'n\&o' classification, 
and parallax errors lower than $20\%$. For each panel, 
we overplotted a $\teff-\logg$ diagram with the location of the plotted stars 
marked in red.}
\end{figure*}

\begin{figure*}
\centering
\includegraphics[width=0.95\linewidth]{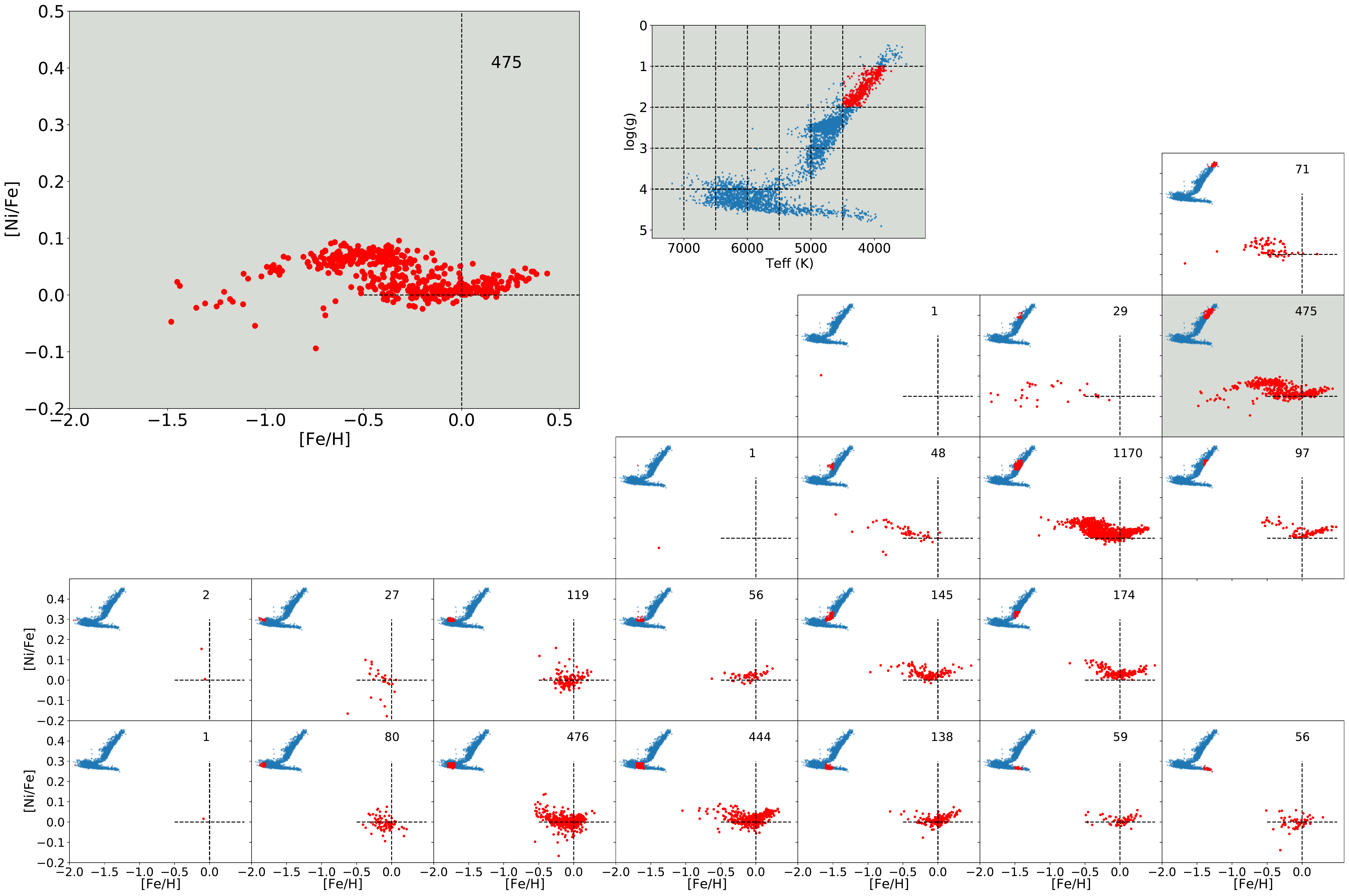}
\includegraphics[width=0.95\linewidth]{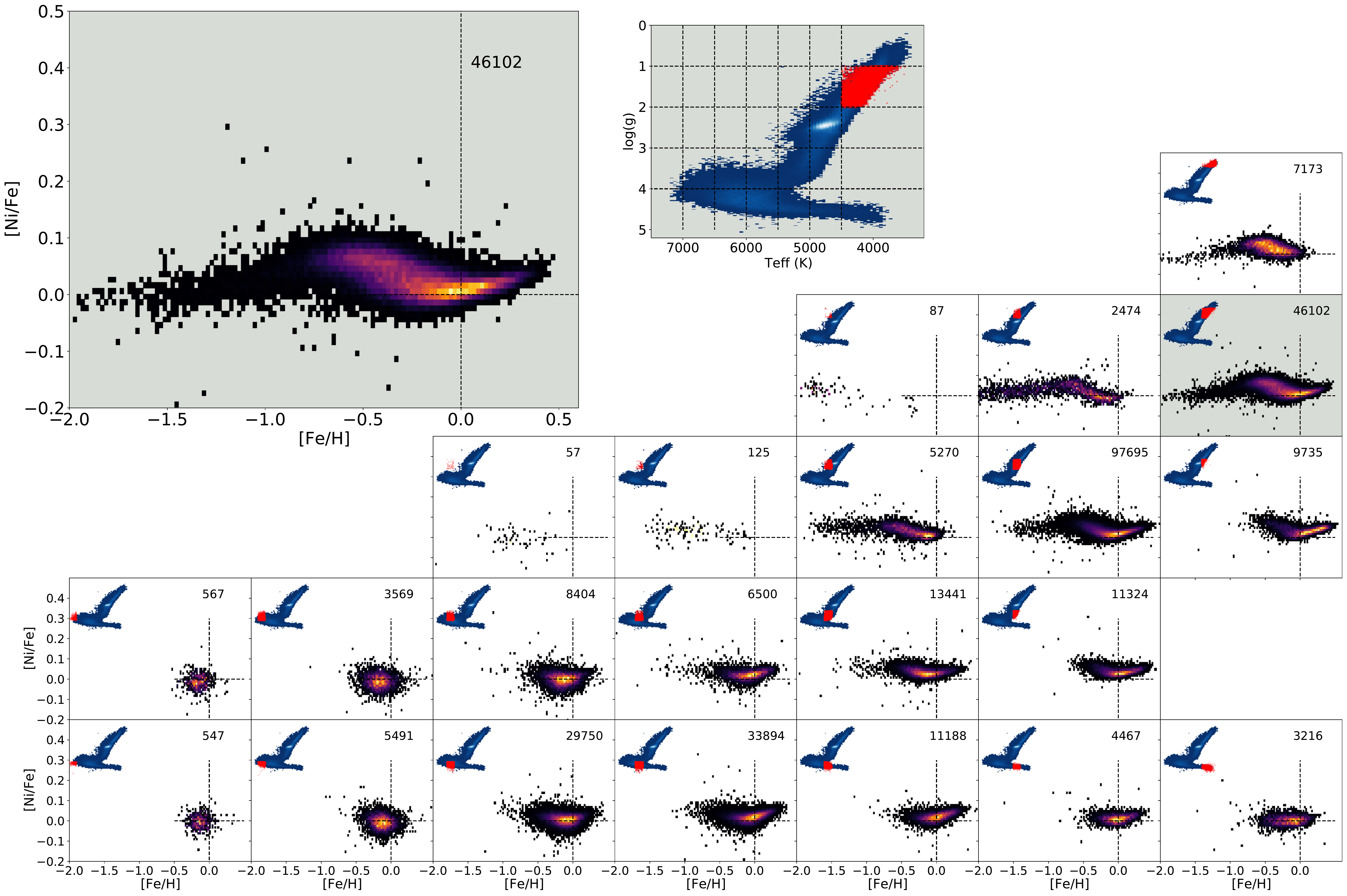}
\caption{\label{kiel_observed_NiFe}Top: $\nife$ vs. $\feh$ for 
the training sample. Bottom: $\nife$ vs. $\feh$ for $301\,076$ stars 
of the observed sample with \snr>30, RAVE DR6 'n\&o' classification, 
and parallax errors lower than $20\%$. For each panel, 
we overplotted a $\teff-\logg$ diagram with the location of the plotted stars 
marked in red.}
\end{figure*}

In this section, we present chemical abundance patterns 
of $\mgfe$, $\sife$, $\alfe$, and $\nife$ as a function of $\feh$ 
in the training and observed samples (\snr>30 and "n" stars). 
\figurename~\ref{kiel_observed_MgFe} and \figurename~\ref{kiel_observed_SiFe} 
present $\mgfe$ and $\sife$ abundances patterns for $301\,076$ stars. 
The trends of both elements look pretty similar to the trends 
of $\alpham$ presented in \figurename~\ref{kiel_observed_AlphaM}, Si and 
Mg being $\alpha$-elements. In \figurename~\ref{kiel_observed_AlFe}, 
we present the chemical abundance patterns of $\alfe$ of the same $301\,076$ stars. 
For $\feh>-1\,$dex, $\alfe$ behaves like an $\alpha-$element 
(consistent with previous findings in the literature, see for example 
\citealt{smiljanic2017}). For $\feh<-1$, we can see that the $\alfe$ ratio 
drops to solar -- and even down to negative ratios. It is mainly driven by the very 
few stars we have in the training sample exhibiting low-$\alfe$ ratios. 
We ought to be particularly careful when using such $\alfe$ abundances. 
In \figurename~\ref{kiel_observed_NiFe}, we present $\nife$ ratios for 
$301\,076$ stars. This ratio is rather flat with $\feh$, as is expected 
for such an Fe-peak element.

\end{document}